# A Comprehensive Review of Techniques, Algorithms, Advancements, Challenges, and Clinical Applications of Multi-modal Medical Image Fusion for Improved Diagnosis

MUHAMMAD ZUBAIR [1], MUZAMMIL HUSSAIN [2], MOUSA AHMAD AL-BASHRAWI [1,3], MALIKA BENDECHACHE [4], MUHAMMAD OWAIS [5*]

[1]*Interdisciplinary Research Center for Finance and Digital Economy, King Fahd University of Petroleum and Minerals, Dhahran, 31261, Saudi Arabia.*
[2]*Department of Software Engineering, Faculty of Information Technology, Al-Ahliyya Amman University, Amman, Jordan.*
[3]*Department of Information Systems and Operations Management, King Fahd University of Petroleum and Minerals, Dhahran, 31261, Saudi Arabia.*
[4]*ADAPT Research Centre, School of Computer Science, University of Galway, H91 TK33 Galway, Ireland.*
[5]*Department of Mechanical and Nuclear Engineering, Khalifa University, Abu Dhabi, 127788, United Arab Emirates.*

[*]*muhammad.owais@ku.ac.ae*

**Abstract:** Multi-modal medical image fusion (MMIF) is increasingly recognized as an essential technique for enhancing diagnostic precision and facilitating effective clinical decision-making within computer-aided diagnosis systems. MMIF combines data from X-ray, MRI, CT, PET, SPECT, and ultrasound to create detailed, clinically useful images of patient anatomy and pathology. These integrated representations significantly advance diagnostic accuracy, lesion detection, and segmentation. This comprehensive review meticulously surveys the evolution, methodologies, algorithms, current advancements, and clinical applications of MMIF. We present a critical comparative analysis of traditional fusion approaches, including pixel-, feature-, and decision-level methods, and delves into recent advancements driven by deep learning, generative models, and transformer-based architectures. A critical comparative analysis is presented between these conventional methods and contemporary techniques, highlighting differences in robustness, computational efficiency, and interpretability. The article addresses extensive clinical applications across oncology, neurology, and cardiology, demonstrating MMIF's vital role in precision medicine through improved patient-specific therapeutic outcomes. Moreover, the review thoroughly investigates the persistent challenges affecting MMIF's broad adoption, including issues related to data privacy, heterogeneity, computational complexity, interpretability of AI-driven algorithms, and integration within clinical workflows. It also identifies significant future research avenues, such as the integration of explainable AI, adoption of privacy-preserving federated learning frameworks, development of real-time fusion systems, and standardization efforts for regulatory compliance. This review organizes key knowledge, outlines challenges, and highlights opportunities, guiding researchers, clinicians, and developers in advancing MMIF for routine clinical use and promoting personalized healthcare.

*keywords:* Multi-modal imaging modalities, Image fusion, Medical imaging, Healthcare, Computer-aided diagnosis, Artificial intelligence, clinical decision support system.

## 1. Introduction

Advanced healthcare has relied on medical imaging because of the non-invasive visualization of internal organs, tissues, and physiological processes. Medical imaging modalities have come a long way since the discovery of X-rays, and now involve a broad range of techniques demonstrating structural and functional insights [1, 2]. Besides aiding in diagnosis, imaging has become essential for directing and planning surgical procedures, assessing therapy efficacy, facilitating real-time monitoring, and advancing personalized medicine. Despite technological advances, many diseases remain too complex for a single imaging technique to fully capture. To address this, multi-modal medical image fusion (MMIF) integrates different imaging data into a single, clinically useful representation [3]. Conventional medical imaging modalities such as Computed tomography (CT) and magnetic resonance imaging (MRI) have been widely used in medical treatment. These modalities have developed and improved their capabilities to provide previously unattainable tissue contrast and spatial resolution [1]. However, currently, structural imaging is frequently augmented with functional imaging techniques like as positron emission tomography (PET) and single-photon emission computed tomography (SPECT), which provide metabolic and molecular findings that improve the diagnosis process. In addition, ultrasound imaging is still vital in cardiology because of its safety and adaptability, point-of-care diagnostics, and prenatal care [4, 5]. When combined, these modalities have significantly advanced our understanding of a variety of conditions, including cardiovascular [6–8], neurological [9–11], and cancerous conditions [12–16]. This has improved patient outcomes by enabling clinicians to detect anomalies earlier, devise therapies with greater precision, and monitor therapy effects more accurately. Regardless of these advancements, reliance solely on a single modality possesses intrinsic constraints. Although CT is an excellent tool for identifying lesions or structural abnormalities and excels at exhibiting anatomical detail, it offers limited insight into functional or metabolic changes [17]. Despite MRI's exceptional soft-tissue contrast, it may overlook tiny vascular lesions or microcalcifications identifiable on CT and fail to capture metabolic dynamics detectable by PET. PET imaging, on the other hand, emphasizes metabolic activity that can identify cancers or follow the course of diseases. However, its spatial resolution is frequently lower than most anatomical modalities [18]. Although accessible and affordable, ultrasound is operator-dependent and limited by acoustic windowing problems. A single approach runs the risk of missing important information required for accurate diagnosis, disease staging, or therapy assessment, and no single modality thoroughly addresses the complex nature of disease pathophysiology. These limitations are particularly evident in complicated illnesses like cancer, where accurate anatomical localization and metabolic characterization of lesions are essential for accurate staging and treatment planning. A clinician's skill to swiftly modify therapies, predict responses, or customize interventions may be hindered by relying only on one modality, which may produce incomplete data. To overcome these constraints, MMIF has emerged as a potent strategy. Instead of looking at isolated data sources, multi-modal data fusion can integrate the respective strengths into a single, enriched dataset. For instance, the combination of PET and CT, now standard in oncological imaging, capitalizes on PET's functional sensitivity to abnormal metabolism and CT's anatomical clarity, producing fused images that have significantly improved tumor detection, characterization, and treatment planning. Similarly, PET/MRI systems provide unparalleled soft-tissue contrast and metabolic insights, expanding applications into neurological and musculoskeletal imaging. The integration of structural and functional data fusion improves diagnostic accuracy, treatment planning, prognostic evaluation, reduces the necessity for invasive procedures, provides intervention guidance, and facilitates more refined therapeutic decisions, such as differentiating viable tumor tissue from post-therapeutic scar tissue or differentiating regions of ischemic tissue from hemorrhage in stroke management. The current advancements in artificial intelligence (AI) and deep learning have substantially improved the robustness and reliability of fusion algorithms, streamlining the integration process, reducing artifacts, quantitative analysis, integration with Big Data and AI pipelines, and ultimately aiding in more objective and reproducible assessments. Multi-modal image fusion helps clinicians navigate intricate clinical scenarios, tailor interventions with surgical precision, and forecast patient-specific outcomes. In the era of precision medicine, the synergy derived from combining modalities facilitates the identification of imaging biomarkers that correlate with specific disease subtypes, genetic profiles, or therapeutic susceptibilities. As machine learning–driven approaches become more sophisticated with advancements in computational power, it is increasingly feasible to incorporate not only multiple imaging sources but also genomic, proteomic, and clinical data streams, forging a comprehensive patient profile that guides highly individualized care. However, challenges remain, such as achieving optimal image registration, managing large and heterogeneous datasets, ensuring cost-effectiveness, and extending access to resource-limited settings. However, improved computational capability has accelerated the optimization of machine learning algorithms, allowing real-time data processing, large-scale population studies, and global health initiatives. The ongoing research and continuous technological refinement hold great promise for bringing multi-modal fusion more fully into everyday clinical practice. The objective of this review is to present a comprehensive and compelling exploration of MMIF, highlighting its techniques, clinical applications, and advancements in revolutionizing modern diagnostic and therapeutic paradigms. It delves into the principles and methodologies of MMIF, discussing the types of fusion techniques and the algorithms that enable the integration of diverse imaging data.

This review systematically explores the many dimensions of MMIF, offering a clear and comprehensive contribution to the field. It aims to provide readers an in-depth, state-of-the-art understanding of MMIF technologies, with a strong emphasis on their transformative potential in enhancing diagnostic precision. Moreover, it critically discusses the MMIF techniques in medical imaging, demonstrating how these approaches support therapeutic planning, and improve disease monitoring. A particular focus is placed on the role of MMIF in advancing computer-aided diagnosis (CAD) systems [19–24], highlighting its ability to integrate fused imaging data to enhance clinical decision-making in complex scenarios. Finally, it delve into the current MMIF challenges, explore their impact on widespread adoption, and provide an in-depth analysis of emerging trends and future directions shaping the evolution of MMIF.

## 2. Overview of medical imaging modalities

Medical imaging has evolved significantly over the past several decades, revolutionizing patient care through ever-improving capabilities to visualize anatomical structures and physiological processes [25]. Medical imaging modalities have become vital technologies in modern healthcare as they allow doctors to gain insight into organs and physiological processes with incredible precision. Each imaging modality, ranging from traditional X-rays to advanced techniques such as diffusion tensor imaging and optical coherence tomography (OCT), has distinct advantages in terms of resolution, tissue contrast, functional information, and therapeutic applicability [26–28].

Each modality has been tailored to capture specific types of functional or anatomical data, satisfying a range of therapeutic and diagnostic needs. Although functional imaging methods such as PET and functional magnetic resonance imaging (fMRI) provide information about neural function and metabolic activity, structural imaging modalities such as CT, MRI, and X-rays offer high-resolution views of anatomical details. Clinicians have become able to recognize, diagnose, and manage a broad spectrum of diseases with formerly uncommon rapidity and preciseness thanks to this variety. Additionally, improvements in disciplines like neurology, cancer, and cardiology, where precise assessment of healthy and sick tissue is crucial for directing interventions have been accelerated by the growing availability of sophisticated scanners and cutting-edge imaging procedures [29–31].

In this section, we provide a comprehensive overview of the most widely employed medical imaging modalities, highlighting their fundamentals, clinical applications, and inherent limitations; the distinctions between structural [32] and functional imaging [33, 34] approaches, as well as specialized techniques that provide additional details about tissue organization and pathology [35]. By highlighting the corresponding benefits of different modalities, we seek to emphasize the justification for multi-modal image fusion, which incorporates complementary datasets to enhance diagnostic confidence and treatment decision-making.

## 2.1. Structural Imaging

Structural imaging modalities remain the foundational tools in diagnostic radiology, offering detailed anatomical visualization essential for disease detection, diagnosis, and treatment planning. Techniques such as X-ray and CT provide high-resolution images of bones and dense tissues, while MRI offers superior soft tissue contrast without ionizing radiation. SPECT, though primarily functional, also contributes to structural insights when combined with CT. Ultrasound imaging, valued for its real-time capability and safety profile, is widely used in obstetrics, cardiology, and abdominal imaging. Together, these modalities form a comprehensive suite that enables clinicians to assess internal structures with precision and guide appropriate interventions.

### 2.1.1. X-Ray

One of the earliest and most popular methods in medical diagnostics is X-ray imaging, which uses the characteristics of electromagnetic radiation to show the inside organs of the human body. X-rays are essentially ionizing radiation with a wavelength between 0.01 and 10 nanometers. These high-energy waves are especially well-suited for taking pictures of bones, tissues by operating on the principle of differential absorption. X-ray image is acquired by directing x-ray beam towards the body, it interacts with tissues in one of three ways: absorption, scattering, or transmission. Dense materials such as bones absorb more X-rays, appearing white or light gray on the resulting radiograph, while softer tissues, which absorb fewer X-rays, appear in shades of gray [36]. Air-filled spaces, such as the lungs, allow most X-rays to pass through, resulting in darker regions on the image. This differential absorption creates a contrast that forms the basis of the diagnostic image. Since its discovery the technique has undergone continuous refinements to improve image quality, reduce patient risk, and broaden clinical applications. Although X-rays involve ionizing radiation, the dose received from a single diagnostic procedure is generally considered low, particularly if modern technology and safety protocols are followed [36]. Protective gear such as lead aprons, thyroid shields, and gonadal shields is standard in many procedures to safeguard sensitive tissues. Recent advances in dosimetry devices have also improved tracking of cumulative exposures, allowing physicians to balance diagnostic yield against potential harm more effectively [37].

Early X-ray systems employed photographic plates, but modern digital radiography relies on sophisticated sensors and digital processing algorithms to enhance image clarity. The use of digital detectors has reduced the repeat exposures by enabling near-instant feedback on image quality. The core physics remain the same, but improvements in detector design and software have elevated diagnostic accuracy and diminished the risks associated with retakes or poor-quality films. The shift from analog film to computed radiography transformed radiological workflows, while the more recent transition to fully digital radiography has further streamlined the process. Current research focuses on high-resolution detectors, dose-reduction algorithms, and portable X-ray units, making the modality increasingly accessible in resource-limited settings [37]. Integration of AI has also assisted clinicians in faster, more consistent X-ray image interpretations [38]. In recent times, X-ray imaging fulfills a variety of purposes, including monitoring lung conditions, guiding surgical procedures, and looking into bone fractures. This imaging technique is notable for its affordability, speed, and diagnostic accuracy, especially in emergency situations. X-ray imaging is particularly effective for diagnosing fractures, joint dislocations, lung infections, dental issues, and certain tumors. It is also used in procedures like fluoroscopy, where real-time X-ray imaging helps guide interventions such as catheter insertions or barium studies. Moreover, it is particularly suited for diagnosing conditions involving dense structures like orthopedic assessments rely on plain radiographs for the initial detection of fractures, joint dislocations, or degenerative changes. Additionally, fluoroscopy, which provides real-time X-ray visualization, aids in guided procedures like cardiac catheterization, gastrointestinal contrast studies, and placement of medical devices. These various uses highlight why X-ray imaging is still essential in almost all hospitals across the globe.

The current research has focused on refining hardware and computational techniques to improve image quality while minimizing radiation exposure. Innovations in contrast agents [39], detector sensitivity [40], and portable technologies [41] are expanding its utility, particularly in remote and underserved areas. The integration of machine learning has further revolutionized image interpretation [38], reducing diagnostic errors and delays. Ongoing developments, supported by advances in AI and sophisticated detection systems, ensure that X-ray imaging remains essential to contemporary healthcare. Its necessity in medical practice will be maintained by striking a balance between these developments, safety concerns, and clinical requirements.

2.1.2. Computed Tomography scan

CT generates three-dimensional reconstructions with superior contrast between soft tissue and bone and is based on X-ray attenuation through tissues [42–44]. The concept behind CT scanning, relies on the attenuation of X-rays as they traverse the body from multiple angles [45]. The CT technology creates cross-sectional images that can be combined to form a three-dimensional view of anatomical structures. It provides far greater detail than conventional radiographs i.e., X-ray imaging, especially for soft tissue differentiation and complex anatomical regions [46].

CT machines operate by using an X-ray tube and a set of rotating detectors to capture detailed images of the body. When X-rays pass through tissues they are absorbed at different levels based on tissue's such as bone, muscle, and organs's density. The detectors measure these variations and convert them into signals, which are processed by software to create highly detailed digital images [42–44]. The technology has advanced significantly, multi-slice or multi-detector CT scanners can now capture entire regions of the body in a fraction of the time. This speed not only increases patient throughput but also reduces motion artifacts, thus improving diagnostic accuracy. Clinically, this imaging methodology has become indispensable as it is routinely used in emergency settings to identify traumatic injuries in the brain, lungs, abdomen, and pelvis. In oncology, CT assists in determining tumor location, size, and involvement of surrounding structures, guiding for surgery and radiation therapy [3]. Cardiology also benefits from CT angiography, enabling detailed visualization of coronary arteries to detect blockages. In addition, CT aids in real-time guidance for biopsy procedures and minimally invasive interventions. Such precision means that clinicians can navigate instruments to the targeted lesion quickly and accurately, reducing procedural risks and improving outcomes.

Advanced CT scanning aimed at reducing radiation exposure by iterative reconstruction algorithms for improved image quality by reducing noise and enhancing clarity, enabling lower radiation doses without sacrificing diagnostic accuracy. Dual-energy CT offers enhanced tissue differentiation by capturing data at two X-ray energy levels. Emerging technologies such as spectral CT and photon-counting CT, still under development, hold the potential to provide highly detailed insights into tissue composition. Cutting-edge methods, such helical or spiral CT, enable quicker imaging with less motion artifacts, which makes them suitable for capturing dynamic processes like pulmonary or cardiac activity [47]. It is a cornerstone for vascular imaging in conjunction with contrast agents, oncologic staging, and trauma evaluation because of its rapid acquisition and wide availability. It is widely used for planning surgical interventions and guiding minimally invasive procedures, such as biopsies and ablations.

The future of CT scanning is being shaped by integrating AI, which is enhancing its diagnostic and clinical capabilities. AI could assist in automated lesion detection, organ segmentation, and quantify disease progression with minimal manual input. Combined with advancements in real-time 3D imaging, these technologies have the potential to revolutionize complex surgeries and interventional procedures by increasing precision and safety. Additionally, ongoing research focuses on identifying CT-based biomarkers to predict patient outcomes and tailor treatments for greater effectiveness. The improvement in resolution, speed, and radiation optimization protocols, CT scanning continues to play a pivotal role in healthcare, driving advancements in both diagnostics and personalized medicine.

### 2.1.3. Magnetic Resonance Imaging

MRI has become indispensable for structural imaging due to its non-invasive nature, superb soft tissue contrast, and absence of ionizing radiation. In contrary to CT, which utilizes X-rays, MRI positions and manipulates hydrogen nuclei in water molecules throughout the body using radiofrequency pulses and a strong magnetic field [48]. The varying concentrations of water and fat in different tissues, MRI generates highly specific contrast, allowing for unparalleled differentiation between soft tissues. The signals that these nuclei generate when they realign with the external magnetic field are converted into incredibly detailed anatomical images. Clinicians can more precisely examine anatomical structures [49], such as the brain's gray and white matter or the soft tissues in musculoskeletal applications, simply by changing the RF pulse sequences that highlight distinct tissue characteristics, to highlight differences in soft tissue composition, water content, and biochemical environment [50]. The versatility and non-ionizing nature of MRI make it a preferred choice for applications ranging from neurological disorders to musculoskeletal injuries. The signals that these nuclei generate when they realign with the external magnetic field are converted into incredibly detailed anatomical images. Advancements in MRI include higher field strengths (e.g., 11.7 Tesla) [51], which provides provide significantly enhanced signal-to-noise ratios and improved spatial resolution, enabling the visualization of fine anatomical details. This has allowed researchers to detect subtle microstructural changes and mapping neural pathways with exceptional precision. Moreover, the novel pulse sequences has reduced scan times and improved motion correction. More precise and repeatable tissue characterization is made possible by quantitative mapping techniques, such as T1 and T2 relaxation time mapping, proton density mapping, and quantitative susceptibility mapping, which aid in the early detection and tracking of conditions like cancer, multiple sclerosis and fibrosis. These developments [52] had pushed the limits, transforming MRI into a valuable instrument in clinical and research setup.

### 2.1.4. Ultrasound

Ultrasound imaging operates by transmitting high-frequency sound waves into the body and interpreting the echoes to construct real-time cross-sectional images of soft tissues [53–55]. These sound waves are harmless and non-invasive, making ultrasound a preferred choice for many diagnostic and therapeutic applications. The concept enables medical practitioners to examine organs, blood flow, and embryonic growth without subjecting patients to ionizing radiation, as in CT or X-ray scans, or requiring the occasionally time-consuming and costly MRI or PET treatments. The technology is particularly effective at showing organs, muscles, and tendons moving, which offers important insights into dynamic processes like joint movements or heartbeats. Additionally, Doppler ultrasound [56] broadens its use by making it possible to evaluate blood flow direction and velocity, which is essential for vascular disease diagnosis and blood circulation monitoring.

The importance of ultrasound in modern healthcare is mostly because of its safety profile, as it does not use ionizing radiation, it can be used on susceptible groups including youngsters and pregnant patients. Its real-time capabilities are advantageous for guided treatments such as fluid drainage and biopsies. Additionally, in emergency or critical care settings, bedside examinations, often known as point-of-care ultrasound, speeding up decision-making enabling the clinicians to quickly check a patient's health. The improved image resolution while low-frequency probes delve deeper into the body, high-frequency transducers provide comprehensive views of surface components. Clinicians are able to measure tissue stiffness using shear-wave and strain elastography, which helps in the detection of medical conditions aiding in the diagnosis of diseases [57]. Additionally, automated measurements and diagnosis have been made possible by the incorporation of AI into ultrasound equipment, speeding up diagnostic procedures in hectic clinical settings. Portable ultrasound devices have also grown in popularity, further democratizing access to diagnostic imaging in remote or resource-limited settings. When compared to other imaging modalities, ultrasound stands out for its cost-effectiveness, ease of use, and lack of ionizing radiation. It can be performed frequently with minimal risk, and tailored to meet evolving clinical questions in real time. Through improved transducer designs, advanced image processing, and integration of machine learning, the utility of ultrasound continues to expand beyond traditional applications.

## 2.2. Functional Imaging

### 2.2.1. functional MRI

Functional magnetic resonance imaging (fMRI) is a powerful neuroimaging method that captures dynamic brain activity by measuring changes in blood oxygenation. Unlike structural MRI, which focuses on the static anatomy of the brain, fMRI capitalizes on the Blood Oxygen Level-Dependent (BOLD) signal, whereby localized alterations in cerebral blood flow and oxygen consumption serve as indirect indicators of neural activation [58]. One central principle of the BOLD mechanism is that the magnetic properties of hemoglobin vary depending on whether it is oxygenated or deoxygenated. Oxygenated hemoglobin is essentially diamagnetic, causing minimal distortion in the surrounding magnetic field, whereas deoxygenated hemoglobin is paramagnetic, generating stronger local magnetic field inhomogeneities [58]. These subtle differences in magnetic susceptibility yield measurable contrasts in the MRI signal, allowing researchers to infer which neuronal populations are engaged at any given time. When a specific brain region becomes more active, the local vascular system compensates with a surge of oxygen-rich blood, creating detectable fluctuations in the magnetic resonance signal. This principle allows researchers and clinicians to map brain functions associated with various cognitive, sensory, and motor tasks [59]. Thus, the BOLD signal acts as an indirect marker of neuronal activity, providing spatiotemporal maps that represent functional changes in the brain. Despite being an indirect measure, BOLD fMRI has proven sufficiently robust and sensitive to capture real-time changes in neural activity across diverse cognitive and sensory domains.

One of the key advantages of fMRI is its non-invasive nature coupled with relatively high spatial resolution [60]. In contrast to PET, fMRI does not rely on radioactive tracers, and it can be repeated multiple times to chart changes in brain activity over weeks or months [61]. Additionally, fMRI has gained clinical relevance for pre-surgical planning, allowing neurosurgeons to identify and preserve functionally critical areas during tumor resection or epilepsy surgery. The method's reproducibility and capacity to integrate with other techniques, such as magnetic resonance spectroscopy or diffusion tensor imaging, underscore its significance in both research and clinical settings. Despite these advantages, fMRI is not without its limitations. Researchers and clinicians must navigate several technical and interpretative challenges to maximize its utility. fMRI faces several challenges that have spurred ongoing research. One issue is susceptibility to motion artifacts, which can distort the BOLD signal and confound interpretations. Another challenge stems from the complexity of data analysis: the statistical pipelines for fMRI involve numerous steps, including pre-processing and modeling, making the technique prone to false positives or erroneous inferences if not performed carefully. Moreover, the BOLD signal itself is only a proxy for neuronal firing, and it is influenced by vascular factors that may not always mirror direct neuronal activity.

Recent advancements in fMRI seek to address these challenges through hardware and software improvements. High-field MRI scanners (e.g., 7, 11.7 Tesla) [51] offer enhanced signal-to-noise ratios, resulting in finer spatial resolution. Advanced pulse sequences also help mitigate artifacts, while novel analysis pipelines integrate machine learning algorithms to detect and classify subtle patterns in large-scale fMRI datasets. Multi-echo acquisitions [62], real-time feedback fMRI, and the combination of fMRI with other imaging modalities [63] further broaden the potential applications.

Table 1. Comparison of structural imaging modalities, highlighting their principle, resolution, radiation exposure, sensitivity, cost, availability, applications, prons, and limitations

| Modality | X-Ray | CT | MRI | Ultrasound |
| --- | --- | --- | --- | --- |
| Sample image | 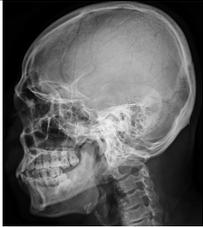 | 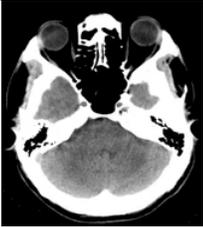 | 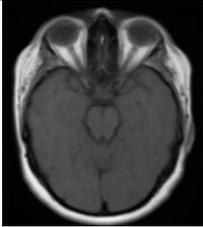 | 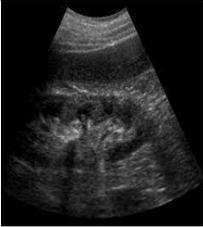 |
| Principle | Absorption of X-rays by tissues; denser tissues absorb more, creating contrast. | Rotating X-rays and detectors create cross-sectional images; data reconstructed digitally. | Uses strong magnetic fields and radiofrequency pulses to create detailed images. | Transmits high-frequency sound waves; echoes create real-time images. |
| Spatial Resolution | Moderate to high | High | Low to moderate. | Low to moderate |
| Temporal Resolution | Very high (real-time imaging). | Moderate (seconds). | Low to moderate | High (real-time imaging). |
| Radiation Exposure | Ionizing radiation exposure. | Higher radiation dose compared to X-rays | None (no ionizing radiation) | None (no ionizing radiation) |
| Sensitivity | Limited for soft tissue imaging | High sensitivity for anatomical imaging | High sensitivity for soft tissue | Moderate sensitivity for soft tissues |
| Cost | Low | Moderate to high | High | Low |
| Availability | Widely available | Available in most medical centers. | Requires advanced medical facilities. | Widely available, even in portable settings. |
| Applications | Bone fractures, chest imaging. | Trauma, cancer detection, vascular diseases. | Neurological, musculoskeletal, and soft tissue imaging. | Obstetrics, cardiology, abdominal imaging. |
| Advantages | Quick, inexpensive, widely available, good for bone imaging. | High spatial resolution, 3D imaging, detailed anatomy visualization. | Excellent soft tissue contrast, no ionizing radiation | Non-invasive, portable, no radiation, real-time imaging. |
| Limitations | Limited soft tissue contrast; ionizing radiation exposure. | High radiation dose; limited for functional imaging. | Longer scan times, not portable, contraindicated for metal implants. | Limited penetration depth, lower resolution, operator-dependent quality. |

Future directions in fMRI research involve personalized and hybrid approaches [64, 65]. Efforts to develop task-free or resting-state fMRI as biomarkers for psychiatric and neurodegenerative disorders are ongoing [66]. The integration of genetic data and novel computational models may accelerate precision medicine initiatives by linking functional brain signatures with specific pathologies. Emerging fields such as connectomics are also utilizing fMRI data to map the complex functional wiring of the brain, complementing traditional structural analyses. Compared to other imaging modalities, fMRI is invaluable due to its balance of spatial resolution, temporal resolution, and non-invasive nature. It avoids the ionizing radiation inherent in PET and yields a richer spatiotemporal map of brain function than techniques such as EEG [67], which, while having high temporal resolution, lacks spatial specificity. As brain research moves further into precision medicine and detailed network analyses, fMRI's ability to resolve functional networks within the living human brain [68, 69] will remain significant for clinical and basic neuroscience investigations [70].

2.2.2. Positron emission tomography

PET is a functional imaging modality that allows the visualization of metabolic processes and biochemical pathways within the human body in real time [71]. PET imaging relies on the administration of radioactive tracers that emit positrons, enabling the detection of specific physiological functions [72]. This approach has become pivotal in neurology for mapping regional brain metabolism, and in cardiology to assess myocardial viability.

The core principle of PET involves labeling biologically active molecules with positron-emitting radionuclides [73]. These radiotracers participate in normal metabolic processes, accumulating in areas of high activity. As positrons emitted by the radionuclide interact with electrons, they annihilate and produce high-energy photons detectable by a ring of gamma detectors. Sophisticated reconstruction algorithms then convert these signals into tomographic images that map tracer distribution throughout the body. This metabolic snapshot reveals essential functional information, particularly relevant for cancer diagnosis, where tumors often exhibit accelerated metabolic rates.

PET holds significant clinical importance due to its exceptional ability to detect subtle biochemical changes at the molecular level, which are often precursors to anatomical alterations. This specificity makes PET an invaluable tool in diagnosing, monitoring, and understanding various diseases. In oncology, for instance, 18F-fluorodeoxyglucose (FDG) PET scanning is widely used as a gold standard for tumor detection, staging, and treatment monitoring. By tracking the uptake of FDG, a glucose analog, PET can identify regions of high metabolic activity that are characteristic of malignant tumors. The modality has also been critical for exploring neurodegenerative disorders, such as Alzheimer's disease, by using tracers that bind to amyloid or tau proteins in the brain. Additionally, cardiac PET offers insights into myocardial perfusion and viability, guiding therapeutic decisions for patients with ischemic heart disease [74]. This versatility across medical specialties underscores PET's unique ability to illustrate functional processes rather than just static anatomy.

Despite its strengths, PET faces challenges i.e., relatively short half-lives of many positron-emitting isotopes, which complicates tracer synthesis and limits the geographical reach of PET scanners to specialized facilities equipped with a cyclotron or generator. Cost is another hindrance: maintaining cyclotrons, producing tracers, and operating PET scanners demand substantial resources [75]. Furthermore, the use of ionizing radiation restricts the frequency of scans that patients can safely undergo, which can be particularly limiting in pediatric populations. In recent years, technological advancements have overcome PET limitations [76]. For instance, PET/CT systems integrate functional and anatomical data [77], providing high-resolution images that correlate metabolic activity with structural context in a single session. Improvements in detector materials have enhanced sensitivity and resolution, enabling lower tracer doses and reduced scanning times. Emerging PET/MRI hybrid systems [78] promise even greater anatomical and soft-tissue contrast while minimizing radiation exposure.

Future directions for PET [79] research encompass the design of novel radiotracers that target specific molecular pathways. AI is revolutionizing PET imaging by enhancing various stages of the image processing pipeline [80, 81], from acquisition to analysis. AI-driven algorithms, particularly those based on deep learning, are being used to denoise images, improve spatial resolution, and correct for motion artifacts, segmentation and quantification, resulting in clearer and more accurate images with reduced scanning times. These advancements are especially beneficial in clinical settings, where time and precision are critical. PET's importance derives from its capacity to detect functional abnormalities before structural changes appear, offering a more proactive diagnostic approach. By complementing anatomical imaging with detailed metabolic and molecular data, PET plays a pivotal role in early disease detection, therapy planning, and monitoring patient response to interventions makes PET an indispensable tool in modern healthcare [82].

### 2.2.3. Single-photon emission computed tomography

SPECT is a functional imaging modality that visualizes physiological processes by detecting gamma-ray photons emitted from radioactive tracers within the body [83]. The fundamental principle behind SPECT involves administering radiopharmaceuticals labeled with gamma-emitting isotopes i.e., technetium-99m [84]. After tracer uptake, gamma cameras rotate around the patient, collecting photons from multiple angles [85]. These projections are then reconstructed into cross-sectional images that reflect tracer distribution. The distribution pattern is indicative of functional status in target tissues, enabling clinicians to evaluate blood flow, metabolic activity, or receptor density with a relatively high level of sensitivity. This functional perspective is especially advantageous in identifying abnormalities that might not be visible through pure structural imaging.

SPECT is widely employed in myocardial perfusion imaging, offering prognostic insights by highlighting areas of reduced blood flow in the heart. It also aids in diagnosing neurological conditions by visualizing cerebral blood flow changes in disorders like dementia or stroke. While it is invaluable for tracking disease progression and treatment outcomes often alongside MRI or CT/SPECT faces certain drawbacks. Its lower spatial resolution, susceptibility to photon scatter and attenuation, and the necessity of ionizing radiation limit frequent scanning. Moreover, the high costs of equipment and the need for specialized radiopharmaceuticals can restrict its broader adoption in resource limited settings.

Recent advancements have sought to enhance SPECT's resolution and quantitative capabilities. Innovations in detector materials, such as cadmium-zinc-telluride [86], have improved sensitivity and image clarity, allowing for shorter scan times and lower tracer doses. Hybrid systems, combining SPECT with CT (SPECT/CT), integrate functional and anatomical information into a single examination, providing a more comprehensive overview of disease. Image processing tools incorporating AI are also emerging to facilitate more accurate interpretation and to automate tasks like segmentation and attenuation correction.

Future directions for SPECT center on developing novel radiopharmaceuticals for molecular imaging, potentially expanding its clinical applications beyond traditional perfusion and metabolic studies [87]. Personalized medicine is set to gain significantly from the development of specialized tracers designed to target specific cellular pathways or genetic markers. These tracers enable tailored imaging that aligns with the unique molecular profile of a patient's condition, improving diagnosis and treatment planning. Alongside these advancements, improvements in hardware design and sophisticated reconstruction algorithms are expected to elevate image quality. Enhanced resolution and clarity will allow for more accurate identification and characterization of disease states, supporting earlier detection, precise monitoring, and more effective therapeutic interventions.

### 2.3. Need of multi-modal imaging

Medical imaging techniques offer non-invasive methods to see physiological processes and internal anatomy with exceptional details. Anatomical details are revealed by structural imaging, which facilitates the diagnosis and monitoring of numerous medical conditions. Meanwhile, functional and advanced modalities delve into physiological and molecular processes, enabling earlier and more precise disease detection by identifying subtle biochemical and functional abnormalities often preceding structural changes. Combined, these techniques enhance the diagnostic process and therapeutic judgment and support individualized treatment planning. A couple of decades of advancements in medical imaging techniques have strengthened their usage in clinical care and research contexts [88, 89]. Their application in clinical care and research by achieving higher spatial and temporal resolutions, developing more intricate contrast mechanisms, and integrating multi-modal imaging techniques. These technological advancements facilitate more accurate diagnostics and effective treatment planning.

## 3. Existing multi-modal imaging combinations

Medical imaging has become an indispensable part of modern diagnostics, allowing clinicians to visualize the interior of the human body noninvasively. However, no single imaging modality can perfectly capture all the necessary information required for a comprehensive diagnosis. Each modality, whether it is X-Ray, MRI, US, SPECT or PET it has its unique strengths and inherent limitations [32, 76, 90–92], as mentioned in Tables 1 and 2. Consequently, combining imaging modalities has emerged as a critical strategy to enhance disease detection, improve anatomical and functional visualization, and support more accurate clinical decision-making.

The combination of imaging modalities holds complementary strengths, allowing for enhanced visualization, more accurate diagnosis, and better characterization of diseases. Figure 1 presents examples of MMIF outcomes from recent studies [93, 94]. The top of the figure 1 illustrates the fusion of MRI and PET images, while the bottom one shows MRI and CT fusion using perceptual (PERCE) loss and structural similarity index measure (SSIM) loss. Unlike conventional pixel-level loss functions such as mean squared error, PERCE, and SSIM losses are better suited to preserving both the structural integrity and perceptual quality of the fused images. These loss functions align more closely with human visual perception, effectively capturing the fine details and content structures that are critical for accurate interpretation. PERCE loss emphasizes content-level features and contributes to maintaining the realism of the fused image, while SSIM loss ensures the structural consistency between the fused result and the source images. As a result, the combined use of PERCE and SSIM losses enhances the overall fidelity of the fusion process, producing images that are not only visually coherent but also clinically meaningful.

Figure 2 visualizes an example of fusion results that involves three different imaging modalities [95]. The methodology incorporates the concept of skewness of pixel intensity alongside a novel adaptive co-occurrence filter-based image decomposition optimization model. This approach is designed to enhance the quality of the fused images by effectively capturing complementary features from each modality while preserving structural and intensity information, ultimately resulting in more informative and visually coherent output.

Table 3 summarizes the existing multi-modal imaging combinations currently utilized for the detection and evaluation of diseases affecting major human organs [93, 94, 96–110]. For instance, CT provides high-resolution anatomical detail but offers limited information about tissue metabolism, whereas PET excels in functional imaging but suffers from lower spatial resolution. Integrating PET with CT (PET/CT) allows clinicians to localize metabolic abnormalities within anatomical structures precisely, significantly enhancing diagnostic accuracy, particularly in oncology [111]. Similarly, MRI combined with PET (PET/MRI) merges superior soft-tissue contrast with molecular imaging capabilities, benefiting neurological and cardiovascular applications [100].

Table 2. Comparison of functional imaging modalities, highlighting their principle, resolution, radiation exposure, sensitivity, cost, availability, applications, prons, and limitations

| Modality | fMRI | PET | SPECT |
|---|---|---|---|
| Sample image | 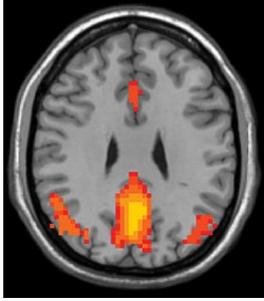 | 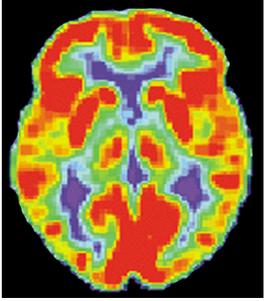 | 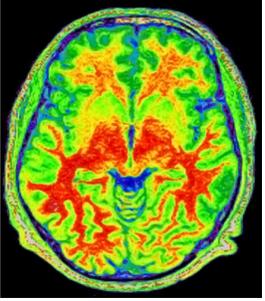 |
| Principle | Measures changes in blood oxygenation (BOLD signal). | Detects gamma rays from injected radiotracer. | Detects gamma rays from a single-photon emitting radiotracer. |
| Spatial Resolution | High (1–3 mm). | Moderate (4–6 mm). | Low to moderate (7–10 mm). |
| Temporal Resolution | Moderate (seconds). | Low (minutes). | Low (minutes). |
| Radiation Exposure | None. | High (due to radiotracer injection). | Moderate (less than PET, but still requires a radiotracer). |
| Sensitivity | Moderate to high for blood flow changes. | High for metabolic and molecular processes. | Moderate to high for functional abnormalities. |
| Cost | High (equipment and operational costs). | Very high (radiotracer production costs). | Moderate to high. |
| Availability | Widely available in advanced medical centers. | Limited due to cyclotron dependency. | More widely available than PET, but less common than fMRI. |
| Applications | Neuroscience, cognitive studies, tumor analysis. | Oncology, cardiology, neurodegenerative diseases. | Cardiology, epilepsy, and some brain disorders. |
| Advantages | Non-invasive, high spatial resolution, real-time imaging. | High sensitivity to metabolic activity, molecular imaging. | Less expensive than PET, suitable for longer imaging durations. |
| Limitations | Sensitive to motion, indirect measurement of neural activity. | Expensive, involves radiation exposure, lower resolution than fMRI. | Lower resolution than PET and fMRI, radiation exposure. |

Beyond PET combinations, ultrasound imaging is frequently paired with other techniques. For example, contrast-enhanced ultrasound can be used alongside CT or MRI to improve lesion characterization in the liver, taking advantage of ultrasound's real-time imaging and CT/MRI's superior anatomical coverage [112]. In interventional procedures, US guidance combined with CT imaging supports both real-time targeting and comprehensive treatment planning.

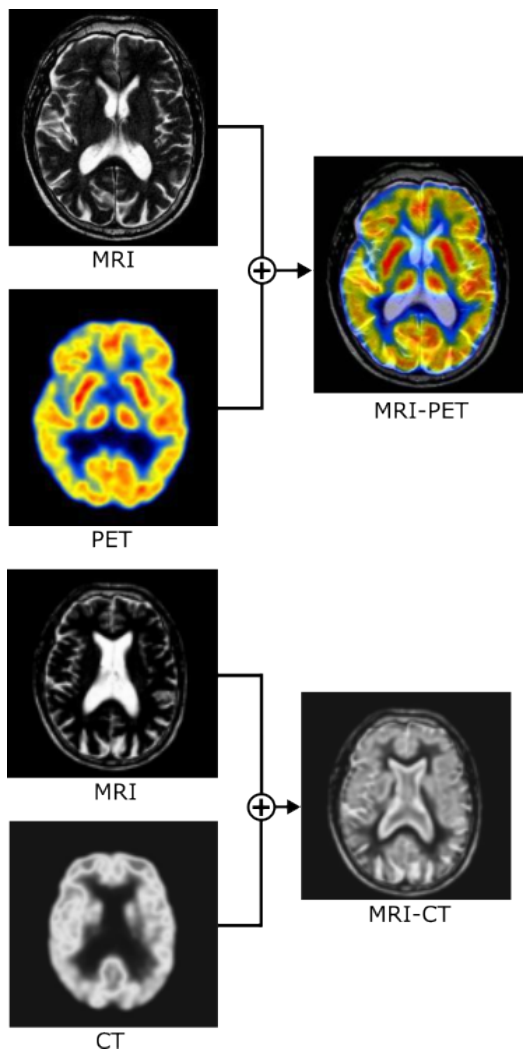

Fig. 1. Visualization examples of multi-modal medical images from recent studies [93, 94].

Hybrid imaging modalities not only improve spatial and functional representation but also address challenges such as artifacts, noise, limited contrast sensitivity, and motion blurring that are characteristic of standalone modalities [113]. For example, in SPECT/CT systems, CT provides attenuation correction to enhance the quantitative accuracy of SPECT imaging, thus improving image quality and diagnostic reliability. The trend toward multi-modal imaging reflects a growing demand for more precise, reproducible, and holistic views of pathology. Moreover, the concept of imaging fusion has expanded into computational approaches, where images acquired from different modalities are registered and combined using advanced algorithms. These fusion techniques can offer even more detailed insights than what hybrid hardware systems alone can provide, further supporting tasks like tumor delineation, treatment planning, and surgical navigation [12].

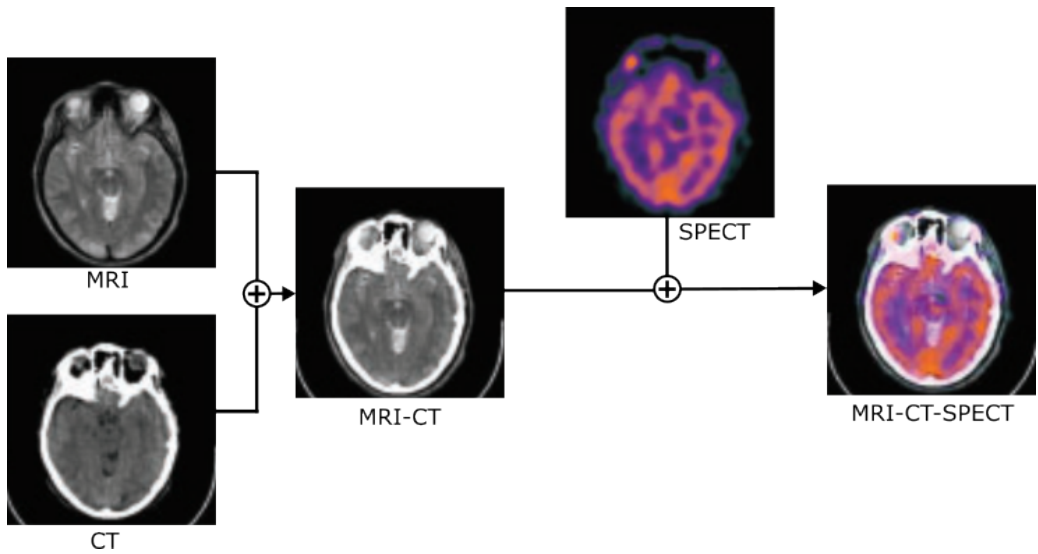

Fig. 2. Sequential fusion process of multi-modal medical images demonstrating the integration of MRI and CT to form MRI-CT, followed by fusion with SPECT to generate a comprehensive MRI-CT-SPECT image [93, 94], enhancing anatomical and functional detail for improved diagnosis.

In summary, the integration of multiple imaging modalities addresses the limitations of single techniques and plays a pivotal role in modern diagnostics and therapeutic monitoring. As technological advances continue, multi-modal imaging is expected to become even more sophisticated, enabling deeper understanding of complex diseases and facilitating precision medicine.

Table 3. Overview of commonly used imaging modalities and their hybrid combinations for enhanced detection of organ-specific abnormalities

| Organ | Commonly used imaging modalities and their functions | Hybrid combination |
|---|---|---|
| Brain | CT: Shows hard tissues of the brain structure | MRI/PET |
| | PET: Assesses brain function by mapping blood flow dynamics and glucose metabolism within active brain tissues. | MRI/SPECT |
| | MRI: Shows the soft tissues of the brain and magnetic activity. | PET/CT |
| Continued on next page | | |

| Organ | Existing imaging modalities and their functions | Hybrid combination |
|---|---|---|
| | SPECT: Measures cerebral blood flow. | CT/SPECT |
| Lungs | X-ray: For cancer, pneumonia, and COPD diagnosis. | PET/CT |
| | CT: Detects tumors, pulmonary embolism, fibrosis in high-risk patients. | |
| | MRI: Identify cystic fibrosis, bronchial carcinoma, and hypertension. | |
| | PET: Diagnoses non-small-cell bronchial carcinoma. | |
| Breast | Mammography: Highlights fibroglandular tissues, masses, calcifications. | Mammography/US |
| | US: Detects breast lesions and abnormalities. | MRI/PET |
| | MRI: Identifies early breast tumors. | US/MRI |
| | PET and SPECT: Evaluates the effectiveness of treatment in breast cancer patients. | |
| Prostate | US: Determines the spread and positioning of cancers within the glands. | MRI/CT |
| | CT: Pre-therapeutic assessment of prostate malignancy and lymph node involve. | PET/CT |
| | PET: Localization of subtle metastatic lesions via metabolic activity imaging. | |
| Liver | US: Non-invasive assessment of suspected metastatic lesions in the liver. | PET/CT |
| | CT: Phase-based evaluation of hypervascular metastatic tumors. | |
| | MRI: High-resolution imaging for lesion detection. | |
| Cardiovascular | CT: Enables imaging of coronary arteries and evaluation of pathological changes. | PET/CT |
| | SPECT and PET: Measures myocardial perfusion, ventricular performance, and detects coronary artery disease. | CT/SPECT |
| Retina | FFA: Visualize retinal blood vessels and detect abnormalities such as leakage, blockage, or neovascularization. | FFA/OCT |
| | OCT: Diagnosis of retinal diseases by detecting structural abnormalities such as edema, thinning, or layer disruptions. | |
| Gastrointestinal | Endoscopoic US: Helps in investigating high-resolution images of the gastrointestinal wall layers and nearby structures for esophageal, gastric, pancreatic, and rectal cancers diagnosis | Endoscopic US/CT |

| Organ | Existing imaging modalities and their functions | Hybrid combination |
|---|---|---|
| | CT: Visualize colorectal, gastric, and pancreatic cancers by evaluating the extent of local invasion and lymph node involvement. | CT/MRI |
| | MRI: Valuable in detailed soft-tissue assessment for colorectal and rectal cancer stage assessment. | MRI/PET |
| | PET: Helps in detecting distant metastases especially in esophageal, gastric, and colorectal cancers. | PET/CT |
| Kidney | US: Helps in detecting cysts, hydronephrosis, and gross renal masses. | US/MRI |
| | CT: Provide detailed anatomical information about lesion size, internal structure (solid vs cystic), vascular invasion, and metastatic spread to assess renal cell carcinoma stage. | CT/MRI |
| | MRI: Provides superior soft tissue contrast to help in detecting vascular involvement and fat-containing tumors. | US/CT |
| | PET: Using novel tracers i.e., 89Zr-DFO-girentuximab for detecting metastatic renal cell carcinoma and evaluating recurrence after nephrectomy | PET/CT |
| Ovary and Uterus | US: Detecting ovarian cysts, fibroids, endometrial abnormalities, and early-stage ovarian and uterine cancers. | US/CT |
| | CT: To assess advanced stages of ovarian or uterine cancers, particularly to evaluate peritoneal spread, lymph node involvement, and distant metastases. | |
| Thyroid | US: Sensitive for detecting thyroid nodules, characterizing benign versus malignant features. | US/CT |
| | CT: Helpful for localizing ectopic parathyroid adenomas in the mediastinum. | PET/CT |
| | MRI: Detects recurrent thyroid cancer or ectopic parathyroid tissue. | MRI/CT |
| | PET: 11C-methionine can help localize recurrent thyroid cancer. | |
| Pancreas | Endoscopic US: Effective for detecting small pancreatic tumors, guiding biopsies, and staging local lymph nodes. | Endoscopic US/CT |
| | CT: Assesses tumor size, vascular involvement, local invasion, and distant metastases for detecting and staging pancreatic adenocarcinoma. | CT/MRI |
| | MRI: Visualize soft-tissue with contrast and functional information without radiation. | MRI/Endoscopic US |

| Organ | Existing imaging modalities and their functions | Hybrid combination |
|---|---|---|
| | MR Cholangiopancreatography (MRCP): visualizes the pancreatic ductal system and biliary tree, which is critical for evaluating ductal obstructions, cystic lesions, and early-stage tumors. | MRI/MRCP |
| | PET: Using 18F-FDG, plays a supportive role in staging, detecting distant metastases, and evaluating recurrence. | PET/CT |
| Bladder | US: For initial assessment of bladder abnormalities. | US/CT |
| | CT: Provides detailed cross-sectional images, aiding in tumor localization and staging. | CT/PET |
| | PET: Evaluate metabolic activity, assisting in detecting metastases. | |

## 4. Types of Multi-modal Image Fusion

The techniques used for medical image fusion can be categorized into the following types based on the level of abstraction at which the fusion occurs: pixel-level, feature-level, hybrid pixel- and feature-level, and decision-level fusion techniques. Figure 3 illustrates the hierarchical categorization of fusion techniques based on the level of abstraction at which the fusion is performed. These techniques are broadly classified into pixel-level, feature-level, and decision-level fusion, each offering distinct advantages and challenges. A detailed discussion of each category, including representative methods and their applications in medical imaging, is provided in the following sections. Table 4 summarizes the characteristics of different fusion levels used in MMIF. It highlights the strengths, limitations, and typical use cases of pixel-level, feature-level, hybrid, and decision-level fusion techniques. Each level offers unique benefits depending on the clinical requirement, ranging from enhanced spatial detail in pixel-level fusion to semantic richness in feature-level fusion, and robustness in decision-level approaches. However, each fusion approach is explored in greater detail in the following sections to offer deeper insights into their working mechanisms and relevance in clinical contexts. The majority of studies primarily focus on pixel-level fusion techniques. However, feature-level fusion and hybrid approaches that combine pixel- and feature-level methods have also gained attention. Moreover, a few studies have achieved a higher level of abstraction by implementing decision-level fusion. Each fusion type is discussed in detail below.

Table 5 provides a comparative overview of various MMIF techniques, outlining their underlying procedures and highlighting their respective advantages. The table categorizes methods such as pixel-level, feature-level, decision-level, hybrid, and several advanced strategies, including guided-filtering, transform-domain, intelligent-based, spatial-domain, and dictionary learning-based fusion. By summarizing how each approach operates and the benefits it offers—such as spatial detail retention, robustness to noise, semantic integration, and computational efficiency. The table serves as a useful reference for selecting appropriate fusion techniques based on clinical or research needs.

### 4.1. Pixel-level fusion

Pixel-level fusion, also referred to as low-level fusion, where raw data from multiple modalities are combined at the earliest stage, typically at the pixel or data level. When performed in the spatial-domain, the pixel values of the images are merged linearly or non-linearly to obtain fusion results [114]. However, in transform domain, nevertheless, the images are first converted to the proper transform domain. This fusion technique enables the integration of detailed, low-level information from diverse modalities to generate a fused representation for subsequent analysis. The primary objective of this fusion technique is to retain maximum amount of original information from all input sources during the creation of a single unified dataset.

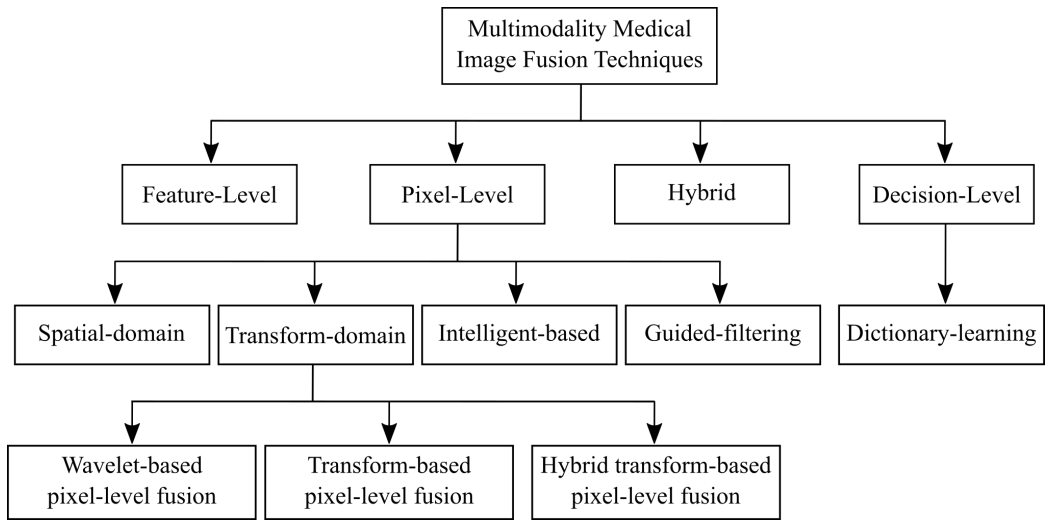

Fig. 3. Types of multi-modality medical image fusion

Pixel-level fusion method is attractive because it preserves essential details from all input images and retains high spatial resolution without significant computational complexity. However, this technique occasionally produce noise or distortions if the source images are not properly pre-processed because to its direct reliance on raw intensity values. Despite this drawback, it is a popular choice in medical image analysis due to its ease of use and capacity to retain important diagnostic information. Various multi-scale decomposition techniques can be used to conduct the pixel-level image fusion process. The discrete wavelet transform, with its various forms and fusion criteria, was embraced by a number of researchers. Other studies focused on medical image fusion using various transform algorithms, such as curvelet transform, non-subsampled shearlet transform (NSST), fuzzy transform, and NSCT. Additionally, some studies have addressed both intelligent-based and guided-filtering-based approaches. Some of these techniques are discussed in the following sections.

4.1.1. Spatial-domain pixel level fusion

Spatial-domain pixel-level fusion involves combining multiple source images directly in the spatial domain by manipulating and integrating their pixel intensities. This method directly processes pixel intensities in the spatial domain, without converting the image to a different representation, such as the frequency domain [115]. This technique of fusion has been widely used in medical imaging, surveillance, and remote sensing applications, where precise detailed information is essential [116]. A primary advantage of this fusion technique is that it often guided by specific criteria, such as preserving salient features, edges, or intensity variations, tailored to the application. Typical techniques include weighted averaging or spatial filtering strategies that emphasize features of interest from each image. Maximum or minimum pixel selection is another method, where the fused image takes the extreme intensity value from the input images at each pixel location. This preserves strong intensity features but may introduce artifacts or disrupt smooth transitions. However, one of the drawback of this technique is that it could be susceptible to noise amplification if the fusion rules are not carefully designed [117]. Additionally, methods like maximum pixel selection may sacrifice spectral or intensity variations, leading to a loss of important image information.

More advanced methods include Principal Component Analysis (PCA), which transforms input images into a new coordinate system to maximize variance along principal axes before fusing the components. Intensity-Hue-Saturation (IHS) fusion is commonly used in multi-spectral image processing. Another approach known as Brovey transform enhances contrast using arithmetic combinations of pixel intensities. Spatial-domain techniques are advantageous for their simplicity, real-time processing capability, and ability to preserve spatial details. Compared to frequency-domain fusion techniques, which process images in transformed domains like Fourier or wavelet domains, spatial-domain methods are simpler and more efficient.

Recent research has explored adaptive weight determination schemes and multi-scale spatial filters that can selectively emphasize critical regions. Region-specific weight allocation or combining spatial pixel-level fusion with edge-preserving smoothing filters to minimize distortions and stabilize the fusion output under varying illumination conditions have been introduced. These innovations demonstrate that, although transform-based methods remain influential, spatial-domain techniques continue to evolve in ways that enhance their robustness and applicability to diverse tasks.

Table 4. Overview of Pixel-, Feature-, Hybrid and Decision-level fusion techniques with their indivisdual sternghts, limistations and applications

| Fusion Level | Strengths | Limitations | Use Cases |
| --- | --- | --- | --- |
| **Pixel-Level** | Preserves spatial details, computationally simple | Sensitive to noise and misregistration, limited semantic capture | Enhancing visual quality for diagnosis (e.g., fusing CT and MRI for better visualization of soft tissue). |
| **Feature-Level** | Captures semantic and contextual information, robust to variations | Computationally complex | Cancer detection, tissue classification (e.g., MRI and PET for tumor delineation). |
| **Hybrid Fusion** | Balances spatial details and semantic information | Computationally intensive | Advanced diagnosis requiring spatial and abstract feature fusion. |
| **Decision-Level** | Modular, robust to modality differences | Loses spatial and feature-level details | Disease classification or risk assessment (e.g., combining MRI and clinical data for Alzheimer's diagnosis). |

4.1.2. Transform-domain pixel-level fusion

Transform-domain pixel-level fusion is an image fusion technique that operates in a transformed domain. Instead of directly manipulating pixel intensities, this approach converts input images into their frequency or spectral representations. The use of transform-domain pixel-level fusion techniques in medical imaging has been widely explored [114, 118]. Following are the fusion techniques discussed that are in the category of transform-domain pixel-level fusion.

**Wavelet-based pixel-level fusion** decomposes each source image into a hierarchy of sub-bands representing varying scales and frequency components. This is the most popular methods in the transform domain due to its ability to analyze images at multiple resolutions. The wavelet transform decomposes an image into a set of sub-bands, including approximation (low-frequency) and detail (high-frequency) components. During the fusion process, approximation components are merged to retain overall intensity, while detail components are combined to preserve edges and textures. Fusion rules such as taking the maximum, averaging, or applying region-based thresholds are applied to these sub-bands to merge the most informative features. The fused sub-bands are then transformed back to the spatial domain to generate the final image. This multi-scale representation effectively balances detail with noise suppression, hence, making it attractive in medical image analysis field. Commonly used methods in wavelet-based fusion is the "maximum selection rule," which selects the coefficient with the highest magnitude from the input images at each pixel position or the "weighted averaging rule," which combines coefficients using predefined or adaptively determined weights.

**Tranform-based pixel-level fusion** extends beyond wavelets to other transforms like Fourier, discrete cosine transform (DCT), and Laplacian pyramids. Contourlet, curvelet, and shearlet transforms, for instance, offer directional and anisotropic features that capture intricate textures or elongated structures more effectively. These methods use the mathematical properties of different transforms to enhance specific image features. For instance, DCT is widely used for its compact representation of frequency information and is commonly applied in multimedia processing. In DCT-based fusion low-frequency components are fused to maintain the overall structure, while high-frequency components are combined to enhance sharpness and detail. Fourier transform-based methods are effective for capturing periodic and global patterns, making them suitable for applications such as texture analysis and frequency-domain filtering. On the other hand, Laplacian pyramid-based fusion works by decomposing an image into multi-scale layers, combining corresponding layers from input images, and reconstructing the fused image. These methods are significantly important in scenarios where spatial structures like edges in MRI scans are critical. By merging the transform coefficients across different frequency bands and directions, transform-based fusion provides enhanced clarity while often mitigating artifacts common in simpler, spatial-domain approaches. These techniques are computationally efficient and work well for both grayscale and multi-spectral image fusion.

**Hybrid transform-based pixel-level fusion** combines multiple transform methods to use their individual strengths [119]. A common strategy is to fuse images in the wavelet domain and then apply morphological operations or other spatial refinement methods to highlight or suppress specific features. This mix of strategies has been shown to boost accuracy and robustness in demanding scenarios, including low-light imaging or multi-sensor detection in highly dynamic environments. For instance, a hybrid approach might involve combining wavelet and DCT transforms, where wavelets capture localized edge information, and DCT provides global frequency representation. Such methods have been shown to produce superior results, particularly in applications requiring both local and global feature preservation.

One of the recent development is the hybrid wavelet-shearlet transformations, which combine the multi-resolution capabilities of wavelets with the directional sensitivity of shearlets. This hybridization improves the ability to handle images with complex geometries, such as medical imaging data. Similarly, hybrid transform techniques often integrate machine learning models, such as CNN, to enhance fusion rule selection and improve adaptability to diverse datasets [120].

### 4.1.3. Intelligent-based pixel-level fusion

Intelligent-based pixel-level fusion utilizes advanced computational techniques to automatically select and integrate image data at the pixel level, thereby enhancing the interpretability of the fused output. These methods utilize advanced algorithms, such as evolutionary optimization, machine learning, and fuzzy logic, to adaptively select and combine the most relevant features from input images. However, traditional methods relies on manually defined rules or transforms, intelligent-based approaches adapt to variations in imaging conditions, noise levels, or domain-specific requirements through learning or optimization processes. The primary advantage of intelligent-based approaches lies in their ability to adapt to different image characteristics, making them effective for medical imaging applications. Techniques such as Particle Swarm Optimization (PSO), Neural Networks, Fuzzy Logic, Gray Wolf Optimization (GWO), and others have been widely explored to improve the quality of fused images.

One technique frequently explored is PSO, inspired by the social behavior of birds and fish. PSO formulates the pixel-level fusion problem as a search for optimal fusion weights or parameter sets in a high-dimensional space. It is widely used in pixel-level fusion for determining optimal weights or coefficients for combining input images. Each particle in the swarm represents a potential solution, and its position is updated based on its own experience and that of its neighbors to minimize or maximize a fitness function. Researchers have reported that PSO-based fusion achieves robust performance, even under challenging conditions like varying illumination or heterogeneous background textures [121]. PSO has been effectively integrated with wavelet transforms for multi-spectral and panchromatic image fusion. PSO improves the sharpness and detail retention in fused images by adaptively selecting fusion rules for each wavelet coefficient. Recent advancements involve PSO hybridization with genetic algorithms, to enhance convergence speed and avoid local optima.

Neural networks, particularly deep learning models, have gained significant attention in image fusion. They have also gained prominence in intelligent-based fusion, especially deep learning architectures that extract discriminative features and fuse them in a data-driven manner. Neural networks are typically trained on diverse datasets to learn fusion rules that generalize across different imaging modalities and found good in identifying high-frequency components such as edges and textures, combining them to produce a result with superior details. Their training often requires carefully curated datasets and sufficient computational resources to perform real-time fusion with impressive accuracy. Recent studies have explored attention mechanisms in neural networks to selectively focus on the most informative regions of input images during fusion [122]. Additionally, pre-trained networks, such as ResNet and VGG, have been adapted for pixel-level fusion tasks, utilizes transfer learning to reduce training time and improve performance [123].

Fuzzy logic is another intelligent technique used for pixel-level fusion, especially in situations involving uncertainty or imprecise data. In fuzzy logic-based fusion, input image pixels are assigned membership values in fuzzy sets based on features such as intensity or gradient. Fusion rules are then applied to combine these membership values, followed by defuzzification to generate the fused image [124]. Unlike traditional logic, which relies on binary true or false values, fuzzy logic operates on varying degrees of truth, making it particularly well-suited for image fusion tasks that demand adaptive decision-making. Fuzzy logic provides a robust framework for handling ambiguous or noisy data by representing pixel relationships through membership functions rather than hard thresholds [125]. This flexibility makes fuzzy logic approaches particularly useful in scenarios where different types of uncertainty, be it sensor noise or inconsistent illumination coexist.

GWO has also drawn attention to intelligent-based fusion [126]. In pixel-level fusion, GWO optimizes fusion parameters, such as weights or thresholds, by minimizing or maximizing fitness functions. Along with other swarm intelligence methods, such as Ant Colony Optimization [127] or Artificial Bee Colony [128–130], GWO illustrates how nature-inspired techniques can effectively converge on robust fusion solutions, even in high-dimensional parameter spaces. It is particularly effective in multi-modal image fusion, such as combining medical images (e.g., CT and MRI scans). Its simplicity and fast convergence make it a preferred choice for high-dimensional problems. Recent advancements include hybrid approaches combining GWO with neural networks or wavelet transforms to further enhance fusion performance. These methods improve spatial resolution while maintaining the spectral integrity of input images.

HSV (Hue, Saturation, and Value) color space-based techniques are commonly used in color image fusion [131, 132]. In these methods, input images are converted from RGB to HSV color space, and the fusion is performed on the value component, as it contains most of the structural and detailed information. The hue and saturation components are usually preserved to maintain the original color characteristics. By fusing intensity information in HSV space, these methods effectively enhance the spatial details of fused images while maintaining color fidelity. HSV-based fusion is widely used in remote sensing and satellite imagery applications.

Apart from the above, several other intelligent-based methods have been explored for pixel-level fusion, i.e., genetic algorithms for optimizing fusion parameters and ant colony optimization [127] to find optimal fusion solutions. Techniques like Adaptive Neuro-Fuzzy Inference Systems combine the strengths of neural networks and fuzzy logic to achieve robust and adaptive fusion [133]. Deep generative models, like Generative Adversarial Networks (GANs), have also shown promising results in image fusion by enhancing visual quality [134]. However, these models require extensive training and computational resources.

### 4.1.4. Guided-filtering-based pixel-level fusion

Guided-filtering-based pixel-level fusion is a technique that employs guided filters to combine multiple input images into a single fused image [135]. This technique uses the edge-preserving properties of guided filters to integrate source images into a single, richly detailed output [136]. The technique's guided filters use a guidance image to influence the filtering process [137], thereby preserving dominant edges and minimizing artifacts. This localized guidance makes it well suited to applications where subtle structural details, such as boundaries in medical scans. They have found widespread applications in multi-modal [138], multi-focus [137], and medical image fusion due to their computational efficiency and adaptability.

Among recent developments, the Rolling Guidance Filter (RGF) has gained traction as a natural extension. In RGF-based methods, the guidance image is iteratively refined to remove progressively smaller-scale details while preserving principal edges. RGF is an iterative edge-preserving filter that eliminates small-scale details while maintaining large-scale structural components. In guided-filtering-based pixel-level fusion, RGF is applied to progressively refine the fusion process by iteratively smoothing the guidance image and input images [139]. The technique has been successfully used in scenarios requiring noise reduction and detail enhancement, such as multi-modal medical images [140]. Its rolling mechanism allows the filter to adapt to varying scales of features, ensuring that significant details are retained in the fused image [141]. This rolling strategy allows for multi-scale fusion at the pixel level, as each iteration emphasizes a different level of detail. Recent advancements in RGF-based fusion include adaptive weighting mechanisms that further enhance the filter's performance in dynamic imaging scenarios. Combining RGF with guided filtering principles can yield a fused image that maintains both macroscopic features (e.g., large contours) and fine textures (e.g., hairline edges), offering a flexible balance between smoothness and sharpness [142, 143].

Another promising avenue involves integrating the Spiking Cortical Model (SCM) into guided-filtering-based fusion workflows. SCM-based guided filtering integrates the principles of visual perception and neural spiking activity to enhance the fusion process [144]. In pixel-level fusion, SCM is often used to pre- or post-process images to enhance features like edges, textures, and contrasts before applying guided filters. When employed in parallel with guided filtering, the SCM can selectively highlight critical edge or texture information while damping noisy, irrelevant features. The SCM-guided filtering approach has been particularly practical in multi-focus image fusion. By encoding pixel intensities into spikes, the model emphasizes spatial and temporal correlations, enhancing edge preservation during the filtering process. Recent experiments indicate that SCM-aided guided filtering can adapt more dynamically to variable illumination or noise conditions, making it suitable for complex environments like hyperspectral imaging [145, 146].

In addition to SCM and RGF, other approaches explore combinations of guided filtering with morphological operators, bilateral filters, or scale-space methods. When carefully combined, these approaches can refine pixel-level fusion by reducing halo artifacts, suppressing unnecessary gradient transitions, and sharpening boundaries. Such hybrid systems, which sometimes integrate rolling guidance or spiking cortical models as modules, highlight a growing trend toward modular, highly adaptive fusion pipelines [147]. Multi-scale guided filtering [137] is another advancement where guided filters are applied at different decomposition scales, such as Gaussian or Laplacian pyramids. This approach allows for the fusion of fine details and large-scale features, resulting in improved clarity and contrast in the fused image. This is important for preserving strong geometric structures, and there is a need for nuanced control over lower-level intensities and textures, ensuring that the fused output is visually coherent and information-rich. Recent research also explores the integration of guided filters with deep learning models to automatically learn optimal guidance images and fusion rules, further enhancing the adaptability of the method [148].

## 4.2. Feature-level fusion

Feature-level fusion methods aim to process and integrate image features rather than individual pixels, since the features are more significant than individual pixels. In this approach, features are extracted separately from each source image, and then a fusion method is applied to these extracted features to generate a unified representation. By using existing image transformation techniques, features like edges, textures, gradients, and frequency components are extracted in feature-level fusion. A unified and enhanced feature set is produced by aligning and combining these features using the proper fusion rules or algorithms. The fusion process improves decision-making accuracy, enhances visual perception, and retains more useful information compared to pixel-level fusion methods. Additionally, the fusion process increases visual perception, improves decision-making accuracy, and preserves more valuable information. This approach enhances the quality of fused data by preserving essential details, improving interpretability, and reducing redundancy. Feature-level fusion is a critical technique in multi-modal image processing, where features extracted from multiple source images are combined to create a more informative and comprehensive representation. It has been widely used in medical imaging as it helps in obtaining a more informative and comprehensive representation of the input data.

Feature-level fusion can be categorized according to the nature of the used methodologies and combined features. These techniques can be broadly categorized into transform-based methods, sparse representation methods, and clustering-based methods. Transform-based methods, such as wavelet transforms and curvelet transforms, decompose images into frequency sub-bands, enabling the fusion of low-frequency (smooth) and high-frequency (detail) components. For instance, the NSST has been used to fuse low- and high-frequency sub-bands using Pulse-Coupled Neural Networks (PCNN), achieving superior detail preservation and energy retention [149]. Similarly, the Uniform Discrete Curvelet Transform (UDCT) employs the FSIM and CCFSIM indices to fuse low-pass and high-pass sub-bands, respectively, ensuring robust fusion results [150]. Sparse representation methods, on the other hand, focus on representing images as sparse vectors in a dictionary. These methods often involve dividing source images into patches, arranging them into vectors, and using a decision map to select optimal vectors for fusion. For example, a study utilized sparse representation to fuse multi-modality medical images, achieving high fusion accuracy by utilizing statistical features such as standard deviation and entropy [151]. Clustering-based methods, such as Quantum Particle Swarm Optimization (QPSO) combined with Fuzzy C-Means (FCM), segment feature spaces and create weighting factors for fusion. These methods are particularly effective in handling heterogeneous features from different imaging modalities [152].

Table 5. A brief comparison of multi-modality image fusion techniques, procedure, and advantages

| Method | Procedure | Advantages |
| --- | --- | --- |
| Pixel-Level Fusion | Direct manipulation of pixel intensities to merge details | Retains spatial details; computationally simple |
| Feature-Level Fusion | Extracts features such as edges, textures, and objects for integration | Robust to noise; integrates high-level semantics |
| Decision-Level Fusion | Combines outputs from classifiers or detection algorithms | Integrates heterogeneous outputs; reduces noise influence |
| Hybrid Pixel-Feature Fusion | Combines pixel-level and feature-level data for complementary advantages | Balances spatial detail and semantic understanding |
| Guided-Filtering Fusion | Uses a guidance image to filter and merge input data while preserving edges | Edge-preserving; computationally efficient |
| Transform-Domain Fusion | Applies frequency transforms (e.g., wavelets, Fourier) to merge frequency components | Multi-resolution detail capture; robust to noise |
| Intelligent-Based Fusion | Uses AI models (e.g., CNNs, PSO) to adaptively learn fusion strategies | Adaptive and robust; handles complex data relations |
| Spatial-Domain Pixel-Level Fusion | Combines pixel intensities directly in the spatial domain using simple averaging, maximum selection, or weighted fusion methods | Simple implementation; preserves spatial location and structural information; low computational cost |
| Dictionary Learning-Based Fusion | Learns sparse representations (dictionaries) of source images and fuses by combining important sparse coefficients | Preserves significant features; enhances sparse details; effective for complex and texture-rich images |

Recent state-of-the-art studies have introduced advanced approaches to feature-level fusion, addressing challenges such as feature heterogeneity and computational complexity. One notable advancement is the integration of attention mechanisms into fusion frameworks [153]. Another recent development is the use of dynamic receptive fields and adaptive feature fusion in federated learning. A study introduced a dynamic receptive field adjustment mechanism to efficiently extract default features from financial credit risk data. By aggregating features from local and global models, this method reduces communication overhead while maintaining high accuracy [154]. Additionally, the FusionMamba framework, which integrates dynamic convolution and channel attention mechanisms, has demonstrated superior performance in multi-modal image fusion tasks. This framework uses selective structured state space models (S4) to handle long-range dependencies efficiently, making it suitable for applications like medical imaging [155]. In the domain of hyperspectral and LiDAR data fusion, a feature-decision level collaborative fusion network has been proposed. This network integrates shared and complementary features through a multilevel interactive fusion (MIF) module, minimizing modality differences and enhancing classification accuracy. The dynamic weight selection (DWS) module further optimizes feature representations, ensuring balanced and effective fusion [156].

Notwithstanding the progress, feature-level fusion still has a number of obstacles to overcome. The fusion process is frequently complicated by the heterogeneity of features from various modalities, necessitating complex alignment and normalization techniques. Furthermore, transform-based and clustering-based approaches can be prohibitively computationally complex, particularly for real-time applications. To overcome these obstacles, future studies might concentrate on creating lightweight fusion frameworks and utilizing cutting-edge machine learning methods like transformers and graph neural networks [155–157]. To sum up, feature-level fusion methods have advanced considerably and now provide reliable ways to integrate multi-modal image data. These strategies improve the caliber and dependability of fusion results by utilizing clustering algorithms, transform-based approaches, and sparse representation. New developments like dynamic feature fusion and attention mechanisms further push the envelope of what is possible, opening the door to multi-modal image processing systems that are more accurate and efficient.

### 4.3. Hybrid pixel- and feature-level fusion

Hybrid pixel- and feature-level fusion combines the strengths of pixel-level and feature-level fusion to create a comprehensive representation of input images. This hybrid approach integrates the low-level spatial information in raw pixel intensities with high-level semantic features extracted from images, achieving a balance between spatial detail preservation and robust feature representation. By uniting these two fusion levels, hybrid methods aim to improve visual clarity and interpretability, making them particularly useful in medical imaging applications [119]. In a hybrid framework, the pixel-level component handles the immediate intensity while mitigating noise and enhancing edges. Concurrently, feature-level fusion processes high-order attributes, including texture descriptors, shape information, or deep learning features. The approach retains the semantic richness needed for downstream tasks like object detection or classification by fusing these feature maps.

A crucial aspect of hybrid pixel- and feature-level fusion is the choice of transformation for decomposing and reconstructing images. Recently, the Discrete Wavelet Frame Transform (DWFT) has gained attention for its effectiveness in multi-scale decomposition, which helps to capture subtle image details. Unlike standard wavelet transforms, DWFT is redundant, translation-invariant, and provides enhanced directional selectivity, making it suitable for retaining spatial and frequency information. DWFT can be applied at the pixel level to separate images into sub-bands, emphasizing edges and textures at varying frequencies. Feature extraction algorithms then operate on these sub-bands or the reconstructed fused image, increasing the overall robustness against noise and artifacts. Research highlights that integrating DWFT with deep feature extraction significantly boosts performance in challenging scenarios. Subsequently, deep networks, such as U-Net variants or attention-based CNN architectures, can further refine these details at the feature level. This combination ensures that the fused image not only exhibits good visual quality but also maintains high discriminative power for tasks like segmentation and classification. With the availability and accessibility of computational resources, this fusion technique employing wavelet-based transforms and deep-learning models is expected to become increasingly practical, leading to more accurate and robust image fusion systems.

## 4.4. Decision-level fusion

Decision-level fusion is a high-level image or data fusion technique that combines decisions made independently by multiple sources or algorithms to produce a final, unified decision. This fusion technique differs from pixel- or feature-level approaches by focusing solely on the "decisions" of each independent model rather than on raw data or extracted features. This approach is instrumental when inputs come from heterogeneous sources, where the formats and characteristics of the data may differ significantly. Real-time systems face dynamically changing conditions, and decision-level fusion focuses on adaptive and context-aware decision fusion, resulting in higher accuracy and better resilience to perturbations. It is commonly used when input data is sensitive to noise or inconsistencies, as the final decision is less likely to be dominated by a single erroneous source. Methods for decision-level fusion include majority voting, Bayesian inference, Dempster-Shafer theory, and ensemble learning techniques, which weigh individual decisions to derive a robust outcome [158].

### 4.4.1. Dictionary-learning-based fusion

Decision-learning-based fusion integrates machine learning and statistical methods to optimize the decision-fusion process. These approaches often involve training a "fusion classifier," such as a random forest or a deep neural network, to produce a final label based on the decisions of lower-level models. A study explored the use of neural networks for decision-level fusion in medical diagnosis by combining independent predictions from classifiers analyzing CT, MRI, and PET scans [159]. This demonstrates the utility of decision-learning-based fusion in scenarios where the relationships between decisions are complex and non-linear. Another advanced technique involves SVM for decision-level fusion to integrate land cover classifications from multi-spectral and hyperspectral images [160], achieving higher classification accuracy and robustness in distinguishing complex terrain features [161].

Ensemble methods such as boosting and bagging are also commonly used in decision-learning-based fusion. For instance, AdaBoost has been applied to combine outputs from multiple classifiers [162] in intrusion detection systems. Similarly, random forest models have been employed in environmental monitoring to integrate decisions from multiple satellite sensors for accurate vegetation and water body detection. A recent study combined CNNs for visual data analysis and RNNs for temporal data in autonomous driving systems. The adaptability of these models allows for dynamic weighting of input decisions, enabling the system to handle diverse scenarios. Deep learning models like CNN and RNN have further advanced decision-level fusion [163].

A comprehensive body of literature has been reviewed and discussed in detail, focusing on various types of MMIF techniques. This discussion covers the fundamental approaches used in fusion, ranging from spatial-domain and transform-based methods to more advanced hybrid strategies. Also highlighting the specific imaging modalities employed, such as CT, US, MRI, PET, and SPECT. Additionally, the applications of each fusion technique in different diagnostic scenarios, including neurological, oncological, and soft tissue evaluations, have been thoroughly analyzed. Table 6 provides a structured summary of this analysis, clearly illustrating the fusion methods, the corresponding imaging modalities, and their targeted clinical applications. It highlights how spatial-domain and transform-based pixel-level fusion approaches are employed using various strategies i.e., such as PCA, IHS, fuzzy transform, and NSCT, across combinations like MRI-PET, CT-MRI, and MRI-SPECT. These fusion techniques aim to enhance diagnostic outcomes, improve image quality, and support disease identification including Alzheimer's, Parkinson's disease, and lesion characterization, thereby contributing to more accurate and efficient medical decision-making. This table serves as a concise map that links fusion techniques to their practical roles in improving diagnostic accuracy and patient care.

Table 6. Multi-modal medical image fusion types with their fusion techniques, and applications using different imaging modalities for improved medical diagnosis

| Method | Technique | Modalities | Applications |
|---|---|---|---|
| Spatial-domain pixel-level fusion | Maximum, minimum, average, PCA, IHS and Borvey [164–166] | CT, MR [167] | Image quality enhancement for improved diagnosis |
| | | MR, PET [99, 100] | Alzheimer's diagnosis |
| | | MR, SPECT [118] | Enhancement of lesion characteristics |
| Transform-based pixel-level fusion | Fuzzy transform [168, 169] | MR, PET [96] | Alzheimer's and Parkinson's disease diagnosis |
| | | MR-T1/T2 | |
| | | MR, CT | |
| | NSCT [170, 171] | MR, CT [172] | Enhances diagnostic accuracy and surgical precision |
| | | MR, PET [173] | |
| | | MR, T1/T2 [174] | |
| | | MR, SPECT [175] | |
| Wavelet-domain pixel-level fusion | DWT [176] | MR, CT [177] | Medical image quality improvement to assist physicians in diagnosis |
| | DWT and SWT [178] | PET, CT [111, 179] | Seizures diagnosis |
| | DFRWT [180, 181] | MR, PET [182] | Improve soft tissue contrast |
| | | MR-T1 [183] | Improves fluid contrast |
| | | MR-T1, MR-T2 [184] | for edema visualization |
| Hybrid-transform-based pixel-level fusion | DWT and PCA [119, 185–187] | MR, CT [119, 188] | Enrich structural details |
| | NSCT and Fuzzy logic [189] | MR, PET [100, 190] | Improved clinical diagnosis |

| Method | Technique | Modalities | Applications |
|---|---|---|---|
|  |  | MR, SPECT [175, 191] | Neuro-degenerative diseases diagnose |
|  |  | MR, T1/T2 [192] | Evaluate memory loss |
|  | NSST and PCNN [193, 194] | MR, PET [194] | Leads to improved lesion segmentation |
|  |  | MR, CT, T1/T2 [195] |  |
| Guided-filtering-based pixel-level fusion | RGF and SCM [196] | CT, MR [197] | Improved clinical diagnosis |
|  |  | US, SPECT [103, 104] | Application in endometriosis surgery |
|  |  | MR-T1, MR-T2 [198] | Enhanced Semantic Segmentation in Brain Tumor |
|  | Guided-filtering and DRC [135] | MR, CT [197] | Enhanced Semantic Segmentation in Brain Tumor |
|  | NSCT and guided filter [199] | CT, MR [135, 200] |  |
| Feature-level fusion | Lifting wavelet transform [201] | MR-T1, MR-T2 [184] | Intertwine medicinal pictures |
|  | Wavelet transform with UDCT [202] | CT, MR [203] | Diagnosis, segmentation, and treatment planning |
|  | DT-CWT and watershed transform [204] | CT, MR [205] |  |
|  | NSST and PCNN [206] | MRI, SPECT [207] |  |
|  | Hadamard transform [208] | MR, SPECT [119, 209] | Functional and structural analysis |
|  | Hybrid CNN-transformer [210] | Infrared Thermal Imaging and US | Enhanced the classification accuracy of thyroid nodules |
| Intelligent-based pixel-level fusion | GMSF and PCNN [211] | MR-T2, CT [207,211] | Improved clinical diagnosis and treatment |
|  | Fourier transform with GWO and wavelet [212] | MR, SPECT [126, 213] |  |

| Method | Technique | Modalities | Applications |
|---|---|---|---|
| | Fourier transform with HCS-GWO [126] | MR, PET [126] | Improved tumor segmentation |
| | Framelet transform and HVS model [211] | MR, PET [214] | |
| Hybrid Pixel- and feature-level fusion | Edge detection and region growing [119] | CT, MR [215] | Enrich details in the fused image |
| | Fuzzy c-means clustering and DWFT [211] | MR, PET [216] | Medical image analysis and AI-based diagnosis |
| Dictionary-learning-based decision-level fusion | Gaussian filter with integrated dictionary learning [217] | CT, MR [218] | Improved fusion quality to enhance diagnostic accuracy and treatment planning |
| | Image patch sampling [219] | MR, PET [100, 220] | Reduce data redundancy, leading to faster processing and lower computational costs |
| | SR and ODL [221] | MR, SPECT [222] | Suitable for real-time medical applications |
| | CS-MCA [223] | MR-T2, CT [223, 224] | Helps reduce noise, distortions, and artifacts in fused image |
| | Local density peak clustering [225] | MR, MRA [106] | Enhances contrast and visibility of soft tissues |
| | MSF, SR, and OMP [226, 227] | MR-T1, MR-T2 [119, 226, 227] | sharp edges, textures, and fine anatomical details |

## 5. Medical Imaging Fusion Algorithms

The process of registering and fusing various images is known as image fusion. The medical image fusion algorithms are categorized into different methods as listed in the studies [228–230]. These algorithms help extract and integrate complementary features [231] to produce an informative fused image for better diagnostic. A comprehensive review of the existing literature broadly categorize medical image fusion algorithms into six major types: morphological-based methods, human visual system (HVS) operator-based methods, sub-band decomposition methods (such as wavelet or contourlet transforms), neural network-based methods, fuzzy logic-based methods, and other hybrid or unconventional techniques. Each category addresses different aspects of fusion, from structural preservation and perceptual enhancement to learning-based adaptability and uncertainty handling. Understanding these categories provides a foundation for selecting suitable fusion approaches tailored to specific medical imaging applications and clinical requirements.

*5.1. Algorithms using morphological operations*

Morphological operations are integral to medical image fusion, particularly for their ability to extract and preserve spatial structures within images. These techniques are especially effective in highlighting edges, contours, and other structural elements, making them valuable in scenarios where precise anatomical detail is crucial. One of the foundational approaches in this domain is the morphological pyramid technique [232]. This method involves creating a multi-scale representation of an image by applying morphological operations such as dilation and erosion. By constructing a pyramid of images at varying resolutions, it becomes possible to analyze and fuse features at different scales, enhancing the overall quality of the fused image. This approach has been shown to improve the visual quality of fused images, particularly in preserving edge information and reducing artifacts.

Another notable method is the morphological difference pyramid, where images are progressively smoothed and subsampled using operations like dilation and erosion to create a multi-scale (pyramidal) representation. This algorithm focuses on capturing the differences between successive levels of the morphological pyramid [233]. This technique emphasizes the unique features present at each scale, allowing for a more detailed and informative fusion of images from different modalities. At each level, spatial features are preserved and abstracted to varying degrees, allowing fusion at different resolutions. This hierarchical decomposition enables the effective merging of fine-grained details from multiple modalities without introducing noise. By highlighting these differences, the method enhances the contrast and clarity of the fused image, which is particularly beneficial in medical diagnostics where subtle variations can be clinically significant. Despite their strengths, morphological methods are not without drawbacks. They can be sensitive to noise and may struggle with images that have significant variability in intensity, resolution, or modality-specific distortions. To mitigate these issues, hybrid approaches have been developed that combine morphological operations with other techniques, such as wavelet transforms or neural networks. These combinations aim to use the strengths of each method, resulting in more robust and accurate image fusion outcomes.

In the last decade, hybrid fusion techniques have emerged, integrating morphological operations with methods such as neural networks [234], wavelet transforms [235], and fuzzy logic systems [236]. These hybrid models capitalize on the structural preservation strengths of morphology through complementary algorithms.

Recent research has shifted towards the integration of morphological decomposition and machine learning techniques. A novel algorithm was proposed that combines multilevel morphological component analysis (MMCA) with a classifier-based decision-making framework [237]. The algorithm initially decomposes source images into natural morphological layers, capturing more intricate details. Feature vectors are then extracted from these layers and fed into a two-class Support Vector Machine trained to classify the relevant features effectively. Finally, image coefficients are fused based on the validated decision matrix, producing a more coherent and information-rich fused image.

In a study, integration of hierarchical decomposition methods with PCA for feature extraction and dimensionality reduction, making it a suitable choice for fusion tasks that require optimal representation of source data [238]. An image fusion algorithm based on hierarchical PCA, which utilizes the strengths of pyramid-based multiresolution decomposition alongside PCA's statistical rigor.

Recent advancements integrate fuzzy logic and multi-rule decision strategies to enhance fusion performance, especially in high-frequency components. A novel approach introduces fusion rules that combine pixel- and region-based decision maps (PDM and RDM), guided by a dissimilarity measure and optimized through a population-based algorithm for low-frequency coefficients [236]. This hybrid strategy successfully integrates the advantages of conventional techniques, resulting in enhanced fused images with improved subjective visual appeal and objective performance measures.

In summary, morphological operations continue to play a vital role in medical image fusion, offering a balance between computational efficiency and the preservation of critical structural information. Their integration into hybrid models further enhances their applicability, making them a valuable tool in the ongoing advancement of medical imaging technologies [223, 239–241].

*5.2. Algorithms using Human Value System operations*

Knowledge-based frameworks are often termed Human Visual System (HVS) operator models. These enable the fusion process to use rich, domain-specific information about anatomical structures, image geometry, and spatial relationships [242]. Integrating expert knowledge into hypothesis generation, these systems can simulate plausible anatomical scenarios, guiding the selection and fusion of multi-modal data for tasks like segmentation and recognition [243]. In practice, HVS and knowledge-based fusion approaches typically follow a three-stage pipeline, hypothesis generation, knowledge-driven integration, and iterative refinement. First, hypothesis generation constructs an initial interpretive model of the scene using prelearned templates or expert-defined rules [244]. Knowledge-driven integration merges regions produced by low-level segmentation, ensuring that fused regions adhere to physiologically plausible shapes and intensities [245]. Finally, iterative refinement uses feedback loops to update segmentations and fused outputs [246].

A diverse range of medical applications has demonstrated the effectiveness of HVS operator models. In mammography, by identifying subtle and localized calcium deposits often one of the earliest indicators of malignancy such systems significantly improve diagnostic sensitivity. This rule-based detection of clustered high-intensity microcalcifications improves early breast cancer screening [247, 248] by fusing complementary modalities [249]. When combined with multi-modal imaging techniques, the fusion of complementary modalities allows for a more comprehensive visualization of breast tissue.

Neural tissue characterization frameworks use anatomical atlases to spatially align and integrate multiple MRI modalities [250, 251] into a unified reference space. This integration improves the resultant image by enhancing the contrast between white and gray matter. This in results in identifying structural boundaries and pathological changes. The tissue segmentation, enables clearer visualization and more precise assessment of neurodegenerative or neurological conditions. This approach significantly aids in the early detection and monitoring of diseases such as Alzheimer's, Parkinson's disease, multiple sclerosis, and brain tumors [250–253].

In neuroimaging, knowledge-based fusion techniques [254] are employed to integrate complementary information from different imaging modalities. For instance, the high-resolution bone detail captured by CT scans [255] and the superior soft-tissue contrast provided by MRI [256]. This fusion enables the clinicians to precisely localize the lesion by investigating a comprehensive anatomically accurate image. By aligning and combining these modalities using anatomical landmarks or statistical models, knowledge-based fusion not only enhances diagnostic clarity but also supports better surgical planning, treatment targeting, and monitoring of neurological conditions.

Hybrid fuzzy rule–based approaches have gained prominence in the classification of abnormal brain tissues [257]. This synergy enhances diagnostic precision, especially in complex cases involving tumors, lesions, or neurodegenerative conditions, and has led to consistently high classification performance across multiple studies [257].

These methods utilize fuzzy logic to manage the imprecision in the medical data, allowing for flexible reasoning that closely mirrors expert clinical judgment. By integrating domain knowledge, such as expert-defined linguistic rules with quantitative features extracted from medical images, hybrid systems can handle overlapping tissue characteristics more effectively. These HVS and knowledge-based methods offer distinct advantages. By benchmarking fusion outputs against established principles of human perception—such as contrast sensitivity and spatial frequency preferences, they help maintain both visual fidelity and clinical relevance. The continued evolution of knowledge-based fusion frameworks, particularly when combined with data-driven methods, holds promise for more interpretable [13] and reliable CAD systems [19, 258–262] in diverse clinical settings.

### 5.3. Algorithms with Neural Networks

Artificial Neural Networks (ANNs) implify the learning process through training mechanisms, their effectiveness heavily depends on the quality of training data and the precision of the learning algorithm. To improve feature quality and enhance the robustness of ANN-based models, researchers often integrate neural networks with other fusion techniques. Numerous ANN-based approaches for MMIF have been explored in the literature, some of which are discussed below.

A pulse-coupled neural network (PCNN)-based technique was introduced for image fusion aimed at improving object recognition [263]. This approach uses PCNNs to merge outputs from various object detection methods, thereby enhancing the accuracy of recognizing objects. Despite its effectiveness, a key limitation of this method is that it maps each pixel to a single neuron, lacking the ability to process multiple pixels per neuron, which can affect its performance in complex visual scenes.

An unsupervised clustering technique known as the Lagrange Constraint Neural Network (LCNN) for the early detection of breast cancer, using two-color mid and long infrared images of the breast [264]. This approach demonstrated the benefits of multi-color mid and long infrared imaging, which is effective for identifying irregularities in the under-skin temperature and pinpointing the precise location of ductal carcinoma. However, a key limitation of this method lies in the clustering of feature pixels.

Another author presented a self-organizing feature mapping wavelet neural network (SOFMWNN) to combine MR/SPECT images [265]. The organization of image pixels in the detail layers provides valuable information that can be integrated efficiently. Another proposed an image fusion method based on clustering analysis for clinical image processing [266]. The analysis divides pixels into feature pixels, which contain valuable medical data, and auxiliary pixels, which hold background information. A partial gradient fusion rule is applied to the feature pixels, while a standard gray-level method is used for the auxiliary pixels [266, 267]. Experimental results show that the proposed fusion technique produces images with higher information content and better quality.

The presence of different pixels for individual neurons is a significant drawback in neural image processing. To address the issue, a study proposed a novel multi-channel display m-PCNN for medicinal image fusion [268]. This approach outlines the mathematical model of mPCNN, followed by a detailed presentation of the dual-channel mPCNN model. Experimental results show that this method outperforms other techniques in both visual quality and objective evaluation metrics.

A new fusion-based approach was introduced, combining Gabor channel and Walsh-Hadamard transform structures through a median absolute deviation (MAD) method [269]. The proposed system consists of three stages and was tested on various real CTy lung images, yielding promising results in the classification of lung disorders. To reduce blocking artifacts in medical images and determine the optimal coefficients for the fused image, an algorithm utilizing Orthogonal Teaching Learning Based Optimization (OTLBO) is employed [270].

A MMIF technique was introduced, combining discrete Tchebichef moments and PCNN. Tchebichef moments effectively capture edge features, while the energy of Tchebichef moments for blocks (ETMB) serves as the external stimulus for PCNN [271]. The proposed method surpasses existing fusion techniques in both subjective and objective performance evaluations.

A novel MMIF method using convolutional neural networks [272, 273] is proposed. The key innovation of this approach lies in its ability to simultaneously perform activity level measurement and weight assignment through network learning, thus addressing the challenges of manual design. In recent years, the application of deep learning techniques to pixel-level image fusion has advanced rapidly.

*5.4. Algorithms using Sub-band Decomposition*

Sub-band decomposition has become a widely adopted technique in MMIF due to its efficiency in extracting and representing high-frequency details. The primary concept involves extracting detailed components from one image and integrating them into another, thereby preserving salient features from both sources. Wavelet transforms are especially effective in this context due to their ability to provide simultaneous localization in both spatial and frequency domains.

Wavelet analysis employs scaled and translated variants of a mother wavelet, enabling a higher resolution of time-frequency information than Fourier analysis, which breaks a signal down into sinusoidal components. The resulting DWT provides a compact representation with inherent localization and directional capabilities in medical image processing tasks such as diagnosis, segmentation, and radiotherapy planning [274].

A study introduces a novel Multi-Level Wavelet CNN (MWCNN) as an innovative solution to overcome limitations in conventional CNNs, particularly in tasks requiring efficient feature extraction and preservation of spatial information [275]. Dilated convolutions have emerged as an alternative, offering a larger receptive field without increasing parameters, yet they suffer from the gridding effect—an artifact that leads to sparse and inconsistent sampling. The authors propose integrating wavelet transforms into the CNN architecture. The MWCNN employs wavelet decomposition to downsample feature maps, thereby expanding the receptive field while preserving important structural details. With the U-Net framework as backbone, the model utilizes inverse wavelet transform for reconstructing high-resolution outputs.

Recent advancements in multi-sensor image fusion, with a focus on methods that enhance feature preservation and accuracy. A fusion approach based on Nonsubsampled Contourlet Transform (NSCT) is proposed for combining Synthetic Aperture Radar (SAR) and Panchromatic (PAN) images [276]. NSCT provides a shift-invariant, multiscale, and multidirectional framework that better captures image details. The fusion process utilize Maximum A Posteriori (MAP) estimation using Rayleigh and Laplacian distributions for SAR despeckling, followed by an edge-based rule to fuse with PAN data which outperforms the existing NSCT-based techniques.

A recent advancement in image fusion aimed at enhancing night-vision context clarity by effectively combining infrared and visible images. The method enhances visible image details using a guided filter, followed by decomposition of both visible and infrared images via curvelet transform [277]. An improved sparse representation technique is applied to fuse the low-frequency components, while the high-frequency components are merged using a parametric adaptation of pulse-coupled neural networks. The final fused image is reconstructed through inverse curvelet transformation.

A Shift-Invariant Multiscale Wavelet (SIMW) method that eliminates downsampling during decomposition using a morphological Haar wavelet, thereby maintaining shift invariance and improving the visual quality of fused images [278]. Similarly [279] developed a fusion algorithm using Redundant Discrete Wavelet Transform (RDWT), a shift-invariant version of DWT. The algorithm show promising results for combining CT and MRI images. Evaluations on the BrainWeb database revealed that RDWT preserves both edge and structural details effectively.

Another study proposes a hybrid MMIF technique that integrates Fuzzy Set Theory with the Undecimated Discrete Wavelet Transform (UDWT) to enhance diagnostic image quality [280].

The UDWT provides multiresolution decomposition without downsampling, preserving spatial details and reducing information loss. Low-frequency subbands in the UDWT domain are fused using a maximum selection criterion, while high-frequency subbands are combined via a Modified Spatial Frequency (MSF) approach to preserve structural and edge details. The final fused image is reconstructed through inverse UDWT. Experimental results show that the proposed method enhances visual clarity and diagnostic value, outperforming existing fusion techniques in entropy, spatial frequency, standard deviation, and edge preservation.

A novel approach for fusing CT and MR images was introduced, utilizing the strengths of the nonsubsampled shearlet transform (NSST) and sparse representation (SR) to enhance fusion quality [281]. This method demonstrated improved results both visually and across various quantitative metrics. However, the technique does not address the fusion of MR with low-dose CT images, as the latter typically exhibits complex noise patterns that cannot be accurately modeled using standard Gaussian distributions.

A new technique was introduced for fusing PET and MRI images, combining the strengths of Stationary Wavelet Transform (SWT) and Nonsubsampled Contourlet Transform (NSCT) [282]. This fusion strategy enhances both spatial detail and color preservation, resulting in a composite image that retains the fine structural details of the MRI while maintaining the original color integrity of the PET image with minimal distortion. Another algorithm was introduced for multiresolution image fusion utilizing the Discrete Sine Transform (DST) [283]. The effectiveness of both DST and its inverse in handling multiresolution image processing tasks was evaluated. Analysis of the resulting error images and performance metrics indicated that applying MDST preserves all relevant information without any noticeable loss.

An image fusion approach was developed combining Nonsubsampled Contourlet Transform (NSCT) with Sparse Representation (SR) to address issues such as contrast degradation, limited decomposition level flexibility, and inadequate feature representation in existing fusion methods [284]. This method produces a fused image that avoids gray-scale inconsistencies and preserves subtle details effectively through the use of a sliding window mechanism. In conclusion, wavelet and sub-band decomposition-based fusion techniques offer significant advantages in terms of preserving critical features across multiple modalities.

### 5.5. Algorithms using Fuzzy Logic Techniques

Algorithms utilizing fuzzy logic techniques are widely applied in various domains, including medical image processing, due to their ability to handle uncertainty and imprecision inherent in real-world data. Fuzzy logic, unlike traditional binary logic, allows for the representation of data in degrees of truth rather than as absolute values. In the context of medical image processing, fuzzy logic algorithms are often used for tasks such as image segmentation, enhancement, and classification. These algorithms are effective in dealing with noisy or incomplete data, improving the accuracy and robustness of image analysis systems. Fuzzy logic can help in extracting meaningful information from images [285], even when the boundaries between different structures are not sharply defined. Fuzzy-based image enhancement methods can improve the visibility of important features in medical images, aiding clinicians in making better diagnostic decisions. Overall, fuzzy logic techniques are valuable tools in medical image processing, offering enhanced performance in handling complex, uncertain, and noisy data.

An advanced generalization of fuzzy set theory known as the intuitionistic fuzzy set (IFS) was introduced [286]. Unlike classical fuzzy sets, IFS accounts for an additional level of uncertainty by explicitly defining both membership and non-membership degrees, which collectively represent the hesitation or uncertainty inherent in real-world scenarios. The non-membership degree particularly captures vagueness effectively, making it well-suited for addressing uncertainties prevalent in various stages of image processing. Employing IFS can therefore significantly reduce ambiguity and lead to improved processing outcomes and diagnostic accuracy.

A fuzzy-based segmentation method, incorporating learning-driven fuzzy models and fusion techniques, was proposed to automatically segment tumor regions from multispectral MRI [287]. Additionally, a content-analysis approach to medical image fusion utilizing fuzzy inference principles was proposed [288]. In their method, the content of two distinct source images is analyzed using histogram data guided by carefully structured fuzzy inference rules. These rules are explicitly tailored based on fusion objectives and derived from the histograms of CT and MRI images.

An approach based on classifier fusion was proposed to enhance the performance of pattern recognition systems. The author developed a hybrid random subspace fusion method, assembling multiple fuzzy classifiers by utilizing diverse subsets of both feature space and sample regions [289]. Complementary imaging modalities such as CT, SPECT, and nuclear magnetic resonance (NMR) effectively encapsulate human anatomical and functional information. Another author proposed pixel-level medical image fusion using neuro-fuzzy logic [290]. To optimize the parameters of the membership functions, a hybrid algorithm integrating back-propagation and the least mean square method was implemented. Another study proposed an innovative multisensor, multi-modal fusion technique employing multiscale geometric analysis based on non-subsampled contourlet transform and a fuzzy-adaptive reduced pulse coupled neural network [291].

In image processing, fuzzy set theory enhances image contrast, smooths important regions, and emphasizes edges and subtle details in medical images. By employing a membership function [292]], fuzzy sets address and mitigate vagueness inherent in medical imagery. However, uncertainties may arise due to incomplete knowledge or human error during the definition of membership functions, resulting in an additional type of uncertainty known as hesitation degree. Combining the degrees of membership, non-membership, and hesitation yields the concept of Intuitionistic Fuzzy Sets [293].

Various intuitionistic fuzzy set algorithms [294, 295] effectively manage uncertainties in multi-modal images and dynamically improve contrast during fusion. Images produced by the proposed approach exhibit superior visual quality characterized by clearly defined edges, smooth textures, refined details, and freedom from artifacts. Consequently, the MMIF technique provides clinicians with enriched diagnostic information, significantly aiding in accurate diagnosis and effective treatment planning.

Performance evaluation of image fusion algorithms typically involves two approaches: objective and subjective assessments. Objective evaluations use quantitative metrics, including Peak Signal to Noise Ratio, Universal Quality Index, Structural Similarity Index Measure, Correlation Coefficient, Entropy, Spatial Frequency, Edge-based image fusion metric, and Standard Deviation, to measure the quality of fused images numerically. These metrics provide a straightforward, consistent, and convenient way to objectively compare algorithm performance and are discussed in detail [see Section 7. On the other hand, subjective evaluation relies on human perception to assess the visual quality of fused images directly. Since the ultimate goal of medical imaging is often clinical interpretation, subjective assessments are particularly valuable, as human judgment ensures that the fused images meet practical diagnostic requirements effectively.

## 6. Contributions of multi-modal image fusion in enhancing medical diagnosis and treatment strategies

In recent years, multi-modal image fusion has demonstrated significant contributions to improving the precision of medical diagnostics and the effectiveness of treatment planning. To systematically capture these advancements, this section provides a detailed exploration of the state-of-the-art research. Building on these developments, MMIF techniques and applications have advanced considerably, largely driven by the growing demand for integrating complementary information captured by different imaging modalities. By combining these distinct sources, clinicians can gain a more comprehensive perspective on both structural and functional aspects of patient anatomy, thereby enhancing diagnostic precision and treatment planning.

In support of this discussion, Table 7 presents a structured summary of notable studies from recent literature. The table categorizes the fusion techniques employed, the specific imaging modalities combined (such as MRI, CT, PET, and SPECT), and highlights the distinctive contributions made by each approach toward clinical diagnosis and therapeutic outcomes. By synthesizing this information, the table allows readers to quickly grasp which fusion strategies have been effective for particular clinical needs, such as lesion characterization, functional assessment, tumor localization, and surgical planning. Additionally, it serves as a practical reference for researchers and clinicians seeking to understand the strengths and limitations of different multi-modal combinations and the technical innovations applied. This comprehensive overview not only contextualizes current research directions but also identifies gaps and opportunities for future advancements in multi-modal medical imaging.

Wavelet transformations and Laplacian pyramids are two examples of earlier fusion techniques that laid a strong foundation by making it possible to extract important low- and high-frequency features from several images and then combine them into a single representation. These methods effectively preserved crucial anatomical details while incorporating functional signals. Over time, continued research has expanded upon these core ideas, introducing more sophisticated algorithms and data-driven frameworks that can adapt to a variety of clinical requirements [229]. Although wavelet-based fusion algorithms have been relatively successful at preserving crucial image details, they often encounter issues such as edge blurring or diminished contrast. To counter these shortcoming, more advanced methods, including dictionary learning and sparse representation, were introduced [296]. These adaptive strategies focus on selectively retaining salient image components from different modalities, thereby producing fusion results that are not only visually coherent but also more faithful to the original anatomical structures. In recent years, deep learning has revolutionized MMIF by enabling the automatic learning of optimal fusion rules directly from data, thereby minimizing the need for extensive handcrafted feature engineering. Notably, CNNs and GANs have been employed to capture local details and global contextual information, which translates into sharper fused images [297]. Such improvements are critical in clinical settings, as they help to accentuate subtle pathological features that might otherwise be overlooked.

The clinical relevance of these fusion methods is readily apparent in numerous scenarios, notably in neuroimaging. By combining the high spatial resolution of MRI with the functional or metabolic data from PET or fMRI, clinicians gain a more precise view of lesion localization and underlying pathophysiological mechanisms, which is particularly useful in diagnosing brain tumors and evaluating epilepsy [298–300]. This integrated perspective assists neurosurgeons in determining the extent of tumor resection or mapping seizure foci, ultimately leading to more targeted interventions. In oncology, PET-CT and PET-MRI fusion similarly bring structural clarity together with metabolic insights, enabling more accurate tumor identification and staging [301]. This eventually brought treatment planning, with greater precision, and effective follow-up assessments.

In cardiovascular care, MMIF techniques such as SPECT-CT and PET-MRI support the assessment of myocardial perfusion, wall motion, and vascular integrity in a single combined view [302, 303]. By overlaying functional information onto anatomical maps, cardiologists can identify ischemic territories and quantify viability more reliably. Moreover, real-time image-guided interventions often employ multi-modal integration of ultrasound with preoperative CT or MRI to enhance precised surgical navigation [304, 305]. Such fused images bolster surgeons' spatial awareness during procedures and help reduce complications, all while fitting smoothly into existing workflows. Thus, fully automated fusion pipelines are a growing interest in robust registration algorithms and unsupervised learning strategies tailored for heterogeneous or incomplete clinical datasets [306].

Looking to the future, there is an increasing emphasis on explainability, interpretability, and trustworthiness in deep learning-based fusion models. Clinicians and researchers actively explore attention mechanisms and interpretable network architectures to understand how each modality contributes to the fused output [307]. Such insights are crucial for validating diagnostic conclusions and ensuring the fusion process does not obscure critical features. Radiologists, computer scientists, and biomedical engineers must continue to work together to overcome obstacles related to varied imaging techniques, a lack of annotated data, and processing expenses. With these concerted efforts, MMIF is emerging as a key tool for personalized diagnosis and treatment across diverse clinical fields.

Table 7. Outlined literature on multi-modal image fusion techniques, used imaging modalities, and their contributions in medical diagnosis and treatment strategies

| Modalities | Technique | Contributions |
| --- | --- | --- |
| MRI, CT | Hybrid optimal CNN with a modified discrete wavelet transform [308] | Enhance diagnostic accuracy across various medical modalities |
| CT and 3D T1-Gd MRI | Unsupervised image fusion via GAN [309] | Optimize radiotherapy with detailed structural fusion |
| MR, CT, PET, SPECT | Texture and structure separation with CNN [246] | Integrity of detailed features is maintained |
| MRI, SPECT, PET, CT | Optimum weighted average fusion with PSO and DWT [310] | Optimize and robustly retain critical medical information |
| MRI, CT, PET | Adolescent identity search algorithm in the non-subsampled Shearlet transform domain [311] | image optimization and to reduce the computational cost and time |
| X-Ray, CT | Proposed a modified AlexNet (MAN) deep-learning framework to evaluate lungs abnormality [108] | An ensemble-feature technique that integrates deep and hand-crafted features to extract informative feature set. |
| MRI, PET, SPECT | Region-based image fusion based on superpixel segmentation and a post-processing optimization [312] | Performance achievement in spectral and spatial details and objective evaluation |
| MRI, CT | A dual attention fusion network using an ensemble Monte Carlo dropout strategy to effectively combine features from multiple imaging modalities [109] | Improved performance over state-of-the-art methods across five publicly available datasets |
| CT, MRI | CNN-based Muti-scale Mixed Attention Network [313] | preserve more textual details, clearer edge information and higher contrast |
| MRI, PET, CT, SPECT | Neural Network and modified Laplacian with guided image filtering [314] | Achieved better performance in terms of image fusion with improved evaluation indexes |

| Modalities | Technique | Contributions |
|---|---|---|
| CT, MRI, SPECT | Convolutional sparse representation and mutual information correlation algorithm [315] | Better results in subjective vision and objective evaluation metrics |
| MRI, CT | Integrates multi-modal deep learning features with clinical and radiological data using SVM [102] | Enhance predictive accuracy, prognostic value and multi-modal integration |
| PET, CT, MRI | Hybrid model using CNN and transformers was used to extract features from both imaging modalities. [97] | Aids in preoperative decision-making by enhancing the accuracy of pathology predictions |
| MRI, CT, PET, SPECT | Intuitionistic fuzzy set-based image fusion [96] | Enhanced fused medical images with improved quality, evaluated both subjectively and objectively |
| MRI, CT | Xception model is used for feature extraction and image fusion using non-subsampled contourlet transform [316] | optimal feature selection for fused image generation |
| CT, PET, SPECT, MRI | Convolutional extreme learning machine developed by incorporating extreme learning machine [317] | Enhanced lesion detection and localization with improved subjective visual performance |
| MRI, CT | Fusion model utilizing modified DWT with an Arithmetic Optimization Algorithm approach [94] | Optimal selection of the fusion rule parameters resulting in better image fusion |
| MRI, CT | A nonsubsampled contourlet transform-based fusion scheme employed to fuse MRI and CT [318] | Secure fused medical images using novel watermarking algorithm using redundant discrete wavelet transform |
| MRI, SPECT, PET | Multi-scale fusion network using multi-dimensional dynamic convolution and a residual hybrid transformer [93] | Enhance feature extraction, context modeling, and improved fusion performance |
| CT coronary angiography (CTCA), cardiac MR (CMR) | Researchers developed a method to fuse CT and MRI data into a single 3D image, combining anatomical and functional information [110] | The fused 3D images enhanced diagnostic accuracy for coronary artery disease by correlating areas of arterial narrowing with myocardial ischemia, aiding in better clinical decision-making |
| Raman Chemical Imaging, Histopathology | Machine learning models were developed to fuse features from RCI and histopathology images [319] | The approach significantly improved the differentiation between Gleason grade 3 and grade 4 prostate cancer tissues |

| Modalities | Technique | Contributions |
|---|---|---|
| CT, MRI | Gabor representation of multi-CNN combination and fuzzy neural network [320] | Significantly better in objective evaluation and visual quality ultimately improving fused image quality |
| MRI modalities | At pixel-level, convolutional PIF-Net and at feature-level an attention-based modality selection feature fusion module [321, 322] | Applying both pixel-level and feature-level fusion effectively improve the fusion quality |
| US, CT, MRI | A deep learning model that integrates features from US, CT, and MRI for ovarian cancer detection [101] | Achieved diagnostic performance comparable to multidisciplinary team evaluations, especially in resource-limited settings. |
| CT, MRI, PET, SPECT | Hybrid NSCT and DTCWT approach [323] | Develop hybrid methods for superior diagnostic accuracy |
| fMRI, MRI | Deep learning model based on generative models rooted in a modular approach and separable convolutional blocks [324] | Effectively fuse multiple 3D neuroimaging modalities on a voxel-wise level |
| PET, CT | A multi-scale calibration mechanism to fuse decoder semantic feature information from different modalities [98] | Enhanced segmentation performance by addressing challenges in delineating pancreatic tumors |
| T1- and T2-weighted MRI | Used image registration for alignment and basic UNets with different fusion points [107] | Provides insights into optimal fusion strategies, lead to statistically significant improvements in segmentation accuracy |
| MRI, PET | An auto-encoder for feature extraction, an information preservation weighted channel-spatial attention model for fusion, and a decoder for image reconstruction [325] | improved the quality of fused images and decreased the fusion time effectively |
| US and Infrared Thermal Imaging (IRT) | Combines features from US and IRT images using feature-level fusion [210] | Enhanced the classification accuracy of thyroid nodules |
| T2-Weighted MRI and Dynamic Contrast-Enhanced | A Self-Attention Multi-Modal Fusion framework that employs multiscale attention modules to extract rich features from T2WI and DCE sequences [105] | Enhance diagnosis and grading performance for the diagnosis of bladder cancer grading |
| CT, MRI | A structure preservation-based two-scale multi-modal fusion algorithm [167] | preserve characteristic of the iterative joint bilateral filter |

| Modalities | Technique | Contributions |
|---|---|---|
| MRI, CT, SPECT | A novel adaptive co-occurrence filter-based optimization model [95] | Outperforms image fusion methods in terms of five objective indices and subjective evaluation |
| MRI, CT | Fusion algorithm based on pulse coupled neural networks and the nonsubsampled contourlet transform methods [326] | Enhance structural and functional information in fusion processes |
| MRI, CT, PET, SPECT | A semantic-guided architecture with a mask-optimized framework [327] | A global refinement and edge enhancement module to boost the edge textures |
| MR, CT, PET | A general learning-based decomposition model using coupled dictionary learning [328] | Enhanced contrast-resolution without intensity attenuation |
| CT, MRI, SPECT | Initial weight construction using Dense SIFT and gradient magnitude is devised [329] | Improved qualitatively and quantitatively results |
| MRI modalities | Dual tree complex wavelet transform and atom search sine cosine algorithm-based deep CNN [330] | Weighted fusion model preserves critical features across various modalities |
| MRI, CT | hybrid fusion method integrating the advantages of bottom-hat–top-hat along with gray-PCA, LSIST, and CNN [331] | Preserving image details with better contrast, less noise and no information loss |
| PET, SPECT, CT, MRI | fusion rule based on modified Laplace and local energy [332] | preserve the energy information of all modalities |
| CT, MRI, SPECT | Contrast-based fusion by calculating the local facts of high-frequency subband modified sum-modified Laplacian [333] | Excellent visual preservations i.e., edge, texture and more information |
| CT, MRI | water wave optimized non-subsampled shearlet transformation technique [334] | enhanced visual feature quality, contour, and computational performance |
| CT, MRI, PET | An adaptive whale optimization algorithm with long short-term memory [335] | Better classification for Alzheimer's encephalopathy, hypertensive encephalopathy and glioma |
| PET, CT | Mirror U-Net by factorizing the multi-modal representation into modality-specific decoder branches [336] | Effective results for Brain tumor segmentation |

| Modalities | Technique | Contributions |
|---|---|---|
| CT, MR | Fusion-based CNN model using a feed-forward propagation [337] | Improved result using histogram matching, equalization, and contrast histogram equalization |
| SPECT, MRI, CT | Target information enhanced fusion using cross-modal learning and information enhancement techniques [338] | Resulting fused image could highlight tumor and edematous features |
| CT, MRI | A multi-branch and multi-scale architecture integrating attention mechanisms to refine fusion processes [339] | preserve image brightness, texture and resembles the source images in both content and structure |
| MRI, PET, SPECT | Multilevel Guided edge-preserving filtering decomposition rule [340] | Increased the visual contrast and received positive subjective review |
| MRI, CT, PET, | dynamic feature-enhanced Mamba image fusion model [152] | enhance fine-grained texture features and detect differential features |

## 7. Evaluation metrics for fusion quality

### 7.1. Average gradient

The average gradient metric evaluates the overall clarity and detail of the image by computing the average magnitude of the gradients in the image. This metric is used to assess the sharpness and clarity of an image and quantifies the overall contrast by measuring the rate of intensity change across adjacent pixels. The equation for the average gradient is illustrated in equation 1

$$AG = \frac{1}{(M)(N)} \sum_{i=1}^{M} \sum_{j=1}^{N} \sqrt{\left(\frac{\partial I(i,j)}{\partial x}\right)^2 + \left(\frac{\partial I(i,j)}{\partial y}\right)^2} \quad (1)$$

where $I(i, j)$ is the intensity of the pixel at location $(i, j)$, and $M$ and $N$ are the dimensions of the image. A higher $AG$ indicates better image sharpness and detail retention.

### 7.2. Standard deviation

Standard deviation (SD) measures the contrast or spread of the pixel intensity values around the mean intensity. A higher SD indicates more variation and, therefore, more detail and contrast. The equation for the SD is illustrated in equation 2

$$SD = \sqrt{\frac{1}{MN} \sum_{i=1}^{M} \sum_{j=1}^{N} (I(i,j) - \mu)^2} \quad (2)$$

where $\mu$ is the mean intensity of the image. A larger standard deviation implies a higher contrast, suggesting a more visually appealing fused image.

### 7.3. Edge intensity

Edge intensity measures the prominence of edges in the image. It is typically computed using edge detection operators such as Sobel or Canny, the sum of the absolute gradient magnitudes at edge pixels provides a measure of the overall edge intensity. It can be computed using equation 3.

$$EI = \sqrt{(S_x^2 + S_y^2)} \qquad (3)$$

Where, $S_x = h_x \otimes f$, $S_y = h_y \otimes f$

and $h_x$, $h_y$ are shown below:

$$\begin{bmatrix} 1 & 0 & 1 \\ -2 & 0 & 2 \\ -1 & 0 & 1 \end{bmatrix}, \qquad \begin{bmatrix} 1 & 0 & 1 \\ -2 & 0 & 2 \\ -1 & 0 & 1 \end{bmatrix}$$

### 7.4. Image Entropy

Image entropy (IE) quantifies the amount of information or randomness in an image. A higher entropy value indicates a richer and more complex image. It also quantifies the amount of information or randomness in an image and is computed using equation 4.

$$IE = -\sum_{k=0}^{L-1} p_k \log_2(p_k) \qquad (4)$$

where $p_k$ is the probability of occurrence of the $k$-th intensity level, and $L$ is the number of possible intensity levels. Higher entropy values indicate more detailed and informative images.

### 7.5. Peak signal-to-noise ratio

Peak signal-to-noise ratio (PSNR) measures the quality of a fused image compared to a reference image. It is expressed in decibels (dB), and a higher PSNR indicates better quality.

$$\text{PSNR} = 10 \log_{10}\left(\frac{\text{MAX}_I^2}{\text{MSE}}\right) \qquad (5)$$

where, $MAX_I$ is the maximum possible pixel value, and $MSE$ is the mean squared error between the fused and reference images. Higher PSNR values indicate better fidelity to the original image.

### 7.6. Xydeas and Petrovic Metric

This metric assesses the structural similarity and information content in fused images by analyzing edge information preservation. It calculates a weighted edge preservation index for each edge pixel, which is then averaged over the entire image. The higher value represents better edge information with higher information conversion from the source image.

$$Q_f^{ab} = \frac{\sum_{i=1}^{M} \sum_{j=1}^{N} \left(Q_{(i,j)}^{af} W_{(i,j)}^{af} + Q_{(i,j)}^{bf} W_{(i,j)}^{bf}\right)}{\sum_{i=1}^{M} \sum_{j=1}^{N} \left(W_{(i,j)}^{af} + W_{(i,j)}^{bf}\right)} \qquad (6)$$

### 7.7. Spatial frequency

Spatial frequency (SF) measures the overall activity level or texture of an image, combining the row and column frequency components.

$$SF = \sqrt{RF^2 + CF^2} \qquad (7)$$

where *RF* row frequency and *CF* (column frequency) are computed based on differences between adjacent rows and columns, respectively. Higher SF values indicate more texture and detail.

*7.8. Structural Similarity Index Measure*

SSIM assesses the similarity between the fused image and a reference image by considering luminance, contrast, and structure. SSIM evaluates the perceptual quality of an image by comparing luminance, contrast, and structure. It is given by:

$$\text{SSIM}(x, y) = \frac{(\mu_x^2 + \mu_y^2 + C_1)(\sigma_x^2 + \sigma_y^2 + C_2)}{(2\mu_x\mu_y + C_1)(2\sigma_{xy} + C_2)} \tag{8}$$

where $\mu_x$ and $\mu_y$ are mean intensities, $\sigma_x^2$ and $\sigma_y^2$ are variances, $\sigma_{xy}$ is the covariance, and $C_1$ and $C_2$ are small constants. SSIM values close to 1 indicate high structural similarity.

## 8. Recent and Emerging Trends in Medical Image Fusion for enhanced diagnosis

The field of MMIF has evolved significantly, fueled by improved algorithm developments alongside advancements in hardware and software technologies. Progress in deep learning algorithms, particularly CNN, has had a profound impact on the fusion process. These algorithms are capable of learning hierarchical features from multi-modal data, leading to enhanced image quality and enabling earlier and more accurate abnormality detection.

Simultaneously, the landscape of MMIF is undergoing rapid transformation, driven by technological innovations, rising clinical demands for higher diagnostic accuracy, and the maturation of machine learning and AI technologies. While traditional pixel-, feature-, and decision-level fusion methods have laid a solid foundation, the field is now shifting toward more adaptive, intelligent, and clinically robust approaches. Emerging trends reveal a clear movement toward deep learning-driven methods, transformer architectures, explainable AI models, real-time and point-of-care fusion systems, privacy-preserving solutions, and the integration of multi-omics data, all aimed at improving diagnostic precision and facilitating patient-specific therapy planning.

*8.1. Deep Learning and End-to-End Fusion Architectures*

Deep learning has revolutionized medical image fusion by replacing hand-crafted fusion rules with data-driven optimization. CNN have been widely adopted for extracting high-level semantic features from multi-modal images and performing fusion in an end-to-end trainable manner. Recent works integrate deep residual learning and multi-scale attention mechanisms to selectively emphasize important structures while suppressing noise and artifacts [94, 109]. In particular, fusion networks such as the Muti-scale Mixed Attention Network [313] and dual-attention fusion strategies [109] have demonstrated superior performance across diverse imaging combinations including MRI, PET, CT, and SPECT. Moreover, dynamic receptive field adjustment, as implemented in federated learning-based fusion frameworks [154], offers promising avenues to adaptively fuse features while respecting local anatomical variability.

GAN are another transformative development, providing the ability to generate highly realistic fused images by learning cross-modal distributions [309]. These models reduce the need for explicit feature engineering and can learn complex mappings between vastly different modalities, i.e., MRI and PET.

*8.2. Transformer-Based Fusion Models*

Transformer architectures, known for their ability to capture long-range dependencies, are emerging as strong alternatives. Recently, hybrid CNN-transformer models [210] have been employed to use local and global feature integration, resulting in fused images with better contextual coherence. Similarly, the FusionMamba model [155], based on selective structured state space models, illustrates how dynamic convolution combined with transformer principles enhances multi-modal feature extraction and fusion, especially for heterogeneous datasets. Transformers' self-attention mechanisms enable the network to prioritize the most informative regions across modalities, offering a new dimension to fusion strategies beyond simple spatial overlays.

### 8.3. Real-Time and Point-of-Care Fusion Systems

A significant trend is the development of lightweight, real-time fusion architectures suitable for point-of-care settings. Advances in AI hardware, including GPU acceleration and edge AI chips, have made it feasible to deploy MMIF models at bedside, in operating rooms, or in mobile clinics. Real-time fusion models prioritize not only accuracy but also speed and minimal computational burden, essential in acute care scenarios such as stroke diagnosis or intraoperative navigation. The development of convolutional sparse representation frameworks [315] and lightweight hybrid models [101] marks important steps toward this goal.

### 8.4. Explainable and Interpretable Fusion Models

Despite impressive technical performance, deep learning models often function as "black boxes," raising concerns regarding clinical trustworthiness. To address this, researchers are embedding interpretability into fusion pipelines. Attention-based feature fusion modules [323], saliency-guided fusion [153], and semantic-guided architectures [327] aim to visualize the contribution of each modality to the final decision-making process. This transparency not only supports clinical validation but also aligns with evolving regulatory standards that demand explainability in AI-driven healthcare tools [341].

### 8.5. Privacy-Preserving and Federated Fusion Learning

Data sharing restrictions due to regulations like The Health Insurance Portability and Accountability Act (HIPPA) [342] and General Data Protection Regulation (GDPR) [343] have hampered large-scale multi-center MMIF research. Federated learning approaches offer a promising solution by enabling models to learn across distributed datasets without transferring sensitive patient data. Early studies combining privacy-preserving federated frameworks with dynamic feature fusion mechanisms [154] show that MMIF algorithms can achieve robustness and generalization without centralizing patient images, a development critical for equitable and large-scale clinical adoption.

### 8.6. Multi-Modal and Multi-Sensor Integration

Beyond conventional imaging modalities, there is an increasing push toward integrating data from advanced sensors such as OCT, photoacoustic imaging, functional near-infrared spectroscopy (fNIRS), and even wearable biosensors. This broader multi-modal sensor fusion aims to bridge structural, functional, and molecular domains, offering a more holistic view of patient physiology. However, the integration challenges, dealing with varying resolutions, noise patterns, and dynamic acquisition conditions, necessitate more sophisticated fusion frameworks capable of heterogeneous data alignment [210].

### 8.7. Cloud-Based and Distributed Fusion Infrastructures

Handling the voluminous and computationally intense nature of multi-modal datasets requires robust infrastructures. Emerging cloud-based MMIF platforms use distributed computing and storage, enabling collaborative research, faster model training, and scalable clinical deployments [152]. Combined with edge computing for real-time applications, these architectures provide a seamless pipeline from image acquisition to fused diagnostic output.

*8.8. Advanced Quality Assessment and Benchmarking*

The absence of standard evaluation protocols has historically hindered objective comparisons across fusion methods. Recent work emphasizes comprehensive benchmarking using quantitative metrics such as SSIM, PSNR, entropy, average gradient, and clinical interpretability scores [see Section 7]. Some initiatives also explore task-specific metrics like lesion segmentation accuracy or diagnostic confidence gains [341], recognizing that fusion performance must ultimately be judged by its impact on clinical decision-making [96, 323].

## 9. Challenges in Multi-modal Medical Image Fusion

MMIF has the potential to revolutionize clinical diagnostics and treatment planning by integrating complementary information from various imaging modalities. However, MMIF also brings several challenges [344, 345], including computational complexity, required development in strong fusion algorithms and their validation are further hampered by the scarcity of high-quality, annotated multi-modal datasets. Additionally, the heterogeneity of imaging modalities, coupled with a lack of standardization in data formats and acquisition protocols, complicates the fusion process. Limited hardware resources in underfunded clinical settings amplify these challenges, leading to inequities in access to advanced MMIF technologies. Most significantly data privacy is critical in facilitating medical data sharing, privacy concerns and stringent regulatory requirements and the necessity for rigorous clinical validation, underscore the urgent need for secure, reliable, and efficient solutions in advancing the future of MMIF.

*9.1. Computational complexity and time constraints*

One of the most significant challenges in MMIF is the intensive computational effort required to process large scale, and high dimensional data. Advancements in imaging technologies have led to the capture of medical images with progressively higher resolutions, resulting in an exponential growth in data volume and imposing significant computational challenges. Advanced algorithms like deep neural networks and iterative optimization for improved accuracy and diagnostic utility. However, these approaches require significant computational resources, with deep learning models demanding substantial time and hardware for both training and inference, particularly when processing large or multiple image volumes. Real-time image fusion is crucial in intervention-based procedures i.e., surgeries, where timely and accurate imaging could help in critical decisions. Delays from computational overhead can impact patient care, making it essential to balance algorithm complexity and speed. Efficient techniques like parallelization, GPU acceleration, or simplified models are preferred for delivering accurate results quickly. Moreover, imaging technology advancements acquire high resolution images which intensifies the computational demands, resulting in scalable fusion algorithms and robust hardware. Sophisticated high-parameter algorithms can yield highly refined fused images [346], however, these gains must be evaluated against the time constraints of practical clinical workflows. The computational complexity and time constraints in MMIF stem from the interplay of large-scale, high-dimensional data, the use of advanced algorithms, and need for real-time accurate results.

*9.2. Data availability and quality*

The development and validation of robust MMIF algorithms depends on the accessibility of high-quality, properly annotated datasets. However, acquiring these datasets in sufficient volume and diversity is challenging due to several factors, including cost, logistical complexities, data privacy and sharing, variability in imaging protocols, scarcity of multi-modal datasets. Healthcare data sharing is adhere to tight regulations [347] by legal authorities such as HIPPA [342] in the United States and GDPR [343] is a European Union, given the sensitivity of patient information.

The data must comply with stringent protocols for de-identification, anonymization, and secure transmission to ensure the protection of patient confidentiality. These regulations often restrict the shareability of multi-modal datasets, resulting in fragmented and smaller data collections that hinder large-scale collaborative research efforts. Moreover, different medical institutions employ diverse imaging protocols, equipment vendors, and scanner configurations, creating inconsistencies that hinder standardized datasets. This limits algorithm generalization, reducing the reproducibility and reliability of fusion methods across diverse imaging conditions.

Single-modality datasets (e.g., large MRI, X-Ray or CT cohorts) are increasingly available, however, comprehensive multi-modal datasets remain relatively rare. Simultaneous capture of various modalities is more difficult because to high expense, logistical challenges, and privacy constraints. Consequently, relying on smaller, domain-specific datasets or synthetic data, limits the ability to rigorously test and validate the generalizability of new fusion algorithms hindering fair benchmarking of algorithms due to inconsistent datasets and evaluation metrics. Addressing data challenges requires institutional collaboration through initiatives like federated learning and secure repositories, enabling training on distributed data while protecting privacy. In addition, standardized data formats and generalized annotation protocols can further unify multi-modal datasets.

### 9.3. Heterogeneity and standardization

Multi-modal medical imaging involves combining information from distinct technologies that each rely on different factors indicated in Tables 1, 2 complicates image registration and necessitating modality-specific corrections to address noise and artifacts. This leads to challenges in image registration, which increases computational demands and affects fusion reliability. Differences in acquired data formats, and vendor-specific calibrations, despite DICOM standards create non-uniform dataset. Achieving seamless fusion demands precise calibration and normalization procedures [348]. Global standards, shared protocols, and advanced machine learning models, and open-source software toolkits can greatly reduce heterogeneity to improve multi-modal image fusion reliability.

### 9.4. Hardware and infrastructure limitation

High-performance hardware machine requirements i.e., GPUs, large-memory servers, and high-bandwidth networking often underpins the computationally demanding algorithms used in MMIF. Outside well-funded research institutes or major medical centers, limited resources restrict access to advanced fusion technologies for many facilities. Deep learning models and advanced image registration routines greatly benefit from GPU acceleration [349], but acquiring and maintaining such hardware can strain budgets and technical resources [350, 351]. These resource limitations can exacerbate inequalities in smaller clinics or rural hospitals providing healthcare may find it prohibitive to implement real-time multi-modal image fusion. The inability to rapidly fuse multiple imaging modalities in these settings might delay diagnosis and limit the effectiveness of intervention strategies. Even if the advanced hardware is available, supporting infrastructure like reliable internet connectivity for cloud-based processing can be equally critical. High-bandwidth networks are essential for transferring large imaging datasets to remote servers, yet many facilities, especially those in rural or underdeveloped regions, may not possess the network infrastructure to support such data-intensive operations. As a result, hardware and infrastructure limitations remain significant obstacles to the broader adoption and equitable distribution of sophisticated MMIF technologies.

### 9.5. Reliability and validation in clinical settings

Advanced image fusion techniques require rigorous validation through costly and time-consuming multi-center trials to ensure accuracy, safety, and generalizability across diverse populations and scanners. Errors in fused images during clinical trials can carry serious consequences, including misdiagnosis or inappropriate treatment planning. Therefore, the burden of proof for accuracy and reproducibility is high. Validation protocols need to confirm that the technology reliably performs under a range of real-world conditions such as different patient anatomies, imaging devices, and clinical workflows. Adding to these challenges, fusion algorithms face stringent regulatory hurdles as they must comply with medical device regulations to ensure patient safety. This requirement often lengthens the translation process from research prototype to clinical product, making it crucial that developers and clinicians collaborate early on to align technical advances with clinical needs and regulatory frameworks.

### 9.6. Data privacy and security

Medical data contains personal information, making privacy a critical concern. Regulations such as HIPAA [342] and GDPR [343] stringent requirements for data protection. These rules complicate large-scale data sharing, hindering collaborative initiatives to develop and refine multi-modal fusion algorithms. Balancing the need for privacy-preserving data practices with the requirements of algorithm training and validation remains a complex challenge. Ensuring secure data transmission and storage is paramount to preventing unauthorized access. Techniques like de-identification, encryption, and secure authentication protocols can help protect patient confidentiality. However, these measures can also introduce technical and logistical hurdles that slow down research collaborations and add complexity to clinical workflows. Additionally, robust cybersecurity infrastructure is essential. As fusion algorithms gain traction in healthcare, institutions must adopt and continually update best practices for data protection.

## 10. Future Perspectives in Multi-modal Medical Image Fusion

The future of MMIF promises to reshape clinical practice by pushing the boundaries of diagnostic precision, real-time decision-making, and personalized treatment planning. As the complexity and volume of medical imaging data continue to grow, there is an increasing need for more intelligent, efficient, and interoperable fusion techniques. A major frontier lies in the integration of AI, where deep learning and advanced machine learning models offer the potential to automate fusion processes, optimize feature extraction, and enhance interpretability without sacrificing diagnostic reliability. Alongside AI integration, the development of real-time and point-of-care fusion solutions is critical for enabling faster, bedside decision support, particularly in emergency and resource-constrained environments [352]. Standardization and interoperability efforts will also become increasingly important, ensuring that multi-modal systems from different manufacturers can communicate seamlessly, facilitating broader clinical adoption. Moreover, the integration of multi-modal sensor technologies, beyond traditional imaging modalities, is anticipated to expand the functional capabilities of fusion systems, bridging anatomical, physiological, and molecular information in a unified framework. Cloud-based and distributed computing infrastructures are likely to play a pivotal role as well, providing scalable platforms for storing, processing, and fusing massive datasets while supporting collaborative, multi-institutional research. Finally, as MMIF technologies mature, robust quality assessment frameworks and clear regulatory pathways will be essential to validate clinical utility and ensure patient safety. Together, these future directions highlight an exciting transformation in the field, moving toward smarter, faster, and more accessible imaging solutions that will redefine the standards of modern healthcare.

### 10.0.1. Integration of artificial intelligence

AI is revolutionizing MMIF by addressing longstanding challenges such as misalignment, noise, and partial volume effects. Conventional fusion techniques often rely on handcrafted features and manual parameter tuning, which can be prone to error and time consuming. However, deep learning models learn complex relationships between imaging modalities such as CT, MRI, PET, and ultrasound largely without extensive human intervention. This ability to automatically adapt to different data distributions holds significant promise for improving diagnostic accuracy and efficiency.

Transformer-based models [93, 323] and GANs [297, 306] are key drivers of this paradigm shift. Transformers excel at capturing long-range dependencies in data, making them suitable for identifying subtle anatomical correspondences across modalities. GANs, on the other hand, can synthesize realistic images from different input domains, facilitating the creation of high-fidelity fused images. The advance fusion systems likely involve minimizing manual preprocessing steps and would streamlining the entire process. Moreover, privacy preserving techniques such as transfer learning and federated learning are also set to play a pivotal role. Overall, the integration of AI in MMIF presents a clear path toward automated, high precision imaging solutions [353]. As these technologies mature, clinicians can expect more accurate diagnoses, faster patient throughput, and improved treatment planning.

### 10.0.2. Real-time and point-of-care fusion

Real-time and point-of-care fusion represents an emerging frontier in MMIF, aiming to deliver immediate, high-quality fusion results required in the operating rooms or mobile diagnostic units. Advances in specialized AI chips and high-performance GPUs are driving these capabilities, allowing clinicians to access fused images in real time or near real time. This shift not only reduces the dependency on large, centralized systems but also places critical diagnostic information directly in the hands of healthcare professionals during patient care. Achieving reliable performance in clinical environments requires hardware and algorithmic innovation. Next-generation hardware accelerators are becoming more powerful providing clinicians on spot fused images, enabling faster decision-making. Moving forward, continued research and development will likely produce even more portable and capable systems, cementing the role of real-time, point-of-care fusion as a vital tool for precision medicine.

### 10.0.3. Standardization and interoperability efforts

Standardization and interoperability are key factors in the widespread adoption and success of MMIF. Now, industry stakeholders, healthcare providers, and regulatory agencies are working together to develop common protocols for image acquisition, storage, and processing workflows. These standards will help ensuring that fusion algorithms behave reliably across different clinical environments and equipment. Multi-center trials can be conducted with fewer logistical hurdles, accelerating the development of both new algorithms and practical clinical applications. As a result, innovations in MMIF can transition more quickly from the research lab to routine clinical use.

### 10.0.4. Multi-modal sensor integration

Multi-modal sensor integration is expanding the horizons of MMIF by incorporating data sources beyond the classical modalities of CT, MRI, PET, and ultrasound. Technologies such as OCT, photoacoustic imaging, and wearable biosensors offer detailed insights into physiological and anatomical processes at varying spatial and temporal scales. By blending these different perspectives, clinicians can gain indepth insights of patient health, potentially leading to earlier diagnoses and more precise therapies. One major challenge in integrating diverse sensors lies in reconciling their varying resolutions, signal intensities, and data formats. As these technologies become more prevalent, future fusion algorithms must be equipped to handle heterogeneous streams of data, merging them effectively into a cohesive representation.

### 10.0.5. Cloud-based and distributed computing

Cloud-based and distributed computing are becoming increasingly important for MMIF, particularly given the computational intensity and large data volumes involved. By migrating data processing to remote servers or distributed networks, healthcare facilities can tap into virtually limitless computing power and storage capabilities, potentially overcoming local hardware constraints. This arrangement also streamlines collaboration, secure sharing of large datasets, robust algorithm development and their validation. Edge computing, which processes data closer to the point of collection, complements these cloud-based models by addressing bandwidth and latency issues. For time-sensitive applications edge-based systems can execute critical tasks locally, reducing the delay associated with transmitting data back and forth to the cloud. This hybrid approach allows on-premises processing and real-time decision support. Moreover the cloud offers scalable storage, analytical tools, and shared computing power for longer-term or larger-scale tasks. These frameworks will support increasingly sophisticated MMIF algorithms that can operate efficiently in a variety of clinical settings, from major hospitals to smaller outpatient centers and mobile diagnostic units.

### 10.0.6. Advanced quality assessment and regulatory approvals

The regulatory licensing process and quality assessment are expected to be crucial to the broad use of MMIF. As fusion algorithms grow more sophisticated, so does the need for rigorous methods to evaluate their performance. Developing standardized benchmarking datasets and metrics will be essential for quantifying quality on multiple fronts. Governments and healthcare agencies require robust evidence that MMIF solutions meet established standards for safety and efficacy. By engaging regulatory authorities early in the development process, researchers and developers can gain clearer insights into tailoring their studies, datasets, and validation procedures to meet approval requirements. Overall, a structured framework for quality assessment coupled with harmonized regulatory oversight will foster innovation while maintaining high standards.

## 11. Discussion

The field of MMIF has witnessed significant evolution over the past decades, transitioning from simple pixel-level operations to sophisticated deep learning and transformer-based frameworks. This progression reflects a growing recognition of the limitations inherent in single-modality imaging and the complementary strengths offered by different imaging techniques such as CT, MRI, PET, SPECT, and ultrasound [1, 32, 71]. As this review has systematically outlined, integrating anatomical, functional, and molecular information through MMIF provides a more comprehensive, accurate, and clinically actionable representation of patient health, which is critical for early diagnosis, precise intervention planning, and effective monitoring.

Initially, classical fusion strategies focused predominantly on spatial-domain or transform-domain techniques [71, 118] utilizing mathematical transformations like discrete wavelet transforms, Laplacian pyramids, and contourlet decompositions to enhance visual clarity and preserve salient structural features. However, while these methods demonstrated considerable utility, they often struggled with challenges such as noise amplification, registration errors, and insufficient robustness against modality-specific artifacts [71, 168]. The advent of intelligent-based and hybrid fusion methods marked a pivotal step forward, enabling adaptive fusion strategies through optimization algorithms like PSO [121], fuzzy inference systems [124], and spiking cortical models [144]. These approaches provided improved flexibility and resilience, particularly in handling heterogeneous and noisy datasets, yet their reliance on manual feature engineering and rule definition limited their scalability in complex clinical scenarios.

The recent and ongoing integration of deep learning into MMIF represents a transformative paradigm shift. By autonomously learning hierarchical representations from large multi-modal datasets, deep CNNs have drastically improved the quality, robustness, and automation of the fusion process [313, 314]. Advanced architectures such as Muti-scale Mixed Attention Networks [313] and dual attention fusion frameworks [109] have demonstrated remarkable capabilities in selectively emphasizing modality-specific information while preserving critical anatomical and functional features. Moreover, the incorporation of transformer-based architectures [155, 210] has further refined the fusion process by capturing long-range dependencies and subtle cross-modal interactions that traditional CNNs might overlook. This evolution underscores the shift towards fully end-to-end, data-driven fusion pipelines that require minimal human intervention while offering superior diagnostic reliability.

Nevertheless, as MMIF methodologies become increasingly complex and powerful, new challenges have emerged. Computational demands have escalated, necessitating high-performance hardware and cloud-based infrastructures to enable real-time or near-real-time fusion, particularly for point-of-care and intraoperative applications [152]. Furthermore, data availability and standardization remain significant bottlenecks. Despite growing open-access initiatives, truly large-scale, multi-center, multi-modal datasets are still scarce, and data heterogeneity across imaging platforms complicates the training and validation of generalizable models [99, 154]. Addressing these challenges will require concerted efforts toward federated learning frameworks [154], standardized data acquisition protocols, and more robust, domain-specific data augmentation strategies.

Another critical concern pertains to the interpretability and trustworthiness of deep learning-based fusion systems. In clinical settings, the black-box nature of many AI-driven models poses barriers to adoption, as clinicians demand transparency and explainability to justify diagnostic or therapeutic decisions. Encouragingly, emerging trends toward explainable MMIF, including attention-based visualizations [321], semantic-guided fusion architectures [327], and interpretable feature attribution techniques, show promise in bridging this gap. Ensuring that AI-enhanced fusion models are not only accurate but also understandable to medical practitioners will be key to their successful clinical translation.

Moreover, privacy and security concerns associated with medical imaging data continue to grow, especially given the sensitivity of patient information and stringent regulations like HIPAA and GDPR [154]. Privacy-preserving methods such as federated learning and encrypted computation are rapidly gaining attention, enabling collaborative model development across institutions without compromising patient confidentiality. Integrating such techniques into MMIF pipelines will be indispensable for enabling large-scale, multi-institutional research and facilitating equitable access to advanced diagnostic technologies.

From a clinical perspective, MMIF has already shown substantial impact across diverse medical specialties. In oncology, fusion of PET and CT has become standard practice, significantly enhancing tumor localization, staging, and therapy monitoring [111, 301]. Neurological applications have similarly benefited from MRI-PET and MRI-SPECT fusion, improving the diagnosis and management of conditions such as epilepsy, Alzheimer's disease, and brain tumors [96, 100, 298]. Cardiovascular imaging has also been revolutionized by fusion systems, enabling better assessment of myocardial perfusion and structural integrity [6, 7, 302]. These real-world successes validate the enormous potential of MMIF to revolutionize clinical workflows and patient outcomes.

Looking forward, the future of MMIF will likely be shaped by several converging trends. Multi-modal sensor integration is broadening the scope of fusion beyond conventional imaging to include functional, molecular, and biomechanical data from emerging modalities like OCT, photoacoustic imaging, and biosensors [105, 210]. The rise of cloud and edge computing infrastructures [152] will facilitate the deployment of MMIF models in resource-limited settings, democratizing access to precision imaging. Furthermore, hybrid fusion systems that integrate imaging data with genomic, proteomic, and clinical metadata are on the horizon, promising truly comprehensive, individualized patient profiles for precision medicine [3, 88]. However, achieving these ambitious goals will require not only technical innovation but also interdisciplinary collaboration among clinicians, engineers, computer scientists, and regulatory bodies to ensure that emerging solutions are clinically relevant, ethically sound, and rigorously validated.

Overall, MMIF stands at the forefront of technological and clinical innovation. The evolution from simple pixel-based methods to AI-driven, real-time, and interpretable fusion architectures represents a significant stride toward holistic, patient-centric care. By addressing the current challenges and utlizing the opportunities presented by emerging trends, MMIF can profoundly transform the landscape of medical imaging, paving the way for earlier diagnoses, improving therapeutic outcomes, and achieving truly personalized medicine.

## 12. Conclusion

Medical imaging has long been a cornerstone of modern healthcare, yet the inherent limitations of single-modality imaging have increasingly necessitated the development of MMIF techniques. This review systematically examines how MMIF synthesizes complementary data from diverse imaging modalities, including CT, MRI, PET, SPECT, and US, thereby offering clinicians a comprehensive perspective on anatomical structures, functional processes, and molecular activities. This synergy between modalities has become vital for enhancing diagnostic confidence, supporting precise therapeutic interventions, and advancing the realization of personalized medicine.

This study has traced the evolution of fusion methodologies from classical pixel-, feature-, and decision-level techniques, through the adoption of transform-domain approaches, intelligent optimization strategies, and hybrid frameworks, to the recent paradigm shift driven by deep learning and transformer-based models. These advancements have significantly improved the robustness, efficiency, and interpretability of the fusion process, enabling more accurate lesion detection, disease characterization, and treatment monitoring across various clinical domains including oncology, neurology, cardiology, and orthopedics.

At the same time, the field has entered a dynamic new phase characterized by emerging trends such as the integration of real-time and point-of-care fusion systems, privacy-preserving federated learning architectures, explainable AI-driven fusion frameworks, and multi-modal sensor fusion extending beyond traditional imaging domains. These innovations promise to expand the clinical reach of MMIF, democratizing access to precision imaging in both advanced healthcare settings and resource-limited environments.

Nevertheless, significant challenges remain. The computational complexity of deep learning models, the lack of extensive, standardized multi-modal datasets, the variability of imaging protocols, and issues related to data privacy and regulatory compliance persist in hindering the comprehensive clinical adoption of MMIF. Moreover, the "black-box" nature of many AI algorithms underscores the urgent need for transparent, interpretable models that can gain clinician trust and regulatory approval.

Looking forward, the effective integration of MMIF into standard clinical practice can rely on ongoing multidisciplinary collaboration among radiologists, computer scientists, engineers, and policymakers. It would require the development of standardized benchmarking frameworks, scalable cloud and edge computing infrastructures, and rigorous clinical validation pipelines to ensure safety, reliability, and generalizability. Furthermore, the integration of multi-omics data with fused imaging information holds the potential to usher in a new era of deeply personalized, predictive medicine.

Ultimately, MMIF finds itself at a strategic crossroads with far-reaching implications. The foundations laid by decades of innovation are now being accelerated by the convergence of AI, advanced sensing technologies, and big data analytics. By continuing to address existing challenges while embracing emerging trends, MMIF is poised to fundamentally reshape diagnostic imaging, improving clinical decision making, improving patient outcomes, and contributing profoundly to the future of precision healthcare.

## Competing interests

The authors declare that there are no financial, personal, or professional conflicts that could have appeared to influence the work reported in this paper.

## References

1. H. A. Ahmad, H. J. Yu, and C. G. Miller, "Medical imaging modalities," Med. imaging clinical trials pp. 3–26 (2014).
2. S. M. S. Islam, M. A. A. Nasim, I. Hossain, et al., "Introduction of medical imaging modalities," in Data Driven Approaches on Medical Imaging, (Springer, 2023), pp. 1–25.
3. B. Rajalingam, R. Priya, and R. Scholar, "Review of multimodality medical image fusion using combined transform techniques for clinical application," Int. J. Sci. Res. Comput. Sci. Appl. Manag. Stud. **7**, 1–8 (2018).
4. J. Joy, I. Cooke, and M. Love, "Is ultrasound safe?" The Obstet. & Gynaecol. **8**, 222–227 (2006).
5. M. H. Wink, H. Wijkstra, J. J. De La Rosette, and C. A. Grimbergen, "Ultrasound imaging and contrast agents: a safe alternative to mri?" Minim. Invasive Ther. & Allied Technol. **15**, 93–100 (2006).
6. S. Ederhy, N. Mansencal, P. Réant, et al., "Role of multimodality imaging in the diagnosis and management of cardiomyopathies," Arch. cardiovascular diseases **112**, 615–629 (2019).
7. G. Muscogiuri, V. Volpato, R. Cau, et al., "Application of ai in cardiovascular multimodality imaging," Heliyon **8** (2022).
8. M. Lemasle, Y. Lavie Badie, E. Cariou, et al., "Contribution and performance of multimodal imaging in the diagnosis and management of cardiac masses," The Int. J. Cardiovasc. Imaging **36**, 971–981 (2020).
9. A. Gautam and P. Komal, "Diagnosis of neurological and non-neurological disorders via bimodal/multimodal imaging with lanthanide based nanoparticles," Coord. Chem. Rev. **532**, 216527 (2025).
10. J. Rokicki, T. Wolfers, W. Nordhøy, et al., "Multimodal imaging improves brain age prediction and reveals distinct abnormalities in patients with psychiatric and neurological disorders," Hum. brain mapping **42**, 1714–1726 (2021).
11. M. Grossman, "Integrated multimodal imaging in neurodegenerative disease," The Lancet Neurol. **14**, 973–975 (2015).
12. M. Haribabu, V. Guruviah, and P. Yogarajah, "Recent advancements in multimodal medical image fusion techniques for better diagnosis: an overview," Curr. Med. Imaging **19**, 673–694 (2023).
13. M. Zubair, M. Owais, T. Hassan, et al., "An interpretable framework for gastric cancer classification using multi-channel attention mechanisms and transfer learning approach on histopathology images," Sci. Reports **15**, 13087 (2025).
14. T. E. Yankeelov, R. G. Abramson, and C. C. Quarles, "Quantitative multimodality imaging in cancer research and therapy," Nat. Rev. Clin. Oncol. **11**, 670–680 (2014).
15. V. V. Tuchin, J. Popp, and V. Zakharov, Multimodal optical diagnostics of cancer (Springer, 2020).
16. M. Zubair, M. Owais, T. Mahmood, et al., "Enhanced gastric cancer classification and quantification interpretable framework using digital histopathology images," Sci. Reports **14**, 22533 (2024).
17. M. J. Gollub, R. Hong, D. M. Sarasohn, and T. Akhurst, "Limitations of ct during pet/ct," J. Nucl. Med. **48**, 1583–1591 (2007).
18. A. Rahmim, J. Qi, and V. Sossi, "Resolution modeling in pet imaging: theory, practice, benefits, and pitfalls," Med. physics **40**, 064301 (2013).
19. M. Zubair, M. Umair, R. A. Naqvi, et al., "A comprehensive computer-aided system for an early-stage diagnosis and classification of diabetic macular edema," J. King Saud Univ. Inf. Sci. **35**, 101719 (2023).
20. M. Zubair, J. Ahmad, F. Alqahtani, et al., "Automated grading of diabetic macular edema using color retinal photographs," in 2022 2nd International Conference of Smart Systems and Emerging Technologies (SMARTTECH), (IEEE, 2022), pp. 1–6.
21. M. Zubair, S. A. Khan, and U. U. Yasin, "Classification of diabetic macular edema and its stages using color fundus image," J. Electron. Sci. Technol. **12**, 187–190 (2014).
22. M. Zubair, M. Umair, and M. Owais, "Automated brain tumor detection using soft computing-based segmentation technique," in 2023 3rd International Conference on Computing and Information Technology (ICCIT), (IEEE, 2023), pp. 211–215.
23. S. Iqbal, A. N. Qureshi, M. Alhussein, et al., "A novel reciprocal domain adaptation neural network for enhanced diagnosis of chronic kidney disease," Expert Syst. **42**, e13825 (2025).
24. A. Shabbir and M. Zubair, "Interpretable deep learning classifier using explainable ai for non-small cell lung cancer," in 2024 Horizons of Information Technology and Engineering (HITE), (IEEE, 2024), pp. 1–6.
25. E. Bercovich and M. C. Javitt, "Medical imaging: from roentgen to the digital revolution, and beyond," Rambam Maimonides medical journal **9** (2018).
26. H. Huang, D. R. Aberle, R. Lufkin, et al., "Advances in medical imaging," Ann. internal medicine **112**, 203–220 (1990).

27. B. Grignon, L. Mainard, M. Delion, et al., "Recent advances in medical imaging: anatomical and clinical applications," Surg. Radiol. Anat. **34**, 675–686 (2012).
28. E. Elyan, P. Vuttipittayamongkol, P. Johnston, et al., "Computer vision and machine learning for medical image analysis: recent advances, challenges, and way forward," Artif. Intell. Surg. **2**, 24–45 (2022).
29. O. Ayo-Farai, B. A. Olaide, C. P. Maduka, and C. C. Okongwu, "Engineering innovations in healthcare: a review of developments in the usa," Eng. Sci. & Technol. J. **4**, 381–400 (2023).
30. E. Munari, A. Scarpa, L. Cima, et al., "Cutting-edge technology and automation in the pathology laboratory," Virchows Arch. **484**, 555–566 (2024).
31. A. Haleem, M. Javaid, R. P. Singh, and R. Suman, "Medical 4.0 technologies for healthcare: Features, capabilities, and applications," Internet Things Cyber-Physical Syst. **2**, 12–30 (2022).
32. E. D. Bigler, "Structural imaging." (2005).
33. M. Ingvar, "Pain and functional imaging," Philos. Trans. Royal Soc. London. Ser. B: Biol. Sci. **354**, 1347–1358 (1999).
34. H. L. Gallagher and C. D. Frith, "Functional imaging of 'theory of mind'," Trends cognitive sciences **7**, 77–83 (2003).
35. T. J. Fuchs and J. M. Buhmann, "Computational pathology: challenges and promises for tissue analysis," Comput. Med. Imaging Graph. **35**, 515–530 (2011).
36. T. Bashore, "Fundamentals of x-ray imaging and radiation safety," Catheter. cardiovascular interventions **54**, 126–135 (2001).
37. A. O. Olatunji, J. A. Olaboye, C. C. Maha, et al., "Revolutionizing infectious disease management in low-resource settings: The impact of rapid diagnostic technologies and portable devices," Int. J. Appl. Res. Soc. Sci. **6**, 1417–1432 (2024).
38. J. H. Thrall, X. Li, Q. Li, et al., "Artificial intelligence and machine learning in radiology: opportunities, challenges, pitfalls, and criteria for success," J. Am. Coll. Radiol. **15**, 504–508 (2018).
39. H. Chen, M. M. Rogalski, and J. N. Anker, "Advances in functional x-ray imaging techniques and contrast agents," Phys. Chem. Chem. Phys. **14**, 13469–13486 (2012).
40. K. Hellier, E. Benard, C. C. Scott, et al., "Recent progress in the development of a-se/cmos sensors for x-ray detection," Quantum Beam Sci. **5**, 29 (2021).
41. G. L. Bosco, "Development and application of portable, hand-held x-ray fluorescence spectrometers," TrAC Trends Anal. Chem. **45**, 121–134 (2013).
42. P. R. Patel and O. De Jesus, "Ct scan," (2021).
43. L. W. Goldman, "Principles of ct and ct technology," J. nuclear medicine technology **35**, 115–128 (2007).
44. T. Donath, F. Pfeiffer, O. Bunk, et al., "Toward clinical x-ray phase-contrast ct: demonstration of enhanced soft-tissue contrast in human specimen," Investig. radiology **45**, 445–452 (2010).
45. G. N. Hounsfield, "Computerized transverse axial scanning (tomography): Part 1. description of system," The Br. journal radiology **46**, 1016–1022 (1973).
46. C. H. McCollough, S. Leng, L. Yu, and J. G. Fletcher, "Dual-and multi-energy ct: principles, technical approaches, and clinical applications," Radiology **276**, 637–653 (2015).
47. W. R. Webb and C. B. Higgins, Thoracic imaging: pulmonary and cardiovascular radiology (Lippincott Williams & Wilkins, 2011).
48. D. Weishaupt, V. D. Köchli, B. Marincek, et al., How does MRI work?: an introduction to the physics and function of magnetic resonance imaging, vol. 2 (Springer, 2006).
49. M. Zubair, S. R. Murris, K. Isa, et al., "Divergent whole brain projections from the ventral midbrain in macaques," Cereb. Cortex **31**, 2913–2931 (2021).
50. M. Garwood and K. Uğurbil, "Rf pulse methods for use with surface coils: Frequency-modulated pulses and parallel transmission," J. Magn. Reson. **291**, 84–93 (2018).
51. N. Boulant and L. Quettier, "Commissioning of the iseult cea 11.7 t whole-body mri: Current status, gradient–magnet interaction tests and first imaging experience," Magn. Reson. Mater. Physics, Biol. Med. **36**, 175–189 (2023).
52. L. J. O'Donnell, A. Daducci, D. Wassermann, and C. Lenglet, "Advances in computational and statistical diffusion mri," NMR Biomed. **32**, e3805 (2019).
53. G. A. Santoro, G. Di Falco, and G. Santoro, "Fundamental principles of ultrasound imaging," Benign Anorectal Dis. Diagn. with Endoanal Endorectal Ultrasound New Treat. Options pp. 3–9 (2006).
54. F. O. Walker, "Basic principles of ultrasound," in Neuromuscular ultrasound, (Elsevier, Philadelphia, 2011), pp. 1–23.
55. S. K. Bhargava, Principles and practice of ultrasonography (Jaypee Brothers Medical Publishers, 2020).
56. H. F. Routh, "Doppler ultrasound," IEEE Eng. Med. Biol. Mag. **15**, 31–40 (1996).
57. J. Bamber and M. Tristam, "Diagnostic ultrasound," The physics medical imaging pp. 319–388 (1988).
58. S. Ogawa, T.-M. Lee, A. R. Kay, and D. W. Tank, "Brain magnetic resonance imaging with contrast dependent on blood oxygenation." proceedings National Acad. Sci. **87**, 9868–9872 (1990).
59. N. K. Logothetis, "What we can do and what we cannot do with fmri," Nature **453**, 869–878 (2008).
60. J. Goense, Y. Bohraus, and N. K. Logothetis, "fmri at high spatial resolution: implications for bold-models," Front. computational neuroscience **10**, 66 (2016).
61. W. Mier and D. Mier, "Advantages in functional imaging of the brain," Front. human neuroscience **9**, 249 (2015).
62. P. Gowland and R. Bowtell, "Theoretical optimization of multi-echo fmri data acquisition," Phys. Med. & Biol. **52**,

1801 (2007).

63. L. Lemieux, K. Whittingstall, and K. Uludağ, "Combining fmri with other modalities: multimodal neuroimaging," fMRI: From Nucl. Spins to Brain Funct. pp. 739–768 (2015).
64. D. A. Orringer, D. R. Vago, and A. J. Golby, "Clinical applications and future directions of functional mri," in Seminars in neurology, vol. 32 (Thieme Medical Publishers, 2012), pp. 466–475.
65. K. Specht, "Current challenges in translational and clinical fmri and future directions," Front. psychiatry **10**, 924 (2020).
66. M. H. Lee, C. D. Smyser, and J. S. Shimony, "Resting-state fmri: a review of methods and clinical applications," Am. J. neuroradiology **34**, 1866–1872 (2013).
67. S. M. Usman, S. Khalid, A. Tanveer, et al., "Multimodal consumer choice prediction using eeg signals and eye tracking," Front. Comput. Neurosci. **18**, 1516440 (2025).
68. J. D. Power, D. A. Fair, B. L. Schlaggar, and S. E. Petersen, "The development of human functional brain networks," Neuron **67**, 735–748 (2010).
69. H.-J. Park and K. Friston, "Structural and functional brain networks: from connections to cognition," Science **342**, 1238411 (2013).
70. P. M. Matthews, G. D. Honey, and E. T. Bullmore, "Applications of fmri in translational medicine and clinical practice," Nat. Rev. Neurosci. **7**, 732–744 (2006).
71. S. R. Cherry, J. A. Sorenson, and M. E. Phelps, Physics in nuclear medicine (Soc Nuclear Med, 2013).
72. S. Basu, T. C. Kwee, S. Surti, et al., "Fundamentals of pet and pet/ct imaging," Ann. New York Acad. Sci. **1228**, 1–18 (2011).
73. M. M. Khalil et al., "Basic science of pet imaging," Tech. rep., Springer (2017).
74. B. S. Santos and M. J. Ferreira, "Positron emission tomography in ischemic heart disease," Revista Portuguesa de Cardiol. **38**, 599–608 (2019).
75. S. F. Barrington and M. J. O'doherty, "Limitations of pet for imaging lymphoma," Eur. J. Nucl. Med. Mol. Imaging **30**, S117–S127 (2003).
76. N. Christian, J. A. Lee, A. Bol, et al., "The limitation of pet imaging for biological adaptive-imrt assessed in animal models," Radiother. oncology **91**, 101–106 (2009).
77. T. Beyer, G. Antoch, S. Müller, et al., "Acquisition protocol considerations for combined pet/ct imaging," J. Nucl. Med. **45**, 25S–35S (2004).
78. J. G. Mannheim, A. M. Schmid, J. Schwenck, et al., "Pet/mri hybrid systems," in Seminars in nuclear medicine, vol. 48 (Elsevier, 2018), pp. 332–347.
79. L. Saint-Aubert, L. Lemoine, K. Chiotis, et al., "Tau pet imaging: present and future directions," Mol. neurodegeneration **12**, 1–21 (2017).
80. K. Matsubara, M. Ibaraki, M. Nemoto, et al., "A review on ai in pet imaging," Ann. Nucl. Med. **36**, 133–143 (2022).
81. A. Al-Ghraibah, M. Altayeb, and F. A. Alnaimat, "An automated system to distinguish between corona and viral pneumonia chest diseases based on image processing techniques," Comput. Methods Biomech. Biomed. Eng. Imaging & Vis. **11**, 2261575 (2024).
82. M. Tauseef, E. Rathod, S. Nandish, and M. Kushal, "Advancements in pet care technology: A comprehensive survey," in 2024 4th International Conference on Data Engineering and Communication Systems (ICDECS), (IEEE, 2024), pp. 1–6.
83. J. Petersson, A. Sánchez-Crespo, S. A. Larsson, and M. Mure, "Physiological imaging of the lung: single-photon-emission computed tomography (spect)," J. applied physiology **102**, 468–476 (2007).
84. A. Boschi, L. Uccelli, and P. Martini, "A picture of modern tc-99m radiopharmaceuticals: Production, chemistry, and applications in molecular imaging," Appl. Sci. **9**, 2526 (2019).
85. M. W. Groch and W. D. Erwin, "Spect in the year 2000: basic principles," J. Nucl. Med. Technol. **28**, 233–244 (2000).
86. S. Dorbala, M.-A. Park, S. Cuddy, et al., "Absolute quantitation of cardiac 99mtc-pyrophosphate using cadmium-zinc-telluride–based spect/ct," J. Nucl. Med. **62**, 716–722 (2021).
87. P. Ritt, "Recent developments in spect/ct," in Seminars in Nuclear Medicine, vol. 52 (Elsevier, 2022), pp. 276–285.
88. D. W. McRobbie, E. A. Moore, M. J. Graves, and M. R. Prince, MRI from Picture to Proton (Cambridge university press, 2017).
89. D. W. Townsend, J. P. Carney, J. T. Yap, and N. C. Hall, "Pet/ct today and tomorrow," J. Nucl. Med. **45**, 4S–14S (2004).
90. S.-A. Zhou and A. Brahme, "On the limitations and optimisation of high-resolution 3d medical x-ray imaging systems," Nucl. Instruments Methods Phys. Res. Sect. A: Accel. Spectrometers, Detect. Assoc. Equip. **648**, S284–S287 (2011).
91. M. E. Moseley and G. H. Glover, "Functional mr imaging: capabilities and limitations," Neuroimaging Clin. North Am. **5**, 161–191 (1995).
92. C. J. Price and K. J. Friston, "Functional imaging studies of neuropsychological patients: applications and limitations," Neurocase **8**, 345–354 (2002).
93. W. Wang, J. He, H. Liu, and W. Yuan, "Mdc-rht: Multi-modal medical image fusion via multi-dimensional dynamic convolution and residual hybrid transformer," Sensors **24**, 4056 (2024).
94. A. A. Alzahrani, "Enhanced multimodal medical image fusion via modified dwt with arithmetic optimization algorithm," Sci. Reports **14**, 19261 (2024).
95. R. Zhu, X. Li, S. Huang, and X. Zhang, "Multimodal medical image fusion using adaptive co-occurrence filter-based

decomposition optimization model," Bioinformatics **38**, 818–826 (2022).
96. M. Haribabu and V. Guruviah, "An improved multimodal medical image fusion approach using intuitionistic fuzzy set and intuitionistic fuzzy cross-correlation," Diagnostics **13**, 2330 (2023).
97. H. Lin, F. Yao, X. Yi, et al., "Prediction of adverse pathology in prostate cancer using a multimodal deep learning approach based on [18f] psma-1007 pet/ct and multiparametric mri," Eur. J. Nucl. Med. Mol. Imaging pp. 1–12 (2025).
98. F. Wang, C. Cheng, W. Cao, et al., "Mfcnet: A multi-modal fusion and calibration networks for 3d pancreas tumor segmentation on pet-ct images," Comput. Biol. Med. **155**, 106657 (2023).
99. J. Zhou, X. Xing, M. Yan, et al., "A fusion algorithm based on composite decomposition for pet and mri medical images," Biomed. Signal Process. Control. **76**, 103717 (2022).
100. H. Sotoudeh, A. Sharma, K. J. Fowler, et al., "Clinical application of pet/mri in oncology," J. Magn. Reson. Imaging **44**, 265–276 (2016).
101. J. Yu, P. Hu, R. Dai, et al., "Integrating ultrasound-ct-mr for preoperative multi-task prediction in ovarian cancer: Achieving diagnostic parity with multidisciplinary team consensus," SSRN (2025).
102. F. Wang, Q. Chen, Y. Chen, et al., "A novel multimodal deep learning model for preoperative prediction of microvascular invasion and outcome in hepatocellular carcinoma," Eur. J. Surg. Oncol. **49**, 156–164 (2023).
103. Y. El Bennioui, A. Halimi, A. Basarab, and J.-Y. Tourneret, "Fusion of magnetic resonance and ultrasound images using guided filtering: Application to endometriosis surgery," in 2024 32nd European Signal Processing Conference (EUSIPCO), (IEEE, 2024), pp. 1631–1635.
104. M. Freesmeyer, T. Winkens, T. Opfermann, et al., "Real-time ultrasound and freehand-spect," Nuklearmedizin-NuclearMedicine **53**, 259–264 (2014).
105. T. Tao, Y. Chen, Y. Shang, et al., "Smmf: a self-attention-based multi-parametric mri feature fusion framework for the diagnosis of bladder cancer grading," Front. Oncol. **14**, 1337186 (2024).
106. S. Li, H. Yin, and L. Fang, "Group-sparse representation with dictionary learning for medical image denoising and fusion," IEEE Trans. on biomedical engineering **59**, 3450–3459 (2012).
107. L. W. Remedios, H. Liu, S. W. Remedios, et al., "Influence of early through late fusion on pancreas segmentation from imperfectly registered multimodal mri," arXiv preprint arXiv:2409.04563 (2024).
108. A. Bhandary, G. A. Prabhu, V. Rajinikanth, et al., "Deep-learning framework to detect lung abnormality–a study with chest x-ray and lung ct scan images," Pattern Recognit. Lett. **129**, 271–278 (2020).
109. J. Dhar, N. Zaidi, M. Haghighat, et al., "Multimodal fusion learning with dual attention for medical imaging," in 2025 IEEE/CVF Winter Conference on Applications of Computer Vision (WACV), (IEEE, 2025), pp. 4362–4371.
110. O. F. Donati, H. Alkadhi, H. Scheffel, et al., "3d fusion of functional cardiac magnetic resonance imaging and computed tomography coronary angiography: accuracy and added clinical value," Investig. radiology **46**, 331–340 (2011).
111. D. W. Townsend, "Combined positron emission tomography–computed tomography: the historical perspective," in Seminars in Ultrasound, CT and MRI, vol. 29 (Elsevier, 2008), pp. 232–235.
112. L. J. Salomon, J.-P. Bernard, A.-E. Millischer, et al., "Mri and ultrasound fusion imaging for prenatal diagnosis," Am. journal obstetrics gynecology **209**, 148–e1 (2013).
113. J. H. Siewerdsen and D. A. Jaffray, "Cone-beam computed tomography with a flat-panel imager: magnitude and effects of x-ray scatter," Med. physics **28**, 220–231 (2001).
114. S. Li, X. Kang, L. Fang, et al., "Pixel-level image fusion: A survey of the state of the art," information Fusion **33**, 100–112 (2017).
115. P. Jagalingam and A. V. Hegde, "Pixel level image fusion—a review on various techniques," in 3rd World Conf. on Applied Sciences, Engineering and Technology, (2014).
116. Z. Shao and J. Cai, "Remote sensing image fusion with deep convolutional neural network," IEEE journal selected topics applied earth observations remote sensing **11**, 1656–1669 (2018).
117. C. Wang, J. Wu, A. Fang, et al., "An efficient frequency domain fusion network of infrared and visible images," Eng. Appl. Artif. Intell. **133**, 108013 (2024).
118. L. Wei, R. Zhu, X. Li, et al., "Pixel-level structure awareness for enhancing multi-modal medical image fusion," Biomed. Signal Process. Control. **97**, 106694 (2024).
119. N. Tawfik, H. A. Elnemr, M. Fakhr, et al., "Hybrid pixel-feature fusion system for multimodal medical images," J. Ambient Intell. Humaniz. Comput. **12**, 6001–6018 (2021).
120. A. B. Hassanat, V. S. Prasath, K. I. Mseidein, et al., "A hybridwavelet-shearlet approach to robust digital imagewatermarking," Informatica **41** (2017).
121. G. Xu, "An adaptive parameter tuning of particle swarm optimization algorithm," Appl. Math. Comput. **219**, 4560–4569 (2013).
122. H. Li, X.-J. Wu, and J. Kittler, "Rfn-nest: An end-to-end residual fusion network for infrared and visible images," Inf. Fusion **73**, 72–86 (2021).
123. Y. Liu, X. Chen, H. Peng, and Z. Wang, "Multi-focus image fusion with a deep convolutional neural network," Inf. Fusion **36**, 191–207 (2017).
124. T. Meitzler, D. Bednarz, E. Sohn, et al., "Fuzzy logic based image fusion," US Army TACON, Tech. Rep **13818** (2002).
125. J. Saeedi and K. Faez, "The new segmentation and fuzzy logic based multi-sensor image fusion," in 2009 24th

International Conference Image and Vision Computing New Zealand, (IEEE, 2009), pp. 328–333.
126. E. Daniel, J. Anitha, K. Kamaleshwaran, and I. Rani, "Optimum spectrum mask based medical image fusion using gray wolf optimization," Biomed. Signal Process. Control. **34**, 36–43 (2017).
127. Y. Jiang and Y. Ma, "Application of hybrid particle swarm and ant colony optimization algorithms to obtain the optimum homomorphic wavelet image fusion: introduction," Ann. Transl. Med. **8** (2020).
128. P. K. Sharma, V. Bhavya, K. Navyashree, et al., "Artificial bee colony and its application for image fusion," IJ Inf. Technol. Comput. Sci. **11**, 42–49 (2012).
129. J. Yu and H. Duan, "Artificial bee colony approach to information granulation-based fuzzy radial basis function neural networks for image fusion," Optik-International J. for Light Electron Opt. **124**, 3103–3111 (2013).
130. Ş. Öztürk, R. Ahmad, and N. Akhtar, "Variants of artificial bee colony algorithm and its applications in medical image processing," Appl. soft computing **97**, 106799 (2020).
131. L. Ding and H. Li, "A new improved hsv image fusion method," in LIDAR Imaging Detection and Target Recognition 2017, vol. 10605 (SPIE, 2017), pp. 426–438.
132. M. Manchanda and R. Sharma, "Fusion of visible and infrared images in hsv color space," in 2017 3rd International Conference on Computational Intelligence & Communication Technology (CICT), (IEEE, 2017), pp. 1–6.
133. S. Shukla and R. Raja, "Digital image fusion using adaptive neuro-fuzzy inference system," Int. J. New Technol. Res. **2**, 263508 (2016).
134. T. Zhou, Q. Li, H. Lu, et al., "Gan review: Models and medical image fusion applications," Inf. Fusion **91**, 134–148 (2023).
135. K. He, J. Sun, and X. Tang, "Guided image filtering," IEEE transactions on pattern analysis machine intelligence **35**, 1397–1409 (2012).
136. W. Gan, X. Wu, W. Wu, et al., "Infrared and visible image fusion with the use of multi-scale edge-preserving decomposition and guided image filter," Infrared Phys. & Technol. **72**, 37–51 (2015).
137. Y. Zhang, P. Zhao, Y. Ma, and X. Fan, "Multi-focus image fusion with joint guided image filtering," Signal Process. Image Commun. **92**, 116128 (2021).
138. W. Li, L. Jia, and J. Du, "Multi-modal sensor medical image fusion based on multiple salient features with guided image filter," Ieee Access **7**, 173019–173033 (2019).
139. F. Kou, W. Chen, C. Wen, and Z. Li, "Gradient domain guided image filtering," IEEE Trans. on Image Process. **24**, 4528–4539 (2015).
140. S. Liu, L. Yin, S. Miao, et al., "Multimodal medical image fusion using rolling guidance filter with cnn and nuclear norm minimization," Curr. Med. Imaging Rev. **16**, 1243–1258 (2020).
141. J. Fu, W. Li, A. Ouyang, and B. He, "Multimodal biomedical image fusion method via rolling guidance filter and deep convolutional neural networks," Optik **237**, 166726 (2021).
142. Y. Lin, D. Cao et al., "Adaptive infrared and visible image fusion method by using rolling guidance filter and saliency detection," Optik **262**, 169218 (2022).
143. L. Jian, X. Yang, Z. Zhou, et al., "Multi-scale image fusion through rolling guidance filter," Future Gener. Comput. Syst. **83**, 310–325 (2018).
144. K. Zhan, H. Zhang, and Y. Ma, "New spiking cortical model for invariant texture retrieval and image processing," IEEE Trans. on Neural Networks **20**, 1980–1986 (2009).
145. L. Wen, Y. Kuntao, S. Leilei, and L. Sheng, "Infrared and visible image fusion algorithm based on gaussian fuzzy logic and adaptive dual-channel spiking cortical model," Infrared Technol. **44**, 693–701 (2022).
146. W. Kong, Q. Miao, Y. Lei, and C. Ren, "Guided filter random walk and improved spiking cortical model based image fusion method in nsst domain," Neurocomputing **488**, 509–527 (2022).
147. Q. Zhang, X. Shen, L. Xu, and J. Jia, "Rolling guidance filter," in Computer Vision–ECCV 2014: 13th European Conference, Zurich, Switzerland, September 6-12, 2014, Proceedings, Part III 13, (Springer, 2014), pp. 815–830.
148. X. Wan, C. Zhao, and B. Gao, "Integration of adaptive guided filtering, deep feature learning, and edge-detection techniques for hyperspectral image classification," Opt. Eng. **56**, 113106–113106 (2017).
149. M. Han, K. Yu, W. Li, et al., "Colliding depths and fusion: Leveraging adaptive feature maps and restorable depth recharge for infrared and visible scene fusion," IEEE Trans. on Multimed. (2024).
150. S. Yang, X. Yang, J. Wu, and B. Feng, "Significant feature suppression and cross-feature fusion networks for fine-grained visual classification," Sci. Reports **14**, 24051 (2024).
151. J. Deng, S. Bei, S. Shaojing, and Z. Zhen, "Feature fusion methods in deep-learning generic object detection: A survey," in 2020 IEEE 9th Joint International Information Technology and Artificial Intelligence Conference (ITAIC), vol. 9 (IEEE, 2020), pp. 431–437.
152. X. Xie, Y. Cui, T. Tan, et al., "Fusionmamba: Dynamic feature enhancement for multimodal image fusion with mamba," Vis. Intell. **2**, 37 (2024).
153. Y. Li, L. Zhang, L. Chen, and Y. Ma, "Superpixel guided spectral-spatial feature extraction and weighted feature fusion for hyperspectral image classification with limited training samples," Sci. Reports **15**, 3473 (2025).
154. R. Li, Y. Cao, Y. Shu, et al., "A dynamic receptive field and improved feature fusion approach for federated learning in financial credit risk assessment," Sci. Reports **14**, 26515 (2024).
155. S. Zhang, X. Meng, Q. Liu, et al., "Feature-decision level collaborative fusion network for hyperspectral and lidar classification," Remote. Sens. **15**, 4148 (2023).
156. P. Lueangwitchajaroen, S. Watcharapinchai, W. Tepsan, and S. Sooksatra, "Multi-level feature fusion in cnn-based

human action recognition: A case study on efficientnet-b7," J. Imaging **10**, 320 (2024).
157. Y. Maqsood, S. M. Usman, M. Alhussein, et al., "Model agnostic meta-learning (maml)-based ensemble model for accurate detection of wheat diseases using vision transformer and graph neural networks." Comput. Mater. & Continua **79** (2024).
158. L. Xu, A. Krzyzak, and C. Y. Suen, "Methods of combining multiple classifiers and their applications to handwriting recognition," IEEE transactions on systems, man, cybernetics **22**, 418–435 (1992).
159. Z. Guo, X. Li, H. Huang, et al., "Medical image segmentation based on multi-modal convolutional neural network: Study on image fusion schemes," in 2018 IEEE 15th International Symposium on Biomedical Imaging (ISBI 2018), (IEEE, 2018), pp. 903–907.
160. M. Pal and G. M. Foody, "Feature selection for classification of hyperspectral data by svm," IEEE Trans. on Geosci. Remote. Sens. **48**, 2297–2307 (2010).
161. B. Waske, S. van der Linden, J. A. Benediktsson, et al., "Sensitivity of support vector machines to random feature selection in classification of hyperspectral data," IEEE Trans. on Geosci. Remote. Sens. **48**, 2880–2889 (2010).
162. Z. Ullah, M. I. Mohmand, S. U. Rehman, et al., "Emotion recognition from occluded facial images using deep ensemble model," Cmc-Computers Mater. & Continua **73**, 4465–4487 (2022).
163. G. Abdi, F. Samadzadegan, and P. Reinartz, "Deep learning decision fusion for the classification of urban remote sensing data," J. Appl. Remote. Sens. **12**, 016038–016038 (2018).
164. Y. Yang, C. Han, X. Kang, and D. Han, "An overview on pixel-level image fusion in remote sensing," in 2007 ieee international conference on automation and logistics, (IEEE, 2007), pp. 2339–2344.
165. V. Naidu and J. R. Raol, "Pixel-level image fusion using wavelets and principal component analysis," Def. science journal **58**, 338–352 (2008).
166. B. P. Goud, B. Sushmita, and A. Vijitha, "Evaluation of image fusion of multi focus images in spatial and frequency domain," Int. J. Comput. Eng. Res. (IJCER) **8**, 2250–3005 (2018).
167. S. Liu, M. Wang, L. Yin, et al., "Two-scale multimodal medical image fusion based on structure preservation," Front. Comput. Neurosci. **15**, 803724 (2022).
168. M. Manchanda and R. Sharma, "A novel method of multimodal medical image fusion using fuzzy transform," J. Vis. Commun. Image Represent. **40**, 197–217 (2016).
169. M. Manchanda and R. Sharma, "An improved multimodal medical image fusion algorithm based on fuzzy transform," J. Vis. Commun. Image Represent. **51**, 76–94 (2018).
170. Z. Zhu, M. Zheng, G. Qi, et al., "A phase congruency and local laplacian energy based multi-modality medical image fusion method in nsct domain," Ieee Access **7**, 20811–20824 (2019).
171. G. Bhatnagar, Q. J. Wu, and Z. Liu, "Directive contrast based multimodal medical image fusion in nsct domain," IEEE transactions on multimedia **15**, 1014–1024 (2013).
172. P. Ganasala and V. Kumar, "Ct and mr image fusion scheme in nonsubsampled contourlet transform domain," J. digital imaging **27**, 407–418 (2014).
173. N. Amini, E. Fatemizadeh, and H. Behnam, "Mri-pet image fusion based on nsct transform using local energy and local variance fusion rules," J. medical engineering & technology **38**, 211–219 (2014).
174. J. Xia, Y. Chen, A. Chen, and Y. Chen, "Medical image fusion based on sparse representation and pcnn in nsct domain," Comput. mathematical methods medicine **2018**, 2806047 (2018).
175. T. Li and Y. Wang, "Multiscaled combination of mr and spect images in neuroimaging: a simplex method based variable-weight fusion," Comput. methods programs biomedicine **105**, 31–39 (2012).
176. M. Chandana, S. Amutha, and N. Kumar, "A hybrid multi-focus medical image fusion based on wavelet transform," Int. J. Res. Rev. Comput. Sci. **2**, 948 (2011).
177. S. P. Yadav and S. Yadav, "Fusion of medical images in wavelet domain: a hybrid implementation," Comput. Model. Eng. & Sci. **122**, 303–321 (2020).
178. K. Indira, R. R. Hemamalini, and R. Indhumathi, "Pixel based medical image fusion techniques using discrete wavelet transform and stationary wavelet transform," Indian J. Sci. Technol. **8**, 1–7 (2015).
179. K. Indira and R. R. Hemamalini, "Evaluation of choose max and contrast based fusion rule using dwt for pet, ct images," Indian J. Sci. Technol. **8**, 1–4 (2015).
180. R. Singh and A. Khare, "Multiscale medical image fusion in wavelet domain," The Sci. World J. **2013**, 521034 (2013).
181. X. Xu, Y. Wang, and S. Chen, "Medical image fusion using discrete fractional wavelet transform," Biomed. signal processing control **27**, 103–111 (2016).
182. E. Daniel, "Optimum wavelet-based homomorphic medical image fusion using hybrid genetic–grey wolf optimization algorithm," IEEE Sensors J. **18**, 6804–6811 (2018).
183. M. Nanavati and M. Shah, "Performance comparison of different wavelet based image fusion techniques for lumbar spine images," J. Integr. Sci. Technol. **12**, 703–703 (2024).
184. V. Bhateja, H. Patel, A. Krishn, et al., "Multimodal medical image sensor fusion framework using cascade of wavelet and contourlet transform domains," IEEE Sensors J. **15**, 6783–6790 (2015).
185. M. Haribabu, C. H. Bindu, and K. S. Prasad, "Multimodal medical image fusion of mri-pet using wavelet transform," in 2012 International Conference on Advances in Mobile Network, Communication and Its Applications, (IEEE, 2012), pp. 127–130.
186. B. S. N. Rao, N. V. K. Raju, M. Dhanush, et al., "Mri and spect medical image fusion using wavelet transform,"

in 2022 7th International Conference on Communication and Electronics Systems (ICCES), (IEEE, 2022), pp. 1690–1696.
187. H. O. S. Mishra and S. Bhatnagar, "Mri and ct image fusion based on wavelet transform," Int. J. Inf. Comput. Technol. **4**, 47–52 (2014).
188. S. Sandhya, M. Senthil Kumar, and B. Chidhambararajan, "A hybrid guided filtering and transform-based sparse representation framework for fusion of multimodal medical images," in International Conference on Futuristic Communication and Network Technologies, (Springer, 2021), pp. 267–274.
189. D. E. Nirmala, R. Vignesh, and V. Vaidehi, "Fusion of multisensor images using nonsubsampled contourlet transform and fuzzy logic," in 2013 IEEE International Conference on Fuzzy Systems (FUZZ-IEEE), (IEEE, 2013), pp. 1–8.
190. S. Sivasangumani, "Decision based fuzzy logic approach for multimodal medical image fusion in nsct domain," Int. J. Image Min. **3**, 117–138 (2018).
191. S. I. Ibrahim, G. S. El-Tawel, and M. Makhlouf, "Brain image fusion using the parameter adaptive-pulse coupled neural network (pa-pcnn) and non-subsampled contourlet transform (nsct)," Multimed. Tools Appl. **83**, 27379–27409 (2024).
192. M. Haribabu and V. Guruviah, "Enhanced multimodal medical image fusion based on pythagorean fuzzy set: an innovative approach," Sci. Reports **13**, 16726 (2023).
193. S. D. Ramlal, J. Sachdeva, C. K. Ahuja, and N. Khandelwal, "Multimodal medical image fusion using non-subsampled shearlet transform and pulse coupled neural network incorporated with morphological gradient," Signal, Image Video Process. **12**, 1479–1487 (2018).
194. P. Ganasala and V. Kumar, "Feature-motivated simplified adaptive pcnn-based medical image fusion algorithm in nsst domain," J. digital imaging **29**, 73–85 (2016).
195. M. Das, D. Gupta, P. Radeva, and A. M. Bakde, "Nsst domain ct–mr neurological image fusion using optimised biologically inspired neural network," IET Image Process. **14**, 4291–4305 (2020).
196. H. Lin, Y. Song, H. Wang, et al., "Multimodal brain image fusion based on improved rolling guidance filter and wiener filter," Comput. Math. Methods Med. **2022**, 5691099 (2022).
197. Y. Na, L. Zhao, Y. Yang, and M. Ren, "Guided filter-based images fusion algorithm for ct and mri medical images," IET Image Process. **12**, 138–148 (2018).
198. P. Ganasala and A. Prasad, "Contrast enhanced multi sensor image fusion based on guided image filter and nsst," IEEE Sensors J. **20**, 939–946 (2019).
199. Y. Wang, B. Wu, and Z. Duan, "Image dehazing based on sigmoid-guided filtering-retinex in nsct domain." IAENG Int. J. Appl. Math. **54** (2024).
200. B. Goyal, A. Dogra, R. Khoond, and F. Al-Turjman, "An efficient medical assistive diagnostic algorithm for visualisation of structural and tissue details in ct and mri fusion," Cogn. Comput. **13**, 1471–1483 (2021).
201. S. Kor and U. Tiwary, "Feature level fusion of multimodal medical images in lifting wavelet transform domain," in The 26th Annual International Conference of the IEEE Engineering in Medicine and Biology Society, vol. 1 (IEEE, 2004), pp. 1479–1482.
202. L. Xu, J. Du, Q. Hu, and Q. Li, "Feature-based image fusion with a uniform discrete curvelet transform," Int. J. Adv. Robotic Syst. **10**, 255 (2013).
203. A. Gupta, A. Kumar, and K. Rautela, "Udct: lung cancer detection and classification using u-net and darts for medical ct images," Multimed. Tools Appl. pp. 1–21 (2024).
204. T. Wei, Q. Gao, N. Ma, et al., "Feature-level image fusion through consistent region segmentation and dual-tree complex wavelet transform." J. Imaging Sci. & Technol. **60** (2016).
205. B. Meher, S. Agrawal, R. Panda, and A. Abraham, "A survey on region based image fusion methods," Inf. Fusion **48**, 119–132 (2019).
206. Q. Xie, L. Ma, Z. Guo, et al., "Infrared and visible image fusion based on nsst and phase consistency adaptive dual channel pcnn," Infrared Phys. & Technol. **131**, 104659 (2023).
207. H. Ullah, Y. Zhao, L. Wu, et al., "Nsst based mri-pet/spect color image fusion using local features fuzzy rules and nsml in yiq space," in 2019 IEEE International Symposium on Signal Processing and Information Technology (ISSPIT), (IEEE, 2019), pp. 1–6.
208. Q. Yang and H. Wang, "Image fusion algorithm based on improved k-singular value decomposition and hadamard measurement matrix," J. Algorithms & Comput. Technol. **13**, 1748301818791511 (2018).
209. A. A. Smadi, A. Abugabah, A. Mehmood, and S. Yang, "Brain image fusion approach based on side window filtering," Procedia Comput. Sci. **198**, 295–300 (2022).
210. N. Zhang, J. Liu, Y. Jin, et al., "An adaptive multi-modal hybrid model for classifying thyroid nodules by combining ultrasound and infrared thermal images," BMC bioinformatics **24**, 315 (2023).
211. N. Tawfik, H. A. Elnemr, M. Fakhr, et al., "Survey study of multimodality medical image fusion methods," Multimed. Tools Appl. **80**, 6369–6396 (2021).
212. M. A. Cody, "The fast wavelet transform: Beyond fourier transforms," Dr. Dobb's J. **17** (1992).
213. A. Kumar, S. Singh, A. Kumar et al., "Grey wolf optimizer and other metaheuristic optimization techniques with image processing as their applications: a review," in IOP Conference Series: Materials Science and Engineering, vol. 1136 (IOP Publishing, 2021), p. 012053.
214. C.-I. Chen, "Fusion of pet and mr brain images based on ihs and log-gabor transforms," IEEE Sensors J. **17**, 6995–7010 (2017).

215. T. Zhou, H. Lu, J. Zhang, and H. Shi, "Pulmonary nodule detection model based on svm and ct image feature-level fusion with rough sets," BioMed research international **2016**, 8052436 (2016).
216. M. Ramprasad, M. Z. U. Rahman, and M. D. Bayleyegn, "A deep probabilistic sensing and learning model for brain tumor classification with fusion-net and hfcmik segmentation," IEEE Open J. Eng. Med. Biol. **3**, 178–188 (2022).
217. Q. Li, X. Yang, W. Wu, et al., "Multi-focus image fusion method for vision sensor systems via dictionary learning with guided filter," Sensors **18**, 2143 (2018).
218. V. Anitha, "Brain tumor detection in combined 3d mri and ct images using dictionary learning based segmentation and spearman regression," Sadhana **49**, 221 (2024).
219. M. Kim, D. K. Han, and H. Ko, "Joint patch clustering-based dictionary learning for multimodal image fusion," Inf. fusion **27**, 198–214 (2016).
220. Y. Wang, G. Ma, L. An, et al., "Semisupervised tripled dictionary learning for standard-dose pet image prediction using low-dose pet and multimodal mri," IEEE transactions on biomedical engineering **64**, 569–579 (2016).
221. J. Li, Y. Peng, M. Song, and L. Liu, "Image fusion based on guided filter and online robust dictionary learning," Infrared Phys. & Technol. **105**, 103171 (2020).
222. Y. Liu, X. Chen, A. Liu, et al., "Recent advances in sparse representation based medical image fusion," IEEE Instrum. & Meas. Mag. **24**, 45–53 (2021).
223. Y. Liu, X. Chen, R. K. Ward, and Z. J. Wang, "Medical image fusion via convolutional sparsity based morphological component analysis," IEEE Signal Process. Lett. **26**, 485–489 (2019).
224. B. Meher, S. Agrawal, R. Panda, et al., "A novel region-based multimodal image fusion technique using improved dictionary learning," Int. J. Imaging Syst. Technol. **30**, 558–576 (2020).
225. Z. Zhu, G. Qi, Y. Chai, and Y. Chen, "A novel multi-focus image fusion method based on stochastic coordinate coding and local density peaks clustering," Future Internet **8**, 53 (2016).
226. N. Aishwarya and C. Bennila Thangammal, "A novel multimodal medical image fusion using sparse representation and modified spatial frequency," Int. J. Imaging Syst. Technol. **28**, 175–185 (2018).
227. F. Zhou, X. Li, M. Zhou, et al., "A new dictionary construction based multimodal medical image fusion framework," Entropy **21**, 267 (2019).
228. X. Jiang, J. Ma, G. Xiao, et al., "A review of multimodal image matching: Methods and applications," Inf. Fusion **73**, 22–71 (2021).
229. A. P. James and B. V. Dasarathy, "Medical image fusion: A survey of the state of the art," Inf. fusion **19**, 4–19 (2014).
230. H. Hermessi, O. Mourali, and E. Zagrouba, "Multimodal medical image fusion review: Theoretical background and recent advances," Signal Process. **183**, 108036 (2021).
231. M. Umair, M. Zubair, F. Dawood, et al., "A multi-layer holistic approach for cursive text recognition," Appl. Sci. **12**, 12652 (2022).
232. G. Eichniann, C. Lu, J. Zhu, and Y. Li, "Pyramidal image processing using morphology," in Applications of Digital Image Processing XI, vol. 974 (SPIE, 1988), pp. 30–37.
233. F. Laporterie and G. Flouzat, "The morphological pyramid concept as a tool for multi-resolution data fusion in remote sensing," Integr. computer-aided engineering **10**, 63–79 (2003).
234. Y. Zhang, Y. Liu, P. Sun, et al., "Ifcnn: A general image fusion framework based on convolutional neural network," Inf. Fusion **54**, 99–118 (2020).
235. Y. Zheng, E. A. Essock, and B. C. Hansen, "An advanced image fusion algorithm based on wavelet transform: incorporation with pca and morphological processing," in Image processing: algorithms and systems III, vol. 5298 (SPIE, 2004), pp. 177–187.
236. J. Saeedi and K. Faez, "Infrared and visible image fusion using fuzzy logic and population-based optimization," Appl. Soft Comput. **12**, 1041–1054 (2012).
237. X. Li, L. Wang, J. Wang, and X. Zhang, "Multi-focus image fusion algorithm based on multilevel morphological component analysis and support vector machine," IET Image Process. **11**, 919–926 (2017).
238. U. Patil and U. Mudengudi, "Image fusion using hierarchical pca." in 2011 international conference on image information processing, (IEEE, 2011), pp. 1–6.
239. Y. Jiang and M. Wang, "Image fusion with morphological component analysis," Inf. Fusion **18**, 107–118 (2014).
240. T. Tirupal, B. C. Mohan, and S. S. Kumar, "Multimodal medical image fusion techniques–a review," Curr. Signal Transduct. Ther. **16**, 142–163 (2021).
241. M. A. Azam, K. B. Khan, S. Salahuddin, et al., "A review on multimodal medical image fusion: Compendious analysis of medical modalities, multimodal databases, fusion techniques and quality metrics," Comput. biology medicine **144**, 105253 (2022).
242. A. Smirnov and T. Levashova, "Knowledge fusion patterns: A survey," Inf. Fusion **52**, 31–40 (2019).
243. Q. Huang and G. Li, "Knowledge graph based reasoning in medical image analysis: A scoping review," Comput. Biol. Med. **182**, 109100 (2024).
244. M. A. Saleh, A. A. Ali, K. Ahmed, and A. M. Sarhan, "A brief analysis of multimodal medical image fusion techniques," Electronics **12**, 97 (2022).
245. J. Huang, T. Tan, X. Li, et al., "Multiple attention channels aggregated network for multimodal medical image fusion," Med. Phys. .
246. X. Gu, Y. Xia, and J. Zhang, "Multimodal medical image fusion based on interval gradients and convolutional

neural networks," BMC Med. Imaging **24**, 232 (2024).
247. D. Dave, A. Akhunzada, N. Ivković, et al., "Diagnostic test accuracy of ai-assisted mammography for breast imaging: a narrative review," PeerJ Comput. Sci. **11**, e2476 (2025).
248. A. A. Alhussan, A. A. Abdelhamid, S. Towfek, et al., "Classification of breast cancer using transfer learning and advanced al-biruni earth radius optimization," Biomimetics **8**, 270 (2023).
249. M. R. Ogiela and L. Ogiela, "Application of cognitive information systems in medical image semantic analysis," Electronics **13**, 325 (2024).
250. R. Ranjbarzadeh, A. Bagherian Kasgari, S. Jafarzadeh Ghoushchi, et al., "Brain tumor segmentation based on deep learning and an attention mechanism using mri multi-modalities brain images," Sci. reports **11**, 10930 (2021).
251. N. Tataei Sarshar, R. Ranjbarzadeh, S. Jafarzadeh Ghoushchi, et al., "Glioma brain tumor segmentation in four mri modalities using a convolutional neural network and based on a transfer learning method," in Brazilian Technology Symposium, (Springer, 2021), pp. 386–402.
252. R. Ranjbarzadeh, A. Caputo, E. B. Tirkolaee, et al., "Brain tumor segmentation of mri images: A comprehensive review on the application of artificial intelligence tools," Comput. biology medicine **152**, 106405 (2023).
253. R. Ranjbarzadeh, P. Zarbakhsh, A. Caputo, et al., "Brain tumor segmentation based on optimized convolutional neural network and improved chimp optimization algorithm," Comput. Biol. Med. **168**, 107723 (2024).
254. X. Ran, J. Shi, Y. Chen, and K. Jiang, "Multimodal neuroimage data fusion based on multikernel learning in personalized medicine," Front. Pharmacol. **13**, 947657 (2022).
255. M. Akhter and R. Recker, "High resolution imaging in bone tissue research-review," Bone **143**, 115620 (2021).
256. F. Bruno, F. Arrigoni, S. Mariani, et al., "Advanced magnetic resonance imaging (mri) of soft tissue tumors: techniques and applications," La radiologia medica **124**, 243–252 (2019).
257. H. A. Owida, G. AlMahadin, J. I. Al-Nabulsi, et al., "Automated classification of brain tumor-based magnetic resonance imaging using deep learning approach," Int. J. Electr. Comput. Eng. **14**, 3150–3158 (2024).
258. M. Zubair, A. Yamin, and S. A. Khan, "Automated detection of optic disc for the analysis of retina using color fundus image," in 2013 IEEE International Conference on Imaging Systems and Techniques (IST), (IEEE, 2013), pp. 239–242.
259. M. Zubair, H. Ali, and M. Y. Javed, "Automated segmentation of hard exudates using dynamic thresholding to detect diabetic retinopathy in retinal photographs." J. Multim. Process. Technol. **7**, 109–116 (2016).
260. Z. Shahzadi and M. Zubair, "Multiclass classification of retinal disorders using optical coherence tomography images," in 2024 Horizons of Information Technology and Engineering (HITE), (IEEE, 2024), pp. 1–6.
261. D. Ahmed, M. A. Hassan, and M. Zubair, "Autism detection in children by features extraction and classification using a deep learning model," in 2024 Horizons of Information Technology and Engineering (HITE), (IEEE, 2024), pp. 1–5.
262. R. Pannu, M. Zubair, M. Owais, et al., "Enhanced glaucoma classification through advanced segmentation by integrating cup-to-disc ratio and neuro-retinal rim features," Comput. Med. Imaging Graph. p. 102559 (2025).
263. R. P. Broussard, S. K. Rogers, M. E. Oxley, and G. L. Tarr, "Physiologically motivated image fusion for object detection using a pulse coupled neural network," IEEE Trans. on Neural networks **10**, 554–563 (1999).
264. H. Szu, I. Kopriva, P. Hoekstra, et al., "Early tumor detection by multiple infrared unsupervised neural nets fusion," in Proceedings of the 25th Annual International Conference of the IEEE Engineering in Medicine and Biology Society (IEEE Cat. No. 03CH37439), vol. 2 (IEEE, 2003), pp. 1133–1136.
265. Q. Zhang, M. Liang, and W. Sun, "Medical diagnostic image fusion based on feature mapping wavelet neural networks," in Third International Conference on Image and Graphics (ICIG'04), (IEEE, 2004), pp. 51–54.
266. X. Lu, B. Zhang, and Y. Gu, "Medical image fusion algorithm based on clustering neural network," in 2007 1st International Conference on Bioinformatics and Biomedical Engineering, (IEEE, 2007), pp. 637–640.
267. H. Zhang, X.-n. Sun, L. Zhao, and L. Liu, "Image fusion algorithm using rbf neural networks," in Hybrid Artificial Intelligence Systems: Third International Workshop, HAIS 2008, Burgos, Spain, September 24-26, 2008. Proceedings 3, (Springer, 2008), pp. 417–424.
268. Z. Wang and Y. Ma, "Medical image fusion using m-pcnn," Inf. fusion **9**, 176–185 (2008).
269. C. Bhuvaneswari, P. Aruna, and D. Loganathan, "A new fusion model for classification of the lung diseases using genetic algorithm," Egypt. Informatics J. **15**, 69–77 (2014).
270. H. Jin and Y. Wang, "A fusion method for visible and infrared images based on contrast pyramid with teaching learning based optimization," Infrared Phys. & Technol. **64**, 134–142 (2014).
271. L. Tang, J. Qian, L. Li, et al., "Multimodal medical image fusion based on discrete t chebichef moments and pulse coupled neural network," Int. J. Imaging Syst. Technol. **27**, 57–65 (2017).
272. G. Litjens, T. Kooi, B. E. Bejnordi, et al., "A survey on deep learning in medical image analysis," Med. image analysis **42**, 60–88 (2017).
273. J. Ker, L. Wang, J. Rao, and T. Lim, "Deep learning applications in medical image analysis," Ieee Access **6**, 9375–9389 (2017).
274. J. Bhardwaj, A. Nayak, C. S. Yadav, and S. P. Yadav, "A review in wavelet transforms based medical image fusion," Evol. Role AI IoMT Healthc. Mark. pp. 199–214 (2021).
275. P. Liu, H. Zhang, W. Lian, and W. Zuo, "Multi-level wavelet convolutional neural networks," IEEE Access **7**, 74973–74985 (2019).
276. D. Anandhi and S. Valli, "An algorithm for multi-sensor image fusion using maximum a posteriori and nonsubsampled

contourlet transform," Comput. & Electr. Eng. **65**, 139–152 (2018).
277. H. Zhang, X. Ma, and Y. Tian, "An image fusion method based on curvelet transform and guided filter enhancement," Math. Probl. Eng. **2020**, 9821715 (2020).
278. L. Wang, B. Li, and L. Tian, "A novel multi-modal medical image fusion method based on shift-invariant shearlet transform," The Imaging Sci. J. **61**, 529–540 (2013).
279. J. Reena Benjamin and T. Jayasree, "Improved medical image fusion based on cascaded pca and shift invariant wavelet transforms," Int. journal computer assisted radiology surgery **13**, 229–240 (2018).
280. T. Tirupal, Y. Pandurangaiah, A. Roy, et al., "On the use of udwt and fuzzy sets for medical image fusion," Multimed. Tools Appl. **83**, 39647–39675 (2024).
281. C. Qiu, Y. Wang, H. Zhang, and S. Xia, "Image fusion of ct and mr with sparse representation in nsst domain," Comput. mathematical methods medicine **2017**, 9308745 (2017).
282. F. Shabanzade and H. Ghassemian, "Combination of wavelet and contourlet transforms for pet and mri image fusion," in 2017 artificial intelligence and signal processing conference (AISP), (IEEE, 2017), pp. 178–183.
283. V. Naidu, M. Divya, and P. Mahalakshmi, "Multi-modal medical image fusion using multi-resolution discrete sine transform," Control. Data Fusion e-Journal **1**, 13–26 (2017).
284. Y. Li, Y. Sun, X. Huang, et al., "An image fusion method based on sparse representation and sum modified-laplacian in nsct domain," Entropy **20**, 522 (2018).
285. J. Ropero, A. Gómez, A. Carrasco, and C. León, "A fuzzy logic intelligent agent for information extraction: Introducing a new fuzzy logic-based term weighting scheme," Expert Syst. with Appl. **39**, 4567–4581 (2012).
286. T. Chaira, Fuzzy set and its extension: The intuitionistic fuzzy set (John Wiley & Sons, 2019).
287. W. Dou, S. Ruan, Y. Chen, et al., "A framework of fuzzy information fusion for the segmentation of brain tumor tissues on mr images," Image vision Comput. **25**, 164–171 (2007).
288. Y. Na, H. Lu, and Y. Zhang, "Content analysis based medical images fusion with fuzzy inference," in 2008 Fifth International Conference on Fuzzy Systems and Knowledge Discovery, vol. 3 (IEEE, 2008), pp. 37–41.
289. A. Assareh and L. G. Volkert, "Fuzzy rule base classifier fusion for protein mass spectra based ovarian cancer diagnosis," in 2009 IEEE Symposium on Computational Intelligence in Bioinformatics and Computational Biology, (IEEE, 2009), pp. 193–199.
290. J. Teng, S. Wang, J. Zhang, and X. Wang, "Neuro-fuzzy logic based fusion algorithm of medical images," in 2010 3rd International Congress on Image and Signal Processing, vol. 4 (IEEE, 2010), pp. 1552–1556.
291. S. Das and M. K. Kundu, "A neuro-fuzzy approach for medical image fusion," IEEE transactions on biomedical engineering **60**, 3347–3353 (2013).
292. K. Gayathri and T. Tirupal, "Multimodal medical image fusion based on type-1 fuzzy sets," J Appl Sci Comput. **5**, 1329–1341 (2018).
293. P. Balasubramaniam and V. Ananthi, "Image fusion using intuitionistic fuzzy sets," Inf. fusion **20**, 21–30 (2014).
294. P. K. Peruru, K. Madhavi, and T. Tirupal, "Multimodal medical image fusion based on undecimated wavelet transform and fuzzy sets," Int. J. Innov. Technol. Explor. Eng. **8**, 7–103 (2019).
295. D. Poornima, S. Gowri, J. Jabez et al., "An effective fusion model of modified wavelet transform for medical diagnosis by using wireless optical communications," in 2023 International Conference on Emerging Research in Computational Science (ICERCS), (IEEE, 2023), pp. 1–6.
296. Y. Liu and Z. Wang, "Multi-focus image fusion based on sparse representation with adaptive sparse domain selection," in 2013 Seventh International Conference on Image and Graphics, (IEEE, 2013), pp. 591–596.
297. J. Ma, H. Xu, J. Jiang, et al., "Ddcgan: A dual-discriminator conditional generative adversarial network for multi-resolution image fusion," IEEE Trans. on Image Process. **29**, 4980–4995 (2020).
298. H. Khotanlou, O. Colliot, J. Atif, and I. Bloch, "3d brain tumor segmentation in mri using fuzzy classification, symmetry analysis and spatially constrained deformable models," Fuzzy sets systems **160**, 1457–1473 (2009).
299. Q. Zhu, J. Yang, B. Xu, et al., "Multimodal brain network jointly construction and fusion for diagnosis of epilepsy," Front. Neurosci. **15**, 734711 (2021).
300. O. Muzik, D. C. Chugani, G. Zou, et al., "Multimodality data integration in epilepsy," Int. J. Biomed. Imaging **2007**, 013963 (2007).
301. H. Chuang and H. Macapinlac, "The evolving role of pet-ct in the management of esophageal cancer," The Q. J. Nucl. Med. Mol. Imaging **53**, 201 (2009).
302. O. Gaemperli, A. Saraste, and J. Knuuti, "Cardiac hybrid imaging," Eur. Heart Journal–Cardiovascular Imaging **13**, 51–60 (2012).
303. D. C. Benz, L. Gaemperli, C. Gräni, et al., "Impact of cardiac hybrid imaging-guided patient management on clinical long-term outcome," Int. journal cardiology **261**, 218–222 (2018).
304. Z. Zhao, J. Poyhonen, X. Chen Cai, et al., "Augmented reality technology in image-guided therapy: State-of-the-art review," Proc. Inst. Mech. Eng. Part H: J. Eng. Med. **235**, 1386–1398 (2021).
305. H. Kehlet and D. W. Wilmore, "Multimodal strategies to improve surgical outcome," The Am. journal surgery **183**, 630–641 (2002).
306. J. Ma, W. Yu, P. Liang, et al., "Fusiongan: A generative adversarial network for infrared and visible image fusion," Inf. fusion **48**, 11–26 (2019).
307. Q. Xie, M. Zhou, Q. Zhao, et al., "Mhf-net: An interpretable deep network for multispectral and hyperspectral image fusion," IEEE Trans. on Pattern Anal. Mach. Intell. **44**, 1457–1473 (2020).

308. M. M. Almasri and A. M. Alajlan, "Artificial intelligence-based multimodal medical image fusion using hybrid s2 optimal cnn," Electronics **11**, 2124 (2022).
309. M. Safari, A. Fatemi, and L. Archambault, "Medfusiongan: multimodal medical image fusion using an unsupervised deep generative adversarial network," BMC Med. Imaging **23**, 203 (2023).
310. S. Shehanaz, E. Daniel, S. R. Guntur, and S. Satrasupalli, "Optimum weighted multimodal medical image fusion using particle swarm optimization," Optik **231**, 166413 (2021).
311. J. Jose, N. Gautam, M. Tiwari, et al., "An image quality enhancement scheme employing adolescent identity search algorithm in the nsst domain for multimodal medical image fusion," Biomed. Signal Process. Control. **66**, 102480 (2021).
312. J. Duan, S. Mao, J. Jin, et al., "A novel ga-based optimized approach for regional multimodal medical image fusion with superpixel segmentation," IEEE access **9**, 96353–96366 (2021).
313. Y. Liu, B. Yan, R. Zhang, et al., "Multi-scale mixed attention network for ct and mri image fusion," Entropy **24**, 843 (2022).
314. L. Li and H. Ma, "Pulse coupled neural network-based multimodal medical image fusion via guided filtering and wseml in nsct domain," Entropy **23**, 591 (2021).
315. P. Guo, G. Xie, R. Li, and H. Hu, "Multimodal medical image fusion with convolution sparse representation and mutual information correlation in nsst domain," Complex & Intell. Syst. **9**, 317–328 (2023).
316. M. Kaur and D. Singh, "Multi-modality medical image fusion technique using multi-objective differential evolution based deep neural networks," J. Ambient Intell. Humaniz. Comput. **12**, 2483–2493 (2021).
317. W. Kong, C. Li, and Y. Lei, "Multimodal medical image fusion using convolutional neural network and extreme learning machine," Front. Neurorobotics **16**, 1050981 (2022).
318. K. N. Singh, O. P. Singh, A. K. Singh, and A. K. Agrawal, "Watmif: Multimodal medical image fusion-based watermarking for telehealth applications," Cogn. Comput. **16**, 1947–1963 (2024).
319. T. Doherty, S. McKeever, N. Al-Attar, et al., "Feature fusion of raman chemical imaging and digital histopathology using machine learning for prostate cancer detection," Analyst **146**, 4195–4211 (2021).
320. L. Wang, J. Zhang, Y. Liu, et al., "Multimodal medical image fusion based on gabor representation combination of multi-cnn and fuzzy neural network," IEEE Access **9**, 67634–67647 (2021).
321. Y. Liu, F. Mu, Y. Shi, et al., "Brain tumor segmentation in multimodal mri via pixel-level and feature-level image fusion," Front. Neurosci. **16**, 1000587 (2022).
322. M. Umair, Z. Saeed, F. Saeed, et al., "Energy theft detection in smart grids with genetic algorithm-based feature selection," Comput. Mater. Continua **74** (2022).
323. W. Tang, F. He, Y. Liu, and Y. Duan, "Matr: Multimodal medical image fusion via multiscale adaptive transformer," IEEE Trans. on Image Process. **31**, 5134–5149 (2022).
324. G. M. Dimitri, S. Spasov, A. Duggento, et al., "Multimodal and multicontrast image fusion via deep generative models," Inf. Fusion **88**, 146–160 (2022).
325. Z. Shi, C. Zhang, D. Ye, et al., "Mmi-fuse: multimodal brain image fusion with multiattention module," IEEE Access **10**, 37200–37214 (2022).
326. S. I. Ibrahim, M. Makhlouf, and G. S. El-Tawel, "Multimodal medical image fusion algorithm based on pulse coupled neural networks and nonsubsampled contourlet transform," Med. & Biol. Eng. & Comput. **61**, 155–177 (2023).
327. J. Li, J. Liu, S. Zhou, et al., "Gesenet: A general semantic-guided network with couple mask ensemble for medical image fusion," IEEE Trans. on Neural Networks Learn. Syst. (2023).
328. F. G. Veshki, N. Ouzir, S. A. Vorobyov, and E. Ollila, "Multimodal image fusion via coupled feature learning," Signal Process. **200**, 108637 (2022).
329. A. Sufyan, M. Imran, S. A. Shah, et al., "A novel multimodality anatomical image fusion method based on contrast and structure extraction," Int. J. Imaging Syst. Technol. **32**, 324–342 (2022).
330. V. Singh, V. D. Kaushik et al., "Dtcwtasodcnn: Dtcwt based weighted fusion model for multimodal medical image quality improvement with aso technique & dcnn," J. Sci. & Ind. Res. **81**, 850–858 (2022).
331. J. A. Bhutto, L. Tian, Q. Du, et al., "Ct and mri medical image fusion using noise-removal and contrast enhancement scheme with convolutional neural network," Entropy **24**, 393 (2022).
332. J. Li, D. Han, X. Wang, et al., "Multi-sensor medical-image fusion technique based on embedding bilateral filter in least squares and salient detection," Sensors **23**, 3490 (2023).
333. M. Diwakar, P. Singh, R. Singh, et al., "Multimodality medical image fusion using clustered dictionary learning in non-subsampled shearlet transform," Diagnostics **13**, 1395 (2023).
334. Amrita, S. Joshi, R. Kumar, et al., "Water wave optimized nonsubsampled shearlet transformation technique for multimodal medical image fusion," Concurr. Comput. Pract. Exp. **35**, e7591 (2023).
335. V. Rai, G. Gupta, S. Joshi, et al., "Lstm-based adaptive whale optimization model for classification of fused multimodality medical image," Signal, Image Video Process. **17**, 2241–2250 (2023).
336. Z. Marinov, S. Reiß, D. Kersting, et al., "Mirror u-net: Marrying multimodal fission with multi-task learning for semantic segmentation in medical imaging," in Proceedings of the IEEE/CVF International Conference on Computer Vision, (2023), pp. 2283–2293.
337. W. El-Shafai, N. El-Hag, A. Sedik, et al., "An efficient medical image deep fusion model based on convolutional neural networks," Comput. Mater. Contin **74**, 2905–2925 (2023).

338. Y. Zhou, X. Yang, S. Liu, and J. Yin, "Multimodal medical image fusion network based on target information enhancement," IEEE Access **12**, 70851–70869 (2024).
339. C. Lin, Y. Chen, S. Feng, and M. Huang, "A multibranch and multiscale neural network based on semantic perception for multimodal medical image fusion," Sci. Reports **14**, 17609 (2024).
340. S. Moghtaderi, M. Einlou, K. A. Wahid, and K. E. Lukong, "Advancing multimodal medical image fusion: an adaptive image decomposition approach based on multilevel guided filtering," Royal Soc. Open Sci. **11**, rsos–231762 (2024).
341. S. Hassan, Q. Li, A. Yasin, and M. Zubair, "Investigating critical factors useful for healthcare data sharing in a blockchain architecture," pp. 215–242 (2025).
342. J. S. Kim, "Legal issues in protecting and utilitizing medical data in united states-focused on hipaa/hitech, 21st century cures act, common law, guidance," The Korean Soc. Law Med. **22**, 117–157 (2021).
343. P. Voigt and A. Von dem Bussche, "The eu general data protection regulation (gdpr)," A practical guide, 1st ed., Cham: Springer Int. Publ. **10**, 10–5555 (2017).
344. D. Lahat, T. Adalỳ, and C. Jutten, "Challenges in multimodal data fusion," in 2014 22nd European Signal Processing Conference (EUSIPCO), (IEEE, 2014), pp. 101–105.
345. J. Liu, X. Cen, C. Yi, et al., "Challenges in ai-driven biomedical multimodal data fusion and analysis," Genom. Proteom. & Bioinform. p. qzaf011 (2025).
346. Z. Liu, H. Yin, Y. Chai, and S. X. Yang, "A novel approach for multimodal medical image fusion," Expert systems with applications **41**, 7425–7435 (2014).
347. C. D. Reddy, "Data sharing principles," in Intelligence-Based Cardiology and Cardiac Surgery, (Elsevier, 2024), pp. 335–343.
348. A. Badano, C. Revie, A. Casertano, et al., "Consistency and standardization of color in medical imaging: a consensus report," J. digital imaging **28**, 41–52 (2015).
349. M. Owais, M. Zubair, L. Seneviratne, et al., "Unified synergistic deep learning framework for multimodal 2-d and 3-d radiographic data analysis: Model development and validation," IEEE Access (2024).
350. D. Lahat, T. Adali, and C. Jutten, "Multimodal data fusion: an overview of methods, challenges, and prospects," Proc. IEEE **103**, 1449–1477 (2015).
351. C. Hao and D. Chen, "Software/hardware co-design for multi-modal multi-task learning in autonomous systems," in 2021 IEEE 3rd International Conference on Artificial Intelligence Circuits and Systems (AICAS), (IEEE, 2021), pp. 1–5.
352. S. Hassan, Q. Li, M. Zubair, et al., "Unveiling the correlation between nonfunctional requirements and sustainable environmental factors using a machine learning model," Sustainability **16**, 5901 (2024).
353. J. Lipkova, R. J. Chen, B. Chen, et al., "Artificial intelligence for multimodal data integration in oncology," Cancer cell **40**, 1095–1110 (2022).

**Highlights:**

- An in-depth review of traditional and advanced multi-modal image fusion techniques.
- Discusses fusion techniques for enhanced diagnosis across multiple organ systems.
- Highlights fusion applications in oncology, neurology, and cardiology imaging.
- Reviews imaging modalities and their roles in enhancing diagnostic accuracy.
- Explores fusion algorithms, recent advances, and key image fusion challenges.